\documentstyle[12pt,epsfig]{report}
\def\Tr{{\rm Tr}}
\def\ln{{\rm ln}}

\def\d {\partial}

\def\pr{Phys. Rev. }
\def\np{Nucl. Phys. }
\def\pl{Phys. Lett. }
\def\zp {Z. Phys. }

\def\dd{\not \hspace{-1.5mm} D}
\def\dpa{\not \hspace{-0.9mm} \partial}

\def\ddp{\not \hspace{-0.6mm} p}
\def\ddq{\not \hspace{-0.6mm} q}
\def\dda{\not \hspace{-1.5mm} A}
\def\dq{\not \hspace{-1mm} q}
\def\dP{\not \hspace{-1mm} p}
\def\beq{\begin{eqnarray}}
\def\eeq{\end{eqnarray}}
\def\jj{J_{J+1/2}}
\def\lan{\langle}
\def\ran{\rangle}

\newcommand{\la}[1]{\label{#1}}

\newcommand{\nn}{\nonumber}
\newcommand{\ed}{\end{document}}
\newcommand{\bq}{\begin{equation}}
\newcommand{\eq}{\end{equation}}
\newcommand{\ba}{\begin{eqnarray}}
\newcommand{\ea}{\end{eqnarray}}
\newcommand{\baz}{\begin{eqnarray*}}
\newcommand{\eaz}{\end{eqnarray*}}
\newcommand{\bb}{\begin{thebibliography}}
\newcommand{\eb}{\end{thebibliography}}
\newcommand{\ct}[1]{${\cite{#1}}$}

\newcommand{\bi}[1]{\bibitem{#1}}

\newcommand{\lsim}{\stackrel{<}{\sim}}
\newcommand{\ben}{\begin{enumerate}}
\newcommand{\een}{\end{enumerate}}

\begin{document}
\baselineskip 0.7cm

\thispagestyle{empty}
\begin{titlepage}

\begin{center}
\vspace*{0.5cm}
\huge
{\bf The Proton Spin Problem\\
     in the Chiral Bag Model}
\vfill

\Large Hee-Jung Lee \vfill

\large
Supervised by \\
Professor Dong-Pil Min \vfill

\large
A Dissertation \\
Submitted to the Faculty of \\
Seoul National University\\
in Partial Fulfillment of \\
the Requirements for the Degree of \\
Doctor of Philosophy\\
Febrary 2002 \vfill

\large
Department of Physics \\
Graduate School \\
Seoul National University
\vfill

\vspace*{1cm}
\end{center}

\end{titlepage}


\pagenumbering{roman}
\setcounter{page}{1}
\chapter*{Abstract}
\addcontentsline{toc}{chapter}{Abstract}
The flavor singlet axial charge has been a source of study in the
last years due to its relation to the so called {\it Proton Spin
Problem}. The relevant flavor singlet axial current is anomalous,
i.e., its divergence contains a piece which is the celebrated
$U_A(1)$ anomaly. This anomaly is intimately associated with the
$\eta^\prime$ meson, which gets its mass from it. When the gauge
degrees of freedom of QCD are confined within a volume as is
presently understood, the $U_A(1)$ anomaly is known to induce
color anomaly leading to ``leakage" of the color out of the
confined volume (or bag). For consistency of the theory, this
anomaly should be cancelled by a boundary term. This ``color
boundary term" inherits part or most of the dynamics of the volume
(i.e., QCD). In this thesis, we exploit this mapping of the volume
to the surface via the color boundary condition to perform a
complete analysis of the flavor singlet axial charge in the chiral
bag model using the Cheshire Cat Principle. This enables us to
obtain the hitherto missing piece in the axial charge associated
with the gluon Casimir effect. The result is that the flavor
singlet axial charge is small independent of the confinement (bag)
size ranging from the skyrmion picture to the MIT bag picture,
thereby confirming the (albeit approximate) Cheshire Cat
phenomenon.
\\
\\
\\
{\bf Key words} : Flavor singlet axial charge, proton, spin,
$U_A(1)$ anomaly, $\eta'$ meson, gluon, Casimir effect
\\
\noindent{\bf Student number} : 95303-823


\tableofcontents
\listoffigures
\listoftables

\newpage
\pagenumbering{arabic}
\setcounter{page}{1}

\chapter{Introduction}

The constituent quark was proposed to explain the structure of the
large number of hadrons being discovered in the sixties
\cite{Gell-Mann}. Soon thereafter deep inelastic scattering of
leptons off protons was explained in terms of point-like
constituents named partons \cite{Feynman}. The analysis of the
data by means of sum rules led to the conclusion that there  was
an intimate relation between the partons and the elementary
quarks. Various models have been developed to understand the
structures of light hadrons and their interactions in terms of
quarks \cite{quarks}. They were built on the basis of low-energy
hadron phenomenologies, particularly, (i) approximate $SU(3)$
flavor symmetry and its explicit breaking, (ii) Okubo-Zweig-Izuka
(OZI) suppression rule for flavor changing processes \cite{OZI},
and (iii) chiral symmetry realization with its spontaneous
breaking pattern.

The birth of Quantum Chromodynamics (QCD) and the proof that it is
asymptotically free set the framework for an understanding of deep
inelastic phenomena beyond the parton model \cite{QCD}. However
the fact that QCD confines does not allow a solution of the theory
in the strong coupling regime and therefore new models had to be
developed to describe hadron structure which realized the
phenomenological principles mentioned before but in a manner
compatible with the  dynamical principles of the theory. This
scheme of confined isolated quarks and gluons, has a strong
relation to the valence quark model for hadrons \cite{models}. The
valence quark model was developed further to a non-relativistic
quark model by De R\'ujula, Georgi and Glashow  \cite{rgg}
initially and exploited phenomenologically by Isgur and Karl
\cite{IsgurKarl}, and into a relativistic quark model framework by
Chodos et al. and De Grand et al. known under the name of MIT bag
model \cite{MIT}.

Although the MIT bag model was successful in describing the
properties of the nucleons, it was not pertinent due to lack of
the chiral symmetry. In order to incorporate chiral symmetry into
the model, the pseudoscalar mesons had to be introduced in this
framework \cite{Chodos75}. The resulting scheme, the so called
chiral bag model, was constructed with  meson fields which are
restricted to be outside the bag \cite{Brown79}.

Chiral symmetry, a property of QCD with massless quarks, has been
instrumental in the description of hadron phenomenology. So much
so, that Skyrme realized its importance and wrote down, much
before QCD, an effective theory for the strong interactions, in
terms of pion fields only, describing a unified theory for baryons
and mesons \cite{Skyrme}. Only many years later his tremendous
intuition was appreciated and his ideas justified from the point
of view of QCD \cite{witten}.

The chiral bag model incorporates in a unified description the
statements above. It is defined by means of a QCD lagrangian
inside the bag and by a Skyrme type theory outside, properly
matched at the surface to preserve the classical and quantum
symmetries.  These formulation leads to a intriguing principle
referred to as the Chesire Cat Principle (CCP) \ct{ccp,ccpph}. The
possibility of formulating a physical theory by means of
equivalent field theories defined in terms of different field
variables, leads to this construction principle for
phenomenologically sensible and conceptually powerful models. This
principle states, that physical observables obtained by means of
equivalent theories defined in a certain space-time geometry,
adequately matched at the boundaries, are independent of the
geometry. In 1+1 dimensions fermionic theories are bosonizable
\ct{two} and the CCP can be made exact and transparent. In the
real four-dimensional world, bosonization with a finite number of
degrees of freedom is not exact. However based on the unproven
``theorem" of Weinberg \ct{wei}, it seems possible to argue that
the CCP should hold also in four dimensions, albeit approximately.

Quantum Chromodynamics (QCD) is the theory of the hadronic
phenomena \ct{QCD}. At sufficiently low energies or long distances
and for a large number of colors $N_C$, it can be described
accurately by an effective field theory in terms of meson fields
\ct{witten,mesons}. In this regime, the color fermionic
description of the theory is extremely complex due to confinement.
However the implementation of the CCP in a two phase scenario
called the Chiral Bag Model (CBM) has proven surprisingly powerful
\ct{cbm}. The CBM is defined by dividing  space-time  in two
regions by a hypertube, that is, the evolving bag. In the interior
of the tube, the dynamics is defined in terms of the microscopic
QCD degrees of freedom, quarks and gluons. In the exterior, one
assumes an equivalent dynamics in terms of meson fields, i.e., one
that respects the symmetries of the original theory and the basic
postulates of quantum field theory \ct{wei}. The two descriptions
are matched by defining the appropriate boundary conditions which
implement the symmetries and confinement \ct{ccp,cbm}. What this
does effectively is to delegate all or part of the principal
elements of the dynamics taking place inside (QCD) the bag to the
boundary. We will see that this strategy works quite efficiently
in the problem at hand. In this scenario the CCP states that the
hadron physics should be approximately independent of the spatial
size of the confinement region or the bag \ct{ccp}. This
realization of the principle has been tested in many instances in
hadronic physics with fair success \ct {ccpph}.

There is one case, however, where the realization of the CCP has
not been as successful as in the other cases, namely, the
calculation of the flavor singlet axial charge (FSAC) of the
nucleon. Indeed in the previous results \ct{pvrb,pv,rv}, the CCP
was realized only partially as it seemed to fail at certain points
such as for zero bag radius. It is the leitmotiv of this work to
remove this apparent failure.\vskip 1ex

Experiments using polarized electrons on polarized targets were
carried out at SLAC \cite{Algard}. Further information came from
the SLAC-Yale group with fascinating implications about the
internal structure of the proton \cite{Baum}. More recently, the
European Muon Collaboration (EMC) obtained very extraordinary
results by the scattering of a polarized muon beam with energy
100-200 GeV on a longitudinally polarized hydrogen target at CERN
\cite{Ashman}. All these results point towards a new scenario in
hadronic structure dominated by a quantum anomaly. To be more
specific, the unexpectedly small asymmetry found by EMC implies a
strong violation of the so-called Ellis-Jaffe sum rule \ct{ej} and
therefore implies that the polarization of the proton is not
carried exclusively by the valence quarks. This problem is called
the {\it Proton Spin Problem} \cite{PSP}.

The EMC result and this problem are now believed to be resolved
through the beautiful relation between the flavor singlet axial
charge and the axial anomaly \ct{abj} \ct{anom}:
\begin{equation}
a^0(Q^2)=\Delta\Sigma-N_F\frac{\alpha_s(Q^2)}{2\pi}\Delta g(Q^2),
\end{equation}
where $a^0(Q^2)$ is the flavor singlet axial charge measured by
EMC, $\Delta\Sigma$ the quark polarization, and $\Delta g(Q^2)$
the gluon polarization.\vskip 1ex

The aim of this thesis is to show the full consistency of the CCP
in the hadronic world for the case of the {\it Proton Spin}, which
was not satisfactorily established in the previous results in this
direction \ct{pvrb,pv,rv}\footnote{Note that in these papers, they
have shown that the CCP holds for non-zero bag radii but it failed
when the bag radius shrank to a point, implying that in the model
studied, the pure skyrmion and the MIT bag did not have the
equivalent structure required by the CCP.}.

In the CBM, the scenario of how the CCP is realized -- which is
the central issue of this thesis -- is very intricate. As stated,
the flavor singlet axial current is associated with the anomaly
and effectively with the $\eta^\prime$ meson. Thus, besides the
pion field of the conventional effective theories which accounts
for spontaneously broken chiral symmetry, the correct treatment of
the flavor singlet axial charge requires minimally the inclusion
of a field describing the $\eta^\prime$ meson.\vskip 1ex

The intricacies of the {\it hedgehog} configuration and its
relevance to the fractionation of baryon charge and other
observables have been extensively discussed \ct{frac} and fairly
well understood \ct{pr,p}. They will be implemented in the present
calculation without much details. Moreover the inclusion of the
$\eta^\prime$ meson carries subtleties of its own. The vacuum
fluctuations inside the bag, that induce the baryon number leakage
into the {\it skyrmion} \ct{frac}, also induce a color leakage if
a coupling to a pseudoscalar isoscalar field is allowed \ct{nrwz}.
This leakage would break color gauge invariance and confinement in
the model unless it is cancelled. As suggested in \ct{nrwz}, this
color leakage can be prevented by introducing into the CBM
Lagrangian a counter term of the form \bq {\cal
L}_{CT}=i\frac{g_s^2}{32\pi^2}\oint_{\Sigma} d\beta\ K^\mu n_\mu
({\Tr}{ \ln} U^\dagger -{\Tr} {\ln} U)\label{lct} \eq where $N_F$
is the number of flavors (here taken to be =3), $\beta$ is a point
on a surface $\Sigma$, $n^\mu$ is the outward normal to the bag
surface, $U$ is the $U(N_F)$ matrix-valued field written as
$U=e^{i\pi/f} e^{i\eta^\prime/f_0}$ and $K^\mu$ the properly
regularized Chern-Simons current
$K^\mu=\epsilon^{\mu\nu\alpha\beta} (G_\nu^a G_{\alpha\beta}^a
-\frac 23 g_s f^{abc} G_\nu^a G_\alpha^b G_\beta^c)$ given in
terms of the color gauge field $G^a_\mu$. Note that the counter
term (\ref{lct}) manifestly breaks color gauge invariance (both
large and small, the latter due to the bag), so the action of the
chiral bag model with this term is not gauge invariant at the
classical level but as shown in \cite{nrwz}, when quantum
fluctuations are calculated, there appears an induced anomaly term
on the surface which exactly cancels this term. Thus gauge
invariance is restored at the quantum level.

The equations of motion for the gluon and quark fields inside and
the $\eta^\prime$ field outside are the same as in \ct{pvrb,pv}.
However the boundary conditions on the surface with the inclusion
of eq.~(\ref{lct}) read \ct{rv} \bq \hat{\bf n}\cdot {\bf
E}^a=-\frac{N_F g_s^2}{8\pi^2 f} \hat{\bf n}\cdot {\bf B}^a
\eta^\prime\label{E} \eq \bq \hat{\bf n}\times {\bf B}^a=\frac{N_F
g_s^2}{8\pi^2 f} \hat{\bf n}\times {\bf E}^a \eta^\prime\label{B}
\eq and \bq \frac 12 \hat{\bf n}\cdot({\bar{\psi}}\mbox{\boldmath
$\gamma$}\gamma_5\psi) =f \hat{\bf n}\cdot\d \eta^\prime +
\frac{N_F g_s^2}{16\pi^2 } \hat{\bf n}\cdot K\label{bc} \eq where
${\bf E}^a$ and ${\bf B}^a$ are, respectively, the color electric
and color magnetic fields. Here $\psi$ is the QCD quark field.

The full Casimir calculation of the gluon modes, which is highly
subtle due to the p-wave structure of the $\eta^\prime$-field, has
to be performed to get the CCP for the flavor singlet axial
charge. Here we would like to side-step this technically difficult
procedure by first assuming the CCP in evaluating the Casimir
contribution with the color boundary conditions (\ref{E}),
(\ref{B}) and (\ref{bc}) taken into account and check a posteriori
that there is consistency between the assumption and the result.
\vskip 1ex

The thesis is organized as follows: In Chapter 2, we introduce the
{\it Proton Spin Problem} via the polarized deep inelastic
scattering experiments and the relation between the spin dependent
structure function, $g_1(x)$, and the flavor singlet axial charge.
A general review of the chiral bag model is given in Chapter 3 for
the next discussion. In Chapter 4, we review the axial anomaly and
present its contribution to the flavor singlet axial charge.
Moreover, we show a derivation of the color anomaly boundary
condition. We address the various static contributions and
calculate the Casimir effect to the flavor singlet axial charge in
Chapter 5. Finally, we discuss our result in Chapter 6.


\chapter{Proton Spin Problem}
Deep inelastic lepton-hadron scattering (DIS) has played an
important role in understanding the internal structure of hadrons.
The discovery of Bjorken scaling in the late nineteen sixties
provided the basis for the idea that hadrons are made up of
point-like constituents. The subsequent development of the Parton
model played an essential role in linking the partons to the
quarks via DIS sum rules. DIS was essential in the discovery of
the missing constituents, identified as gluons, and therefore in
assembling all different pieces of the hadronic puzzle into a
coherent dynamical theory of quarks and gluons, Quantum
Chromodynamics.

Polarized DIS, describes the collision of a longitudinally
polarized lepton beam on a nucleonic target polarized either
longitudinally or transversely to an arbitrary direction. It
provides a  more complete insight into the structure of the
nucleon than unpolarized DIS. Whereas the latter probes the number
density of partons  with a fraction $x$ of the momentum of the
parent nucleon, polarized DIS leads to more sophisticated
information, namely it determines the number density of partons
with given $x$ and given spin projection in a nucleon of definite
polarization.
\footnote{We summarize here the conventions for the Dirac spinors.
With four momentum $p^{\mu}=(E,{\bf p}),$ the Dirac spinors are
normalized as;
$$u^{\dagger}u=2E,\ \ v^{\dagger}v=2E,\ \ \bar{u}u=2M,\ \ \bar{v}v=-2M,$$
for both the massive and massless case. From these, the following
relations can be derived $$\bar{u}(p)\gamma^{\mu}u(p)=2p^{\mu},\
 \ \ \ \bar{v}(p)\gamma^{\mu}v(p)=2p^{\mu}.$$
Incorporating $\gamma_5$ matrix, for a fermion of mass $M$, there
is the relation $$\bar{u}(p,S)\gamma^{\mu}\gamma_5
u(p,S)=-\bar{v}(p,S)\gamma^{\mu}\gamma_5v(p,S)=2MS^{\mu},$$ with
the covariant spin $S^{\mu}$ normalized $S_{\mu}S^{\mu}=-1$.
Additionally, for massless fermion of helicity $\lambda=\pm1/2$,
the above relation is changed to
$$\bar{u}(p,\lambda)\gamma^{\mu}\gamma_5 u(p,
\lambda)=-\bar{v}(p,\lambda)\lambda\gamma_5
v(p,\lambda)=\lim_{M\rightarrow 0} 2MS^{\mu}(\lambda)=4\lambda
p^{\mu}.$$ Moreover, all states are normalized so that $$\langle
P'|P\rangle=(2\pi)^3 2E\ \delta^3({\bf p'}-{\bf p}).$$}

In this chapter, we give a short review of polarized DIS, show the
relation between the spin dependent structure function, $g_1(x)$,
and the flavor singlet axial current, and discuss some relevant
facts about the proton spin problem.

\section{Polarized Deep Inelastic Scattering} In the laboratory
frame the differential cross section for the polarized
lepton-nucleon scattering has the form
\begin{equation}
\frac{d^2\sigma}{d\Omega
dE'}=\frac{1}{2M}\frac{\alpha^2}{q^4}\frac{E'}{E}
L_{\mu\nu}W^{\mu\nu}, \label{sigma}
\end{equation}
where the four momenta of the incoming and the outgoing lepton
with mass $m$ are $k=(E,{\bf k})$ and $k'=(E',{\bf k}')$,
respectively, and the four momentum for the nucleon is $P=(M,0)$.
The momentum transfer is  $q=k-k'$ and $\alpha$ is the fine
structure constant. In eq.~(\ref{sigma}) the leptonic tensor
$L_{\mu\nu}$ is given by
\begin{equation}
L_{\mu\nu}=[\bar{u}(k',s')\gamma_{\mu}u(k,s)]^*[\bar{u}(k',s')\gamma_{\nu}
u(k,s)],
\end{equation}
where $s\ (s')$ is the spin four vector of the incoming(outgoing)
lepton such that $s\cdot k=0=s'\cdot k'$ and $s\cdot s=-1=s'\cdot
s'$. $L_{\mu\nu}$ can be decomposed into symmetric ($S$) and
antisymmetric ($A$) parts under ${\mu},{\nu}$ interchange;
\begin{eqnarray}
L_{\mu\nu}(k,s;k's')&=&L_{\mu\nu}^S(k,k')+iL_{\mu\nu}^A(k,s;k')\nonumber\\
&&+L_{\mu\nu}^{\prime S}(k,s;k',s')+iL_{\mu\nu}^{\prime
A}(k;k',s').
\end{eqnarray}
Explicitly they become
\begin{eqnarray}
L_{\mu\nu}^S(k,k')
&=&k_{\mu}k'_{\nu}+k'_{\mu}k_{\nu}-g_{\mu\nu}(k\cdot k'-m^2),
\nonumber\\
L_{\mu\nu}^A(k,s;k')&=&m\epsilon_{\mu\nu\alpha\beta}s^{\alpha}(k-k')^{\beta},\nonumber\\
L_{\mu\nu}^{\prime S}(k,s;k',s')&=&(k\cdot
s')(k'_{\mu}s_{\nu}+s_{\mu}k'_{\nu}
-g_{\mu\nu}k'\cdot s)\nonumber\\
&&-(k\cdot
k'-m^2)(s_{\mu}s'_{\nu}+s'_{\mu}s_{\nu}-g_{\mu\nu}s\cdot s')
\nonumber\\
&&+(k'\cdot s)(s'_{\mu}k_{\nu}+k_{\mu}s'_{\nu}-(s\cdot
s')(k_{\mu}k'_{\nu} +k'_{\mu}k_{\nu}), \label{L123}
\end{eqnarray}
and
\begin{equation}
L_{\mu\nu}^{\prime
A}=m\epsilon_{\mu\nu\alpha\beta}s'^{\alpha}(k-k')^{\beta}
\label{L4}
\end{equation}
where $m$ is the lepton mass \cite{Anselmino95}. Summation over
$s'$ leads to $2L_{\mu\nu}^S+2iL_{\mu\nu}^A$ . Summation of
$L_{\mu\nu}$ over $s'$ and averaging  over $s$ gives the
unpolarized leptonic tensor, $2L_{\mu\nu}^S$.

Due to the internal structure of hadrons, the hadronic tensor
$W^{\mu\nu}$ is unknown and is defined in terms of four structure
functions as \cite{Dre64Bjo66Car72Hey72} \cite{Manohar};
\begin{equation}
W_{\mu\nu}(q;P,S)=W_{\mu\nu}^S(q;P)+iW_{\mu\nu}^A(q;P,S)
\end{equation}
with
\begin{eqnarray}
\frac{1}{2M}W_{\mu\nu}^S&=&\bigg(-g_{\mu\nu}+\frac{q_{\mu}q_{\nu}}{q^2}\bigg)
W_1(p\cdot q,q^2)\nonumber\\
&&+\left[\bigg(P_{\mu}-\frac{P\cdot q}{q^2}q_{\mu}\bigg)
\bigg(P_{\nu}-\frac{P\cdot q}{q^2}q_{\nu}\bigg)\right]
\frac{W_2(P\cdot q,q^2)}{M^2},\\
\frac{1}{2M}W_{\mu\nu}^A&=&\epsilon_{\mu\nu\alpha\beta}q^{\alpha}\bigg[
MS^{\beta}G_1(P\cdot q,q^2)\nonumber\\
&&+\{(P\cdot q)S^{\beta}-(S\cdot
q)P^{\beta}\} \frac{G_2(P\cdot q, q^2)}{M}\bigg], \label{W}
\end{eqnarray}
where $S^{\mu}$ is the spin four vector of the nucleon. With these
structures eq.~(\ref{sigma}) becomes
\begin{equation}
\frac{d^2\sigma}{d\Omega
dE'}=\frac{1}{2M}\frac{\alpha^2}{q^4}\frac{E'}{E}
\bigg[L_{\mu\nu}^SW^{\mu\nu,S}+L_{\mu\nu}^{\prime S}W^{\mu\nu,S}
-L_{\mu\nu}^AW^{\mu\nu,A}-L_{\mu\nu}^{\prime A}W^{\mu\nu,A}\bigg].
\end{equation}

The individual terms inside the square brackets can be separately
studied by considering cross-sections or differences between
cross-sections with particular initial and final polarizations.
Each of these terms is an observable quantity in terms of the
spin-averaged structure functions $W_1,\ W_2$ and of the
spin-dependent structure functions $G_1,\ G_2$. For example, while
the unpolarized cross-section contains only
$L_{\mu\nu}^SW^{\mu\nu,S}$
\begin{eqnarray}
\frac{d^2\sigma^{\rm unp}}{d\Omega dE'}=\frac{1}{4}\sum_{s,s',S}
\frac{d^2\sigma}{d\Omega dE'}
=\frac{1}{2M}\frac{\alpha^2}{q^4}\frac{E'}{E}\
2L_{\mu\nu}^SW^{\mu\nu,S},
\end{eqnarray}
the difference of cross-section with opposite target spins
contains $L_{\mu\nu}^AW^{\mu\nu,A}$
\begin{eqnarray}
&&\sum_{s'}\left[\frac{d^2\sigma}{d\Omega dE'}(k,s,P,-S;k',s')
-\frac{d^2\sigma}{d\Omega dE'}(k,s,P,S;k',s')\right]\nonumber\\
&&\hspace{2cm}=\frac{1}{2M}\frac{\alpha^2}{q^4}\frac{E'}{E}\
4L_{\mu\nu}^AW^{\mu\nu,A}. \label{sums'}
\end{eqnarray}

In the laboratory frame, the cross-section for the inelastic
scattering of an unpolarized leptons on an unpolarized nucleon,
can be written explicitly as
\begin{equation}
\frac{d^2\sigma^{\rm unp}}{d\Omega dE'}=\frac{4\alpha^2E'^2}{q^4}
\bigg(2W_1\sin^2\frac{\theta}{2}+W_2\cos^2\frac{\theta}{2}\bigg),
\label{1.14}
\end{equation}
where the lepton mass has been neglected. Here $\theta$ is the
scattering angle of the lepton. This cross section provides
information on the unpolarized structure functions $W_1(P\cdot
q,q^2)$ and $W_2(P\cdot q,q^2)$. In the deep inelastic scattering
regime, the Bjorken limit is defined by
\begin{equation}
-q^2=Q^2\rightarrow\infty,\hspace{1cm}\nu=E-E'\rightarrow\infty,\hspace{1cm}
x=\frac{Q^2}{2P\cdot q}=\frac{Q^2}{2M\nu}={\rm fixed}, \label{x}
\end{equation}
and the structure functions obey, so called, scaling for fixed $x$
\cite{Bjo69};
\begin{eqnarray}
\lim_{Q^2\rightarrow\infty}MW_1(P\cdot q,Q^2)&=&F_1(x),\nonumber\\
\lim_{Q^2\rightarrow\infty}\nu W_2(P\cdot q,Q^2)&=&F_2(x),
\label{1.16}
\end{eqnarray}

Similarly, from eq.~(\ref{sums'}), eq.~(\ref{L123}), and
eq.~(\ref{L4}), the difference of the cross-sections with opposite
target polarization can be written as;
\begin{eqnarray}
&&\frac{d^2\sigma^{s,S}}{d\Omega
dE'}-\frac{d^2\sigma^{s,-S}}{d\Omega dE'}
\nonumber\\
&\equiv&\sum_{s'}\left[\frac{d^2\sigma}{d\Omega
dE'}(k,s,P,S;k',s')
-\frac{d^2\sigma}{d\Omega dE'}(k,s,P,-S;k',s')\right]\nonumber\\
&=&\frac{8m\alpha^2}{q^4}\frac{E'}{E}\bigg[\bigg(q\cdot S)(q\cdot
s)+Q^2(s\cdot S)\bigg)MG_1\nonumber\\
&&\hspace{2cm}+Q^2\bigg((s\cdot S)(P\cdot q)-(q\cdot S)(P\cdot
s)\bigg) \frac{G_2}{M}\bigg]. \label{sigma-sigma}
\end{eqnarray}
This expression supplies information on the polarized structure
functions $G_1(P\cdot q,q^2)$ and $G_2(P\cdot q,q^2)$. In the
Bjorken limit, they are also known to obey the scaling,
\begin{eqnarray}
\lim_{Q^2\rightarrow\infty}\frac{(P\cdot q)^2}{\nu}G_1(P\cdot
q,Q^2)&=&g_1(x),
\nonumber\\
\lim_{Q^2\rightarrow\infty}\nu(P\cdot q)G_2(P\cdot
q,Q^2)&=&g_2(x).
\end{eqnarray}
In terms of $g_{1,2}$ the expression for $W_{\mu\nu}^A$ can be
written as
\begin{equation}
W_{\mu\nu}^A=\frac{2M}{P\cdot
q}\epsilon_{\mu\nu\alpha\beta}q^{\alpha}
 \left[S^{\beta}g_1(x,Q^2)+\bigg(S^{\beta}-\frac{S\cdot q}{P\cdot q}\ P^{\beta}
g_2(x,Q^2)\bigg)\right].
\end{equation}

\begin{figure}
\centerline{\epsfig{file=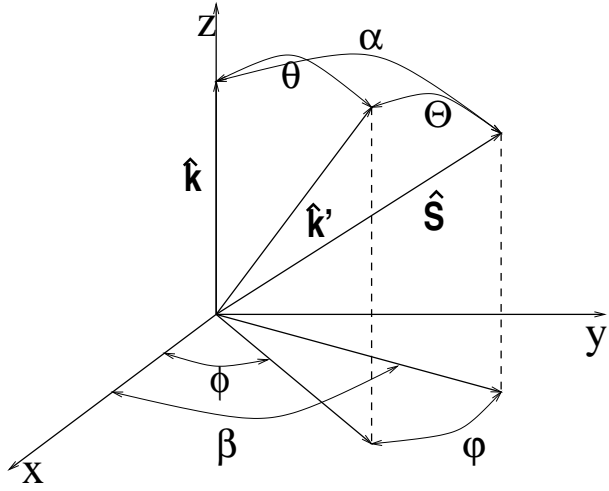,width=8cm}} \caption{The
angles defining the kinematical and spin variables of the studied
polarized cross section are shown}\label{fig1-xyz}
\end{figure}

To get information on the polarized structure functions $G_1,\
G_2$, we need to look at eq.~(\ref{sigma-sigma}) with particular
spin configurations of the incoming leptons and the target
nucleons. We consider firstly the case of longitudinally polarized
leptons. The symbol, $\rightarrow(\leftarrow)$, denotes the spin
of the initial lepton along (opposite) to the direction of motion
and the nucleons at rest are polarized along ($S$) or opposite
($-S$) to an arbitrary direction $\hat{\bf S}$;
$$s^{\mu}_{\rightarrow} = -s^{\mu}_{\leftarrow}=\frac{1}{m}(|{\bf k}|,
\hat{\bf k}E),$$ with $\hat{\bf k}=\frac{{\bf k}}{|{\bf k}|}$, and
\begin{equation}
S^{\mu} = (0,\hat{\bf S}).
\end{equation}
Choosing the $z$-axis along the incoming lepton direction, we have
\begin{eqnarray}
k^{\mu}&=&(E,0,0,|{\bf k}|)\simeq E(1,0,0,1),\nonumber\\
k'^{\mu}&=&(E',{\bf k}')\simeq E'(1,\hat{\bf k'})\nonumber\\
&=&E'(1, \sin\theta\cos\phi, \sin\theta\sin\phi, \cos\theta),\nonumber\\
\hat{\bf S}&=&(\sin\alpha\cos\beta, \sin\alpha\sin\beta,
\cos\alpha).
\end{eqnarray}
This kinematical scenario is depicted in Fig.~\ref{fig1-xyz}

Substituting these vectors into eq.~(\ref{sigma-sigma}) yields
\begin{eqnarray}
&&\frac{d^2\sigma^{\rightarrow,S}}{d\Omega dE'}
-\frac{d^2\sigma^{\rightarrow,-S}}{d\Omega
dE'}=-\frac{4\alpha^2}{Q^2}
\frac{E'}{E}\nonumber\\
&&\times\left[(E\cos\alpha+E'\cos\Theta)MG_1+2EE'(\cos\Theta-\cos\alpha)G_2
\right] \label{rightS}
\end{eqnarray}
where $\Theta$ is the angle between the outgoing lepton direction,
$\hat{\bf k'}$, and $\hat{\bf S}$;
\begin{eqnarray}
\cos\Theta&=&\sin\theta\cos\phi\sin\alpha\cos\beta+\sin\theta\sin\phi\sin\alpha
\sin\beta+\cos\theta\cos\alpha\nonumber\\
&=&\sin\theta\sin\alpha\cos\varphi+\cos\theta\cos\alpha.
\end{eqnarray}
Here $\varphi$ is given by $\varphi=\beta-\phi$. For nucleons
polarized along ($\Rightarrow$) the initial lepton direction of
motion  or opposite ($\Leftarrow$) to it, that is, $\alpha=0,
\Theta=\theta$, eq.~(\ref{rightS}) gives
\begin{equation}
\frac{d^2\sigma^{\rightarrow,\Rightarrow}}{d\Omega dE'}
-\frac{d^2\sigma^{\rightarrow,\Leftarrow}}{d\Omega dE'}
=-\frac{4\alpha^2}{Q^2}\frac{E'}{E}\left[(E+E'\cos\theta)MG_1-Q^2G_2\right].
\label{1.24}
\end{equation}
For transversely polarized nucleons, that is, when the spin of the
nucleon is perpendicular to the direction of the incoming lepton,
$\alpha=\pi/2$, and eq.~(\ref{rightS}) yields
\begin{equation}
\frac{d^2\sigma^{\rightarrow,\Uparrow}}{d\Omega dE'}
-\frac{d^2\sigma^{\rightarrow,\Downarrow}}{d\Omega dE'}
=-\frac{4\alpha^2}{Q^2}\frac{E'^2}{E}\sin\theta\cos\varphi(MG_1+2EG_2).
\label{1.25}
\end{equation}
In the case of $\varphi=\pi/2$, which corresponds to the nucleon
spin being perpendicular to both vectors $\hat{\bf k}$ and $\
\hat{\bf k'}$, the difference of cross-sections, eq.~(\ref{1.25}),
vanishes. The value of such a difference has a maximum when
$\varphi=0,\ {\rm or}\ \pi,$ that is, when the nucleon spin
vector, which is perpendicular to $\hat{\bf k}$, lies in the plane
determined by the two vectors $\hat{\bf k}$ and $\hat{\bf k'}$.

Experimentally, the polarized structure functions $g_1$ and $g_2$
are determined by measuring two asymmetries
\begin{equation}
A_{\|}=\frac{d\sigma^{\rightarrow,\Leftarrow}-d\sigma^{\rightarrow,\Rightarrow}}
{d\sigma^{\rightarrow,\Leftarrow}+d\sigma^{\rightarrow,\Rightarrow}},\hspace{1cm}
A_{\perp}=\frac{d\sigma^{\rightarrow,\Downarrow}-d\sigma^{\rightarrow,\Uparrow}}
{d\sigma^{\rightarrow,\Downarrow}+d\sigma^{\rightarrow,\Uparrow}},
\end{equation}
where the abbreviation $d\sigma$ for $d^2\sigma/d\Omega dE'$ has
been introduced.

Using the fact that the denominator is simply twice the
unpolarized cross-section, from eq.~(\ref{1.14}),
eq.~(\ref{1.24}), and eq.~(\ref{1.25}), the asymmetries become
\begin{eqnarray}
A_{\|}&=&\frac{Q^2[(E+E'\cos\theta)MG_1-Q^2G_2]}
{2EE'[2W_1\sin^2\frac{\theta}{2}+W_2\cos^2\frac{\theta}{2}]},\nonumber\\
A_{\perp}&=&\frac{Q^2\sin\theta(MG_1+2EG_2)}
{2E[2W_1\sin^2\frac{\theta}{2}+W_2\cos^2\frac{\theta}{2}]}\cos\varphi.
\end{eqnarray}
It is convenient to write the asymmetries $A_{\|}$ and $A_{\perp}$
in terms of the virtual Compton scattering asymmetries $A_{1,2}$
given by \cite{models}
\begin{equation}
A_1=\frac{\sigma_{1/2}-\sigma_{3/2}}{\sigma_{1/2}+\sigma_{3/2}},\hspace{1cm}
A_2=\frac{2\sigma^{TL}}{\sigma_{1/2}+\sigma_{3/2}},
\end{equation}
where $\sigma_{1/2}$ and $\sigma_{3/2}$ are the virtual photon
absorption cross sections for $\gamma^*(1)+N(-\frac{1}{2})$ and
$\gamma^*(1)+N(\frac{1}{2})$ scatterings, respectively, and
$\sigma^{TL}$ is the cross section for the interference between
transverse and longitudinal virtual photon-nucleon scatterings.
The asymmetries $A_{1,2}$ have the bounds
\begin{equation}
|A_1|\leq 1,\hspace{1cm}|A_2|\leq \sqrt{R},
\end{equation}
where $R$ is the ratio of the longitudinal to transverse cross
section, $R\equiv \sigma_L/\sigma_T$, with
$\sigma_T\equiv(\sigma_{1/2}+\sigma_{3/2})/2$. The asymmetries can
be written in terms of $A_{1,2}$ as
\begin{equation}
A_\|=D(A_1+\eta A_2),\hspace{1cm} A_{\perp}=D(A_2-\xi A_1),
\end{equation}
where $D$ is a depolarization factor of the virtual photon, $\eta$
and $\xi$ depend only on kinematic variables \cite{Anselmino95}.
The asymmetries $A_{1,2}$ in the virtual photon-nucleon scattering
have relation to the polarized structure functions $g_1$ and
$g_2$;
\begin{equation}
A_1=\frac{g_1-\gamma^2g_2}{F_1},\hspace{1cm}
A_2=\frac{\gamma(g_1+g_2)}{F_1}
\end{equation}
with $\gamma\equiv Q/\nu=Q/(E-E')=2Mx/\sqrt{Q^2}$ and $F_1$ in
eq.~(\ref{1.16}). Since in the Bjorken limit $\gamma$ goes to
zero, one obtains
\begin{equation}
g_1(x,Q^2)\simeq
F_1(x,Q^2)\frac{A_{\|}}{D}=\frac{F_2(x,Q^2)}{2x(1+R(x,Q^2))}
\frac{A_{\|}}{D}, \label{g_10}
\end{equation}
where $F_2(x,Q^2)$ is the unpolarized structure function in the
scaling regime and $R$ is given in terms of the unpolarized
structure functions
\begin{equation}
R=\frac{W_2}{W_1}\bigg(1+\frac{\nu^2}{Q^2}\bigg)-1.
\end{equation}
Note that the last result of the eq.~(\ref{g_10}) is from the fact
that $R$ can be written in terms of the Bjorken scaling functions,
\begin{equation}
R=\frac{F_2(x)}{2xF_1(x)}-1.
\end{equation}
in the limit $\frac{4M^2x^2}{Q^2}\rightarrow 0$.

Experimental results on the polarized structure functions $g_1(x)$
for the nucleon can be found in the Table~\ref{table}.
\footnote{We have quoted this table from ref.~\cite{Cheng96}.}
\begin{table}
\caption{Experiments on the polarized structure functions
$g_1^p(x,Q^2),~g_1^n(x,Q^2)$ and $g_1^d(x,Q^2)$.}\label{table}
{\footnotesize\begin{center}
\begin{tabular}{|c|c|c|c|c||c|c|} \hline
Exper. & Year & Target & $\lan Q^2\ran$ & $x$ range & $\Gamma^{\rm
target}_1$ & Ref.  \\
  & & & (GeV$^2$) & & $=\int^1_0g_1^{\rm target}(x,\lan Q^2\ran)dx$ & \\ \hline
E80/E130 & 1976/1983 & $p$ & $\sim 5$ & $0.1<x<0.7$ & $0.17\pm
0.05^*$ &
\cite{E80,E130} \\
EMC & 1987 & $p$ & 10.7 & $0.01<x<0.7$ & $0.126\pm 0.010\pm
0.015^\dagger$ &
\cite{Ashman} \\
SMC & 1993 & $d$ & 4.6 & $0.006<x<0.6$ & $0.023\pm 0.020\pm 0.015$
&
\cite{SMC93}  \\
SMC & 1994 & $p$ & 10 & $0.003<x<0.7$ & $0.136\pm 0.011\pm 0.011$
&
\cite{SMC94} \\
SMC & 1995 & $d$ & 10 & $0.003<x<0.7$ & $0.034\pm 0.009\pm 0.006$
&
\cite{SMC95} \\
E142 & 1993 & $n$ & 2 & $0.03<x<0.6$ & $-0.022\pm 0.011$ & \cite{E142}  \\
E143 & 1994 & $p$ & 3 & $0.03<x<0.8$ & $0.127\pm 0.004\pm 0.010$ &
\cite{E143a} \\
E143 & 1995 & $d$ & 3 & $0.03<x<0.8$ & $0.042\pm 0.003\pm 0.004$ &
\cite{E143b} \\   \hline
\end{tabular}
\end{center}}
{\footnotesize \noindent $^*$ Obtained by assuming a Regge
behavior $A_1\propto x^{1.14}$
for small $x$.   \\
\noindent $^\dagger$ Combined result of E80, E130 and EMC data.
The EMC data alone give $\Gamma_1^p=0.123\pm 0.013\pm 0.019\,$.}
\end{table}
\vskip 0.45cm
\section{The Structure Functions in the Parton Model} In the
parton model the nucleon is regarded as a collection of almost
free constituents, namely the partons, each carrying a fraction
$x'$ of the nucleon four momentum. Lepton-nucleon DIS can be
understood as the incoherent sum of scatterings between the lepton
and the spin-1/2 partons \cite{models} \cite{ChengLi}. We shall
assume for our description that the charged partons are quarks and
antiquarks, a statement which was proven historically a posteriori
by studying the experimental structure function sum rules. The
hadronic tensor $W_{\mu\nu}$ can be obtained in terms of the
elementary quark tensor $w_{\mu\nu}$ as;
\begin{eqnarray}
W(q;P,S)&=&W_{\mu\nu}^S(q;P)+iW_{\mu\nu}^A(q;P,S)\nonumber\\
&=&\sum_{q,s}e_q^2\frac{1}{2P\cdot q}\int_0^1\frac{dx'}{x'}
\delta(x'-x)n_q(x',s;S)w_{\mu\nu}(x',q,s),
\end{eqnarray}
where $n_q(x',s;S)$ is the number density of quarks with charge
$e_q$. Here $s$ is the spin of the quarks inside a nucleon with
the spin $S$ and four momentum $P$, the sum runs over quarks and
antiquarks, and $x$ is the Bjorken variable given in
eq.~(\ref{x}). The quark tensor has the same form as the leptonic
tensor, eq.~(\ref{L123}) and eq.~(\ref{L4}), with the replacements
$k^{\mu}\rightarrow xP^{\mu}$ and  $k^{\prime\mu}\rightarrow
xP^{\mu}+q^{\mu}$. After summation over the unobserved final quark
spin, $w_{\mu\nu}$ becomes
\begin{equation}
w_{\mu\nu}(x,q,s)=w_{\mu\nu}^S(x,q)+iw_{\mu\nu}^A(x,q,s)
\end{equation}
with the quantities
\begin{eqnarray}
w_{\mu\nu}^S(x,q)&=&2[2x^2P_{\mu}P_{\nu}+xP_{\mu}q_{\nu}+xq_{\mu}P_{\nu}
-x(P\cdot q)g^{\mu\nu}]\nonumber\\
w_{\mu\nu}^A(x,q,s)&=&-2m_q\epsilon_{\mu\nu\alpha\beta}s^{\alpha}q^{\beta},
\end{eqnarray}
where the quark mass has been taken to be $m_q=xM$ for
consistency.

Comparing these equations with the definition of the structure
functions eq.~(\ref{W}), the unpolarized structure functions
become
\begin{eqnarray}
F_1(x)&=&\frac{1}{2}\sum_q e^2_q q(x)\nonumber\\
F_2(x)&=&x\sum_q e^2_ p(x)=2xF_1(x),
\end{eqnarray}
where the unpolarized quark density is defined by
\begin{equation}
q(x)=\sum_s n_q(x,s;S).
\end{equation}
Similarly the polarized structure functions are obtained as
\begin{eqnarray}
g_1(x)&=&\frac{1}{2}\sum_q e^2_q\Delta q(x,S),\nonumber\\
g_2(x)&=&0,
\end{eqnarray}
where $\Delta q(x,S)$ is the difference between the number density
of quarks with the spin parallel ($s=S$) to the nucleon spin and
those with the spin anti-parallel ($s=-S$);
\begin{equation}
\Delta q(x,S)=n_q(x,S;S)-n_q(x,-S;S).
\end{equation}
It is known that in the parton model $\Delta q(x,S)$ cannot depend
on the direction of the nucleon spin $S$, that is, $\Delta
q(x,S)=\Delta q(x)$ \cite{Anselmino95}.

\section{Relation between the Spin Structure Function $g_1$ and
the Flavor Singlet Axial Charge} {}From the previous analysis or
from the operator product expansion (OPE) \cite{Manohar}, the
first moment of the polarized proton structure function defined
by,
\begin{equation}
\Gamma_1^p(Q^2)\equiv\int_0^1g_1^p(x,Q^2)dx
\end{equation}
can be connected to the flavor singlet axial current of the quarks
by the relation
\begin{equation}
\Gamma_1^p(Q^2)\equiv\int_0^1g_1^p(x,Q^2)dx=\frac{1}{2}\sum_q
e^2_q\Delta q(Q^2)=\frac{1}{2}\sum_q e^2_q\langle p,S|\
\bar{q}\gamma_{\mu}\gamma_5 q\ |p,S\rangle S^{\mu},
\end{equation}
where $\Delta q$ is the net helicity of the quark flavor $q$ along
the direction of the proton spin at  momentum transfer $-Q^2$. In
general, the form of $\Delta q$ depends on $Q^2$. For example, in
the infinite momentum frame, it becomes
\begin{equation}
\Delta q=\int_0^1\Delta q
(x)dx\equiv\int_0^1\bigg[q^{\uparrow}(x)+\bar{q}^{\uparrow}(x)
-q^{\downarrow}(x)-\bar{q}^{\downarrow}(x)\bigg]dx.
\end{equation}

\section{The Proton Spin Problem} At the EMC energies $Q^2 \le
10.7{\rm GeV}^2$ \cite{Ashman},  three light flavors are relevant
and the first moment of the polarized proton structure function
has the form
\begin{equation}
\Gamma_1^p(Q^2)=\frac{1}{2}\bigg(\frac{4}{9}\Delta
u(Q^2)+\frac{1}{9}\Delta d(Q^2)+\frac{1}{9}\Delta s(Q^2)\bigg).
\end{equation}
In terms of the form factors
\footnote{Here we used the normalization of the spin vector of the proton:
$$S^{\mu}S_{\mu}=-M^2,$$ instead of the previous one.}
in the forward proton matrix elements of the renormalized axial
currents \cite{shore}, i.e.,
\begin{eqnarray}
&&\langle p,S|\ A_{\mu}^3\
|p,S\rangle=\frac{1}{2}a^3S_{\mu},\hspace{0.5cm}\langle p,S|\
A_{\mu}^8\ |p,S\rangle=\frac{1}{2\sqrt{3}}a^8
S_{\mu},\nonumber\\
&&\hspace{3cm}\langle p,S|\ A_{\mu}^0\ |p,S\rangle=a^0
S_{\mu},
\end{eqnarray}
with
\begin{equation}
A_{\mu}^a=\bar{\psi}\gamma_{\mu}\gamma_5\frac{\lambda^a}{2}\psi,\hspace{1cm}
A_{\mu}^0=\bar{\psi}\gamma_{\mu}\gamma_5\psi,
\end{equation}
the sum rule for the first moment is
\begin{equation}
\Gamma_1^p(Q^2)=\frac{1}{12}C_1^{NS}(\alpha_s(Q^2))\bigg(a^3+\frac{1}{3}a^8\bigg)
+\frac{1}{9}C_1^S(\alpha_s(Q^2))a^0(Q^2),
\end{equation}
where $\alpha_s(Q^2)$ is the perturbatively running QCD coupling
constant and $C_1(\alpha_s(Q^2))$ are first moments of the Wilson
coefficients of the singlet ($S$) and the non-singlet ($NS$) axial
currents given by \cite{Larin}
\begin{eqnarray}
C_1^{NS}&=&1-\frac{\alpha_s}{\pi}-\frac{43}{12}\bigg(\frac{\alpha_s}{\pi}\bigg)^2
-20.22\bigg(\frac{\alpha_s}{\pi}\bigg)^3\nonumber\\
C_1^S&=&1-\frac{\alpha_s}{\pi}-1.10\bigg(\frac{\alpha_s}{\pi}\bigg)^2,
\end{eqnarray}
up to ${\cal O}(\alpha_s^3)$ for three quark flavors. Since there
is no anomalous dimension associated with the axial-vector
currents $A_{\mu}^3$ and $A_{\mu}^8$, the non-singlet form factors
do not evolve with $Q^2$. The non-singlet form factors are related
to the $SU(3)$ parameter $F$ and $D$ by
\begin{equation}
a^3=F+D,\hspace{1cm}a^8=3F-D. \label{FD}
\end{equation}
Their values \cite{PDG98} \cite{goto} are
\begin{equation}
F=0.463\pm
0.008,\hspace{0.5cm}D=0.804\pm0.008,\hspace{0.5cm}\frac{F}{D}=0.576\pm0.016
\end{equation}
and from these $a^3=1.2670\pm0.0035$. From the definitions, the
form factors can be written in terms of the quark polarizations
\begin{eqnarray}
&&a^0(Q^2)=\Delta u(Q^2)+\Delta d(Q^2)+\Delta
s(Q^2)\equiv\Sigma(Q^2),\hspace{0.5cm}a^3=\Delta u(Q^2)-\Delta
d(Q^2),\nonumber\\
&&\hspace{3cm}a^8=\Delta u(Q^2)+\Delta d(Q^2)-2\Delta s(Q^2).
\label{a's}
\end{eqnarray}

Before the EMC measurement of the polarized structure functions, a
prediction for $\Gamma_1^p$ known as the Ellis-Jaffe sum
rule \cite{ej} was based on the assumption that the strange sea
quark in the proton is unpolarized
\begin{equation}
\Gamma_1^p(Q^2)=\frac{1}{12}a^3+\frac{5}{36}a^8,
\end{equation}
without QCD corrections. The measured result of EMC,
$\Gamma_1^p=0.126\pm0.010\pm0.015$, is smaller than what was
expected from the Ellis-Jaffe sum rule: $\Gamma_1^p=0.185\pm
0.003$ without QCD corrections and $\Gamma_1^p=0.171\pm0.006$
with the leading-order correction. From eq.~(\ref{FD}),
eq.~(\ref{a's}), and the EMC result, the quark polarizations are
obtained as
\begin{eqnarray}
&&\Delta u(Q^2)=0.77\pm0.06,\hspace{0.5cm}\Delta
d(Q^2)=-0.49\pm0.06,\nonumber\\
&&\hspace{2.5cm}\Delta s(Q^2)=-0.15\pm0.06,
\label{q-pol}
\end{eqnarray} and
\begin{equation}
\Delta\Sigma=0.14\pm0.17 \label{siGma}
\end{equation}
at $Q^2=10.7{\rm GeV}^2$. The results eq.~(\ref{q-pol}) and
eq.~(\ref{siGma}) reveal two surprising things: The strange quark
sea has negative non-vanishing polarization, and the total
contribution of quark helicities to the proton spin is small and
consistent with zero. These facts raise some puzzles, for example,
from where does the proton get its spin? why is there negative
polarized strange sea quark? how is the total quark spin component
small? These puzzles are sometimes (inappropriately) referred to
as the {\it proton spin problem (or crisis)}.

The proton spin problem arises from the fact that the experimental
results seem to be in contradiction with the naive quark-model.
The non-relativistic $SU(6)$ constituent quark model yields that
$\Delta u=\frac{4}{3}$ and $\Delta d=-\frac{1}{3}$. Therefore,
from these polarizations one gets $\Delta\Sigma=1$ and $g_A^3\
(=a^3)=\frac{5}{3}$, which is larger than the measured value
$1.2670\pm0.0035$ \cite{PDG98}. In a relativistic quark model, the
quark polarizations $\Delta u$ and $\Delta d$ are reduced by the
same factor of $\frac{3}{4}$ to $1$ and $-\frac{1}{4}$, and
$g_A^3$ is reduced to $\frac{5}{4}$ due to the presence of the
lower component of the Dirac spinor. The reduction of the total
quark spin $\Delta\Sigma$ to 0.75 requires that  the orbital
angular momentum of the quark, $L_Q$, contributes to the nucleon
spin as required by the sum rule \cite{Sehgal}
\begin{equation}
\frac{1}{2}=\frac{1}{2}\Delta\Sigma +L_Q.
\end{equation}
Therefore, it is expected that in the relativistic quark model 3/4
of the proton spin arises from the quarks and the quark orbital
angular momentum accounts for the rest of the spin. The MIT bag
model, which is a relativistic model with QCD confinement
incorporated via its boundary conditions, leads to the similar
value: $\Delta\Sigma=2/3$. On the other hand, the Skyrme model for
the baryons yields $\Delta\Sigma=0$ \cite{SchWei}.

One way to understand the experimental value $\Delta\Sigma\sim
0.30$, which is smaller than the expectation of the quark models,
is to introduce a negatively polarized quark sea. The quark
polarization can be decomposed into valence and sea components,
$\Delta q=\Delta q_v+\Delta q_s$. Then, the the total quark spin
of the proton becomes
\begin{equation}
\Delta\Sigma=\Delta \Sigma_v+\Delta \Sigma_s=(\Delta u_v+\Delta
d_v)+(\Delta u_s+\Delta d_s+\Delta s_s).
\end{equation}
The gluons can induce a quark sea polarization through the
$U(1)_A$ anomaly \cite{abj}, which cancels the spin from the
valence quarks when the gluon has negative spin
component \cite{Cheng96}.

Another way is to use the axial anomaly directly in calculating
the flavor singlet axial current. In other words, the
experimentally measured quantity is not merely the quark spin
polarization $\Delta\Sigma$ but rather the singlet form factor
(the flavor singlet axial charge), to which the gluons contribute
through the axial anomaly as
\begin{equation}
a^0(Q^2)=\Delta\Sigma-N_F\frac{\alpha_s(Q^2)}{2\pi}\Delta g(Q^2),
\end{equation}
where $\Delta g$ is the polarization of the gluons and $N_F$ the
number of flavors \cite{Anselmino95}. These explanations, and
possibly others, could be reconciled if one were to establish that
they are gauge dependent statements, while the measured quantity
is gauge-invariant \cite{Cheng00}. Incorporating the gluons, the
spin sum rule becomes \cite{Jaffe96}
\begin{equation}
\frac{1}{2}=\frac{1}{2}\Delta\Sigma+L_Q+\Delta g+L_G
\end{equation}
with the orbital angular momentum of the gluon, $L_G$, and the
integral of the polarized gluon distribution, $\Delta g$.

The analysis of the flavor singlet axial charge and the gluon spin
in the chiral bag model will be discussed and compared with those
of the MIT bag model in Chapter 5 after introducing anomalies in
Chapter 4 .


\chapter{The chiral bag model}
In this chapter we give a general overview of the chiral bag model
as initially  presented \cite{chiralbag}. We review its definition
in terms of quark, gluon and meson degrees of freedom. Here we
shall be dealing with a Lagrangian that is classically
gauge-invariant. We discuss the solution with this Lagrangian
obtained by using the hedgehog ansatz, including the effects on
the vacuum structure. Our discussion will incorporate the
$\eta^{\prime}$ meson since it will be relevant for later
purposes. It turns out that due to color anomaly, this theory is
not gauge invariant at the quantum level. We avoid in here the
complications arising from the quantum structure of the theory,
relegating this subject to the next chapter.

\section{Model Lagrangian} The chiral bag model is a field
theoretic description of hadron structure whose aim is to
represent QCD in the low energy regime. This description separates
space-time into two regions by a surface, the bag, in which
different effective realizations of the underlying theory, QCD,
are used to represent the dynamics. The bag, which is closed in
space, defines an interior region, conventionally called quark
phase, which is described by means of quark and gluon fields. The
exterior region, called mesonic phase, is defined by an effective
mesonic field theory in accord with the requisites of Weinberg's
unproven theorem \cite{wei}. The bag, the surface separating the
two phases, serves to connect the two types of degrees of freedom
through boundary conditions, whose structure resembles the
bosonization relations in two dimensions.

The motivation for this sophisticated description lies in the
properties of the fundamental theory, QCD, which the model
implements in a dynamical fashion. Let us be more precise:

(1) Color Confinement: the bag is responsible for confining the
color degrees of freedom (quarks and gluons) and the boundary
conditions on it implement the non perturbative character of this
property. Despite the apparent weak interaction  between these
fields in the quark phase, the fact that they are represented by
cavity modes satisfying the boundary conditions, confers them a
non perturbative character very different from that of free Fock
states, even in the case of an empty mesonic sector \cite{MIT}.

(2) Asymptotic freedom: It is known that quarks and gluons in QCD
interact very weakly at large momentum transfers, i.e., short
distances. This important property of the theory, associated with
the negative sign of its $\beta$ function \cite{gross} is
responsible for the slow logarithmic deviations from scaling in
deep inelastic scattering. In the bag description it is
implemented by the perturbative treatment of the interaction
between quark and gluons in the interior region.

(3) Spontaneous broken chiral symmetry: Nature, and therefore QCD,
realizes chiral symmetry in a spontaneously broken fashion, i.e.
the flavor symmetry $SU_L(N_f)\times SU_R(N_f)$ of the currents is
broken down to $SU_V(N_f)$ by the vacuum. This phenomena is
implemented by the mesonic effective theory outside, which
incorporates the pseudoscalar mesons, the required Goldstone
bosons in the chiral limit. Through the boundary conditions this
phenomenon transfers to the interior. The ultimate objective of
the model is to encompass both long-wavelength and
short-wavelength regimes, with the Cheshire Cat principle defined
below bridging the two regimes.

In this well defined scenario with a given chiral bag Lagrangian,
the boundary plays a crucial role because it relates the degrees
of freedom of the two phases in a manner which preserves all the
symmetries and their realization.

The chiral bag model as described above can be implemented by the
following Lagrangian density
\begin{equation}
{\cal L}=({\cal L}_Q-B)\Theta_B+{\cal L}_M\bar{\Theta}_B+{\cal
L}_{QM}\Delta_B, \label{Lag}
\end{equation}
where ${\cal L}_Q, {\cal L}_M$ and ${\cal L}_{QM}$ describe the
dynamics for the quark and gluon fields inside the bag, the meson
fields outside, and the interaction between the quark and meson
phases at the bag surface, respectively. Here $\Theta_B$, which is
needed to define the quark phase inside the bag only, it gives 1
inside the bag and zero outside, $\bar{\Theta}_B=1-\Theta_B$, and
the bag delta function $\Delta_B$ is defined by
$\Delta_B=-n^{\mu}\partial_{\mu}\Theta_B$ where $n^{\mu}$
represents the outward normal unit four vector. $B$ is the so
called bag constant and corresponds to the energy density required
for creating the bag in the QCD vacuum.

The Lagrangian density for the quark phase in case of $SU(3)$
flavor symmetry is given by
\begin{equation}
{\cal
L}_Q=\bar{\psi}\bigg(i\gamma^{\mu}\frac{1}{2}(\overrightarrow{\partial_{\mu}}
-\overleftarrow{\partial_\mu})-M\bigg)\ \psi
-g_sG^a_{\mu}\bar{\psi}\gamma^{\mu}\frac{\lambda^a_c}{2}\psi
-\frac{1}{4}G^a_{\mu\nu}G^{a\mu\nu}
\end{equation}
with
\begin{eqnarray}
G^a_{\mu\nu}&=&\partial_{\mu}G^a_{\nu}-\partial_{\nu}G^a_{\mu}
-g_sf^{abc}G^b_{\mu}G^c_{\nu},\\
M&=&{\rm diag}\ (m_u, m_d, m_s).
\end{eqnarray}
Here $\psi$ represents the quark fields, $G^a_{\mu}$ the gluon
fields, $M$ the current quark mass matrix $m_u\simeq m_d\simeq 0,
m_s\approx 150 \ {\rm MeV}$. Asymptotic freedom is realized by
allowing the interaction between the quark and gluon fields to be
treated perturbatively with respect to the effective bagged QCD
coupling constant $g_s$. The Lagrangian has the same form as that
of QCD, but it is only meaningful in the weak coupling regime. The
quark field is arranged into the fundamental representation of
flavor $SU(3)$
\begin{eqnarray}
\psi=\left(\begin{array}{c}
        u\\
        d\\
        s
        \end{array}\right).
\end{eqnarray}
$\lambda^a_c\ (a=1,2,\cdots,8)$ are the Gell-Mann matrices
associated with color and we use the normalization ${\rm
Tr}\lambda^a_c\lambda^b_c=2\delta^{ab}$.

The meson phase is described by the following Lagrangian density
\begin{eqnarray}
{\cal L}_M&=&\frac{f_\pi^2}{4}{\rm
Tr}(\partial_{\mu}U^{\dagger}\partial^{\mu}U) +\frac{1}{32e^2}{\rm
Tr}
([U^{\dagger}\partial_{\mu}U,U^{\dagger}\partial_{\nu}U]^2)+{\cal
L}_{WZW}
\nonumber\\
&&-\sigma{\rm
Tr}(M(U+U^{\dagger}-2))-\frac{f_\pi^2}{16N_F}m_{\eta'}^2
\bigg({\rm Tr}(\ln U-\ln U^\dagger)\bigg)^2,
\end{eqnarray}
where the chiral field $U$ is of the form
\begin{equation}
U=\exp\bigg(i\frac{\eta'}{f_0}+i\frac{\lambda\cdot\pi}{f_{\pi}}\bigg)
\end{equation}
with $f_0=\sqrt{\frac{N_F}{2}}f_\pi$, and
\begin{eqnarray}
2\sigma&=&\langle \bar{u}u\rangle_0=\langle \bar{d}d\rangle_0
\simeq\langle\bar{s}s\rangle_0\nonumber\\
&&\simeq-\frac{m_{\pi}^2f_{\pi}^2}{m_u+m_d}\simeq-\frac{m_K^2f_{\pi}^2}{m_s},
\end{eqnarray}
where $\lambda^a\ (a=1,2,\cdots,8)$ are the Gell-Mann matrices
associated with the flavor symmetry in this case, $f_{\pi}$ is the
pion decay constant\footnote{Unless otherwise specified we assume
$f_{\pi}\approx f_{K}$. The difference appears at higher order in
the chiral counting.}, $m_{\pi}$, $m_K$ and $m_{\eta'}$ represent
the pion kaon and $\eta'$ masses, respectively. The $\eta'$ has
been introduced by extending the chiral field $U$ from $SU(3)$, as
appears in the chiral hyperbag \cite{pr}, to $U(3)$. In this way
the $\eta'$ is decoupled from the other pseudoscalar mesons. It
plays an important role in the flavor singlet axial charge \cite
{Zahed} due to its flavor singlet structure. The second term in
the above ${\cal L}_M$ is the one introduced by Skyrme to
stabilize the embedded $SU(2)$ soliton solution. From an analysis
of the nucleon axial form factor $g_A$ in the Skyrmion model \cite
{Skyrme}, the parameter $e$ can be fixed to 4.75 throughout whole
bag radius.

The last term, the so called Wess-Zumino-Witten term ${\cal
L}_{WZW}$ \cite{wzw}, comes from the requirement that an effective
theory should have the same symmetries and anomalies as the
fundamental theory, at its validity scale. Its explicit form,
which can only be written as an action, is
 \begin{equation}
\Gamma_{WZW}=-i\frac{N_c}{240\pi^2}\int_{\bar{M}}d^5x \
\epsilon^{\mu\nu\lambda\rho\sigma}{\rm Tr}
(U^{\dagger}\partial_{\mu}UU^{\dagger}\partial_{\nu}U
U^{\dagger}\partial_{\lambda}UU^{\dagger}\partial_{\rho}U
U^{\dagger}\partial_{\sigma}U),
 \end{equation}
where the integral is defined on the five dimensional manifold
$\bar{M}=\bar{B} \times S^1\times[0,1]$ with $\bar{B}$, the
three-space volume outside the bag, and $S^1$ the compactified
time. The extension $[0,1]$ is needed to be able to write the
Wess-Zumino-Witten term in a local form.

The interaction between the quark field and meson field on the bag
surface is given by
\begin{equation}
{\cal L}_{QM}=-\frac{1}{2}\bar{\psi}U_5\psi
=-\frac{1}{2}(\bar{\psi}_L U\psi_R+\bar{\psi}_R U^{\dagger}\psi_L)
\end{equation}
with $\psi_{R,L}=\frac{1}{2}(1\pm\gamma_5)$ and
$U_5=\exp\bigg(i\gamma_5(\eta'/f_0+
\lambda\cdot\pi/f_{\pi})\bigg)$. This interaction provides quark
confinement classically in a chirally invariant way. Note that no
interactions between the gluon field and meson field appear at the
classical level.

In case of massless quarks, this Lagrangian is invariant under
flavor $SU_L(3)\times SU_R(3)$ transformations. Noether's theorem
gives the following conserved currents
\begin{eqnarray}
J_{\mu}^{a,R}&=&\frac{1}{2}\bar{\psi}(1+\gamma_5)\gamma_{\mu}
\frac{\lambda^{a}}{2}\psi\Theta_B\nonumber\\
&&+\bigg[-\frac{i}{4}{\rm
Tr}(\lambda^{a}U^{\dagger}\partial_{\mu}U) +\frac{i}{16e^2}{\rm
Tr}([\lambda^{a},U^{\dagger}\partial^{\nu}U]
[U^{\dagger}\partial_{\mu}U,U^{\dagger}\partial_{\nu}U])\nonumber\\
&&+\frac{N_c}{48\pi^2}\epsilon_{\mu\nu\rho\sigma} {\rm
Tr}\bigg(\frac{\lambda^{a}}{2}U^{\dagger}
\partial^{\nu}UU^{\dagger}\partial^{\rho}UU^{\dagger}\partial^{\sigma}U\bigg)
\bigg]\bar{\Theta}_B,\\
J_{\mu}^{a,L}&=&\frac{1}{2}\bar{\psi}(1-\gamma_5)\gamma_{\mu}
\frac{\lambda^{a}}{2}\psi\Theta_B\nonumber\\
&&+\bigg[\frac{i}{4}{\rm
Tr}(\lambda^{a}\partial_{\mu}U^{\dagger}U) -\frac{i}{16e^2}{\rm
Tr}([\lambda^{a},\partial^{\nu}UU^{\dagger}]
[\partial_{\mu}UU^{\dagger},\partial_{\nu}UU^{\dagger}])\nonumber\\
&&+\frac{N_c}{48\pi^2}\epsilon_{\mu\nu\rho\sigma} {\rm
Tr}\bigg(\frac{\lambda^{a}}{2}
\partial^{\nu}UU^{\dagger}\partial^{\rho}UU^{\dagger}\partial^{\sigma}U
U^{\dagger}\bigg) \bigg]\bar{\Theta}_B, \label{LRcurrent}
\end{eqnarray}
where the index $a$ runs over $1,\cdots,8$. The vector and axial
vector currents can be constructed from these currents as
\begin{eqnarray}
V_{\mu}^{a}&=&J_{\mu}^{a,R}+J_{\mu}^{a,L},\nonumber\\
A_{\mu}^{a}&=&J_{\mu}^{a,R}-J_{\mu}^{a,L}. \label{VAcurrent}
\end{eqnarray}
The baryon number current corresponding to the $U_V(1)$ symmetry
of the Lagrangian is
\begin{equation}
B_{\mu}=\bar{\psi}\gamma_{\mu}\psi\Theta_B+\frac{1}{24\pi^2}
\epsilon_{\mu\nu\lambda\rho}{\rm Tr}
(U^{\dagger}\partial^{\nu}UU^{\dagger}\partial^{\lambda}U
U^{\dagger}\partial^{\rho}U)\bar{\Theta}_B. \label{baryoncurrent}
\end{equation}
The last term corresponds to the topological winding number
arising from the Wess-Zumino-Witten term after proper gauging
\cite{wzw}. The conservation of this term is a consequence of
topology in case of the $SU(2)$ symmetry. The $U_A(1)$ symmetry of
the Lagrangian yields a flavor singlet axial vector current of the
form
\begin{equation}
A^{(0)}_{\mu}=\bar{\psi}\gamma_5\gamma_{\mu}\psi\Theta_B
+2f_{\pi}\partial_{\mu}\eta'(x)\bar{\Theta}_B ,
\end{equation}
which is broken through the well known {\it axial anomaly}
\cite{abj} providing the $\eta'$ with its mass. Its role in the
proton spin problem will be discussed in chapter 5.

The Hamiltonian, of the chiral bag model, can be derived from the
Lagrangian and turns out to be
\begin{eqnarray}
&&H=\int_B d^3r\ \psi^{\dagger}\bigg(-i\mbox{\boldmath
$\alpha$}\cdot \mbox{\boldmath $\nabla$}+\beta\bigg)\psi
+\frac{1}{2}\int_{\partial B} d^3r\ \bar{\psi}U_5\psi\nonumber\\
&&+\int_{\bar{B}}d^3r\bigg[\frac{f_{\pi}^2}{4}{\rm Tr}
(\partial_0U^{\dagger}\partial_0U
+\partial_iU^{\dagger}\partial_iU)\nonumber\\
&&\hspace{2cm}+\frac{1}{32e^2}{\rm Tr}\bigg(
2[U^{\dagger}\partial_0U,U^{\dagger}\partial_iU]^2
-[U^{\dagger}\partial_iU,U^{\dagger}\partial_jU]^2\bigg)\nonumber\\
&&\hspace{2cm}+\sigma{\rm
Tr}(UM+M^{\dagger}U^{\dagger}-M-M^{\dagger})\bigg]
+H_{WZW}.
\label{hamiltonian}
\end{eqnarray}

Finally, we should mention the magnitude of the quark-gluon
coupling constant $g_s$, a parameter of the model. Although $g_s$
has a scale dependence, governed by the $\beta$ function, it is
customary to take it at a fixed value throughout the volume, as
corresponds to lowest order perturbation theory. The MIT group
fixed $g_s$ by studying  the masses of the proton and the delta in
their model \cite{MIT}, and obtained $\alpha_s=g_s^2/4\pi=2.2$,
which we will use for computational purposes.

\section{The Hedgehog Solution} Our aim here is to describe a
solution to the equations of motion of the model Lagrangian. Our
first approximation will be to consider the bag as a static sphere
of radius $R$. The symmetries of the Lagrangian are instrumental
in finding the adequate solution. In the quark sector we recall
that the up and down quark masses can be neglected since they are
small, the strange quark mass is neither so small to be neglected
nor so large to allow a heavy quark treatment. Moreover, the
strange quark mass breaks not only chiral symmetry $SU_L(3)\times
SU_R(3)$ down to $SU_V(3)$ but also the symmetry $SU(3)$ down to
$SU_V(2)\times U_Y(1)$, {\it i.e.}, $SU_L(3)\times
SU_R(3)\rightarrow SU_V(2)\times U_Y(1)$.

The symmetry breaking scheme of the  quark phase suggests that the
meson phase can be described by the classical configuration of the
chiral field $U_0$ which is the $SU(2)$ hedgehog solution embedded
in $SU(3)$. The explicit form of $U_0$ is given by \cite{nmp}
\begin{equation}
U_0=\exp(i\lambda_i\hat{r}_i\theta(r)) =\left( \begin{array}{cc}
         e^{i\mbox{\boldmath $\tau$}\cdot\hat{\bf r}\theta(r)} & 0 \\
                        0                     & 1
        \end{array} \right) ,
\label{ansatz}
\end{equation}
where $\tau_i$ are the Pauli matrices. $\theta(r)$ is called the
chiral angle. We do not include now the $\eta'$ meson in our
description, but we will do so later when its presence becomes
relevant.

{}From the Lagrangian of eq.~(\ref{Lag}) and the above hedgehog
ansatz, if the quark-gluon coupling is turned off, the quarks, in
the spherical cavity approximation, satisfy the following equation
of motion and boundary condition
\footnote{Eq.~(\ref{bcforquark}) yields the quark confinement
condition, $i\bar{\psi}\mbox{\boldmath $\gamma$}\cdot\hat{\bf
r}\psi=0$, as can be seen by using that eq. (\ref{bcforquark}) is
changed to $i\bar{\psi}\mbox{\boldmath $ \gamma$}\cdot\hat{\bf
r}=\bar{\psi}U_5$ under the hermitian conjugation.}
\begin{eqnarray}
i\gamma^{\mu}\partial_{\mu}\psi|_{r=R}&=&0, \ \ \ \ r<R,\label{eomofquark} \\
in_{\mu}\gamma^{\mu}\psi|_{r=R}&=&U^5_0\psi|_{r=R}.    \ \ \ \ \
r=R,\label{bcforquark}
\end{eqnarray}
The ansatz eq.~(\ref{ansatz}), the equation of motion
eq.~(\ref{eomofquark}), and the boundary condition
eq.~(\ref{bcforquark}) for the quarks show that the strange quark
is decoupled from the u and d quarks. In addition, due to the
Pauli matrices of $SU(2)$ flavor space in the hedgehog ansatz, the
u and d quarks form the multiplet of the grand spin operator ${\bf
K}$ defined by
\begin{equation}
{\bf K}={\bf J}+{\bf I},
\end{equation}
where ${\bf J}$ is the total spin and ${\bf I}$ is the isospin.
This is called the hedgehog quark state. Denoting the wave
functions of the hedgehog quark state and the strange quark,
respectively, by $\phi^h_n({\bf r})e^{-i\varepsilon_n t}$ and
$\phi^s_n({\bf r})e^{-i\omega_n t}$ with the appropriate quantum
number $n$, they satisfy the following equations of motion and
boundary conditions
\begin{eqnarray}
-i\mbox{\boldmath $\alpha$}\cdot\mbox{\boldmath $
\nabla$}\varphi^h_n({\bf r})&=&\varepsilon_n \varphi^h_n({\bf r}),
\ \ \ \ \ \ \ \ \ r<R,\nonumber\\
-i\mbox{\boldmath $\gamma$}\cdot\hat{\bf r}\varphi^h_n({\bf r})
&=&e^{i\mbox{\boldmath $\tau$}\cdot\hat{\bf r}\theta_s}
\varphi^h_n({\bf r}), \ \ \ \ \ r=R,\label{bchdg}\\
-i\mbox{\boldmath $\alpha$}\cdot\mbox{\boldmath $
\nabla$}\varphi^s_n({\bf r}) &=&\omega_n\varphi^s_n({\bf r}),
\ \ \ \ \ \ \ \ \ r<R,\nonumber\\
-i\mbox{\boldmath $\gamma$}\cdot\hat{\bf r}\varphi^s_n({\bf
r})&=&\varphi^s_n({\bf r}),
 \ \ \ \ \ \ \ \ \ \ \ \ r=R \label{sbc},
\end{eqnarray}
where $\theta_s\equiv\theta(R)$ and
\begin{equation}
\phi^h_n({\bf r})=\left(\begin{array}{c}
                  \varphi^h_n({\bf r})\\0
                      \end{array}\right),
\ \ \ \phi^s_n({\bf r})=\left(\begin{array}{c}
                  0\\ \varphi^s_n({\bf r})
                      \end{array}\right),
\label{phi^0}
\end{equation}
have been used.

Incorporating the color degrees of freedom $|\alpha\rangle$, the
quark field $\psi$ can be expanded in terms of these wave function
as
\begin{eqnarray}
\psi({\bf r},t)&=&\sum_{n,\alpha,\ \varepsilon_n>0}\phi^h_n({\bf
r}) e^{-i\varepsilon_n t} |\alpha\rangle a_n^{\alpha}
+\sum_{n,\alpha,\ \varepsilon_n<0}\phi^{h*}_n({\bf r})
e^{i\varepsilon_n t}|\alpha\rangle b_n^{\alpha\dagger}\nonumber\\
&&+\sum_{n,\alpha,\ \omega_n>0}\phi^s_n({\bf r})e^{-i\omega_n t}
|\alpha\rangle c_n^{\alpha}+\sum_{n,\alpha,\
\omega_n<0}\phi^{s*}_n({\bf r}) e^{i\omega_n t}|\alpha\rangle
c_n^{\alpha\dagger}.
\end{eqnarray}
$a\ (b^{\dagger})$ is the annihilation operator for the positive
(negative) energy hedgehog quark and $c\ (d^{\dagger})$ the
annihilation operator for the positive (negative) energy strange
quark. The operators, $a, b, c, d,$ satisfy the usual
anti-commutation rules:
\begin{equation}
\{a_n^{\alpha\dagger},a_m^{\alpha'}\}=\{b_n^{\alpha\dagger},b_m^{\alpha'}\}
=\{c_n^{\alpha\dagger},c_m^{\alpha'}\}=\{d_n^{\alpha\dagger},d_m^{\alpha'}\}
=\delta_{mn}\delta_{\alpha\alpha'}
\end{equation}
vanishing all other anti-commutators. The quark vacuum is defined
by $a_n^{\alpha}|0\rangle
=b_n^{\alpha}|0\rangle=c_n^{\alpha}|0\rangle=d_n^{\alpha}|0\rangle=0$,
that is, all the negative energy eigenstates of the hedgehog and
the strange quarks are filled with three different colors.

The hedgehog quark state has the following quantum number: $K$,
the grand spin such that the eigenvalue of ${\bf K}^2$ is
$K(K+1)$, $M_K$, the eigenvalue of third component of ${\bf K}$,
$P$, the parity, and finally $n$, the radial quantum number. It is
convenient to introduce  additional quantities such as
$\kappa=P(-1)^K$ and  $\epsilon$, the sign of energy eigenvalue
which makes the radial quantum number a positive integer. The
hedgehog quark state will be denoted by $|\underline{m}\rangle$,
{\it i.e.}, $\underline{m}$ denotes the set of indices
$\{K,M_K,P,n,\kappa,\epsilon\}$.

The eigenstate $|K,M_K\rangle$ of ${\bf K}^2$ and $K_z$ can be
constructed by a linear combination of the eigenstates of the
total spin operator and the eigenstates of the isospin operator.
With the help of the eigenstates of the total spin operator ${\bf
J}={\bf L}+{\bf S}$, there are four combinations for ${\bf
K}\neq0$. Because of the parity, in terms of $|K,M_K\rangle_i$
given in the appendix, the wave function for the hedgehog quark
state can be written as\vskip 0.5ex (i) for $\kappa=+1$
\begin{eqnarray}
\varphi^h_{\underline{m}} &=&\alpha N_1 \left( \begin{array}{c}
                   j_K(\varepsilon_n r)\\
      i\mbox{\boldmath $\sigma$}\cdot\hat{\bf r}j_{K+1}(\varepsilon_n r)
                    \end{array}\right)
|K,M_K\rangle_1\nonumber\\
&&+\beta N_2 \left( \begin{array}{c}
                   j_K(\varepsilon_n r)\\
      -i\mbox{\boldmath $\sigma$}\cdot\hat{\bf r}j_{K-1}(\varepsilon_n r)
                    \end{array}\right)
|K,M_K\rangle_2,
\end{eqnarray}
and\vskip 0.5ex
(ii) for $\kappa=-1$
\begin{eqnarray}
\varphi^h_{\underline{m}} &=&\alpha N_1 \left( \begin{array}{c}
                   j_{K+1}(\varepsilon_n r)\\
     -i\mbox{\boldmath $\sigma$}\cdot\hat{\bf r}j_{K}(\varepsilon_n r)
                    \end{array}\right)
|K,M_K\rangle_3\nonumber\\
&&+\beta N_2 \left( \begin{array}{c}
                   j_{K-1}(\varepsilon_n r)\\
      i\mbox{\boldmath $\sigma$}\cdot\hat{\bf r}j_{K}(\varepsilon_n r)
                    \end{array}\right)
|K,M_K\rangle_4,
\end{eqnarray}
where $j_K(x)'$s are the spherical Bessel functions, $N_1, N_2$
the normalization constants:
\begin{eqnarray}
N_1&=&\bigg(\frac{1}{R^3}\frac{\Omega_n}{\Omega_n(j_K^2(\Omega_n)
+j_{K+1}^2(\Omega_n))-2(K+1)j_K(\Omega_n)j_{K+1}(\Omega_n)}\bigg)
^{\frac{1}{2}},\nonumber\\
N_2&=&\bigg(\frac{1}{R^3}\frac{\Omega_n}{\Omega_m(j_K^2(\Omega_n)
+j_{K-1}^2(\Omega_n))-2Kj_K(\Omega_n)j_{K-1}(\Omega_n)}\bigg)
^{\frac{1}{2}}, \label{1.28}
\end{eqnarray}
with $\Omega_n=\varepsilon_n R$, and $\alpha, \beta$ constants
with the condition, $\alpha^2+\beta^2=1$.

\begin{figure}
\centerline{\epsfig{file=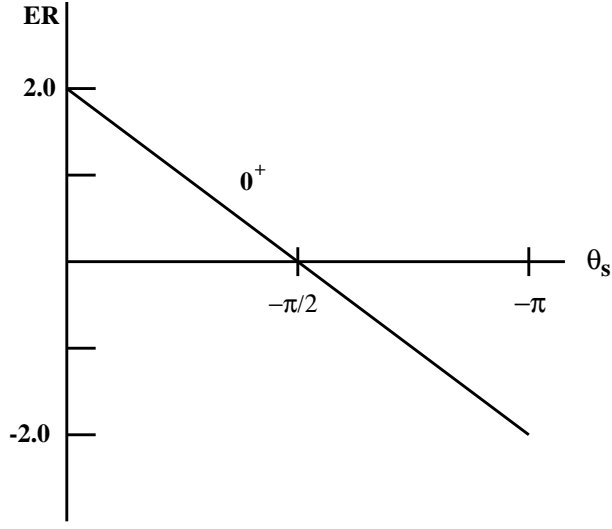,width=8cm}} \caption{The
lowest energy level of the hedgehog quark state for $K=0$.
}\label{helevel}
\end{figure}

Substituting these wave functions into the boundary condition
eq.~(\ref{bchdg}), the energy eigenvalue $\varepsilon_n$ and the
constants $\alpha, \beta$ are determined by the linear equation
\begin{eqnarray}
&&\left(\begin{array}{cc}
\kappa\bigg(1-\frac{\sin\theta_s}{2K+1}\bigg)j_{K+1}-\cos\theta_sj_K
&
-\sin\theta_s\frac{2\sqrt{K(K+1)}}{2K+1}j_{K-1}\\
-\sin\theta_s\frac{2\sqrt{K(K+1)}}{2K+1}j_{K+1} &
\kappa\bigg(1+\frac{\sin\theta_s}{2K+1}\bigg)j_{K-1}+\cos\theta_sj_K
\end{array}\right)\nonumber\\
&&\hspace{4cm}\times
\left(\begin{array}{c} \alpha N_1\\ \beta N_2
\end{array}\right)
=0.
\end{eqnarray}
The energy eigenvalues are obtained from the determinant of this
matrix which is of the form
\begin{eqnarray}
&&\cos\theta_s\{j_K^2(\Omega_n)-j_{K+1}(\Omega_n)j_{K-1}(\Omega_n)\}
-\kappa j_K(\Omega_n)\{\ j_{K+1}(\Omega_n)-j_{K-1}(\Omega_n)\ \}
\nonumber\\
&&\hspace{5cm}+\frac{\sin\theta_s}{\Omega_n}j^2_K(\Omega_n)=0,
\label{energyeq}
\end{eqnarray}
for an arbitrary $K>0$. The equation for $K=0$ is obtained simply
by setting $j_{K-1}=0$:
\begin{equation}
\kappa\cos\theta_s
j_1(\Omega_n)-(1+\kappa\sin\theta_s)j_0(\Omega_n)=0.
\end{equation}
In Fig.~\ref{helevel} the lowest energy level is drawn as a
function of the chiral angle $\theta_s$ for $K=0$ of positive
parity.
By the structure of the eq.~(\ref{energyeq}), the energy levels
are degenerate with respect to the quantum number $M_K$. For
arbitrary $\theta_s$, the energy spectrum is asymmetric with
respect to $\varepsilon_n=0$. The energy spectrum becomes though
symmetric for specific values of the chiral angle, $\theta_s=
n\pi$ and $\theta_s=n\pi+\pi/2$ with $(n=0,\pm 1,\pm 2,\cdots)$.
The case $\theta_s=n\pi$ corresponds to the MIT bag model and all
quark states have partners for negative energy. For
$\theta_s=n\pi+\pi/2$, the energy spectrum is symmetric except for
the zero-energy state. Furthermore, the symmetry of the Dirac
equation and of the boundary condition give the energy spectrum
the following symmetries
\begin{equation}
\varepsilon_{\underline{m}}(\theta_s)=\varepsilon_{\underline{n}}(\pi+\theta_s),
\end{equation}
where $\underline{m}=\{K,M_K,P,n,\kappa,\epsilon\}$ and
$\underline{n}=\{K,M_K,-P, n,-\kappa,\epsilon\}$, and
\begin{equation}
\varepsilon_{\underline{m}}(\theta_s)=-\varepsilon_{\underline{n}}(-\theta_s),
\end{equation}
where $\underline{m}=\{K,M_K,P,n,\kappa,\epsilon\}$ and
$\underline{n}=\{K,M_K,-P, n,-\kappa,-\epsilon\}$.

{}From the Lagrangian for the meson phase with the hedgehog ansatz
(classical configuration), eq.~(\ref{ansatz}),
\begin{equation}
U_0({\bf r})=\exp(i\lambda_i\cdot\hat{r}_i\theta(r)),
\end{equation}
the equation of motion for $\theta(r)$ and the boundary condition
are
\begin{eqnarray}
&&\bigg(1+\frac{2\sin^2\theta}{r^2}\bigg)\frac{d^2\theta}{dr^2}+\frac{2}{r}
\frac{d\theta}{dr}-\frac{\sin2\theta}{r^2}\left[1-\bigg(\frac{d\theta}{dr}\bigg)^2
+\frac{2\sin^2\theta}{r^2}\right]=0,\nonumber\\
&&\hspace{6cm} \ \ {\rm for}\ r>R,\\
&&\hspace{0.5cm} \frac{4\pi
f_{\pi}}{e}(r^2+2\sin^2\theta)\frac{d\theta}{dr}
=-\frac{1}{2}\langle {\rm H}|\int_{\partial B} d^3r\
\bar{\psi}\gamma_5 \vec{\mbox{\boldmath
$\tau$}}\cdot\hat{r}\vec{\mbox{\boldmath $\gamma$}}\cdot\hat{r}
\psi\ |{\rm H}\rangle_0,\nonumber\\
&&\hspace{6cm}\ \  {\rm for}\ r=R.
\end{eqnarray}
Substituting the hedgehog ansatz into the Hamiltonian
eq.~(\ref{hamiltonian}), the contribution from the meson phase
becomes
\begin{equation}
E_{\rm meson}=\frac{2\pi f_{\pi}}{e}\int_R^{\infty}r^2dr\left\{
\bigg(\frac{d\theta}{dr}\bigg)^2+2\frac{\sin^2\theta}{r^2}
+\frac{\sin^2\theta}{r^2}\left[2\bigg(\frac{d\theta}{dr}\bigg)^2+
\frac{\sin^2\theta}{r^2}\right]\right\}, \label{E^0_meson}
\end{equation}
where the mass terms have been omitted. Since for the static case
$E_M=-L_M$, the minimization of the energy with respect to the
variation of $\theta(r)$ leads also to the equations of motion.

The strange quark has the same energy spectrum as that of the MIT
bag model since there is no classical configuration associated
with the kaons. The strange quark states $|\overline{m}\rangle$
are described by four quantum numbers, $j$, the total spin such
that the eigenvalue of ${\bf J}^2$ is $j(j+1)$, $m_j$, the
eigenvalue of the third component of the total spin, $P$, the
parity, and $n$, the radial quantum number and, for convenience,
the two indices, $\kappa$, which has the value $\pm 1$
corresponding to $j=l\pm 1/2$, and $\epsilon$, the sign of the
energy eigenvalue, i.e., $\overline{m}=\{j,m_j,
P,n,\kappa,\epsilon\}$.

Using the eigenstates $|j,m_j\rangle_{\kappa}$ of the total spin
${\bf J}$ appearing in the appendix as basis, the wave functions
for the strange quark become
\begin{eqnarray}
({\rm i})\ {\rm for}\ \kappa=+1&&\bigg(j=l+\frac{1}{2}\bigg)\nonumber\\
&&\varphi^s_{\overline{m}}=N_1\left(\begin{array}{c}
      j_l(\omega_n r)\\ i\mbox{\boldmath $\sigma$}\cdot\hat{\bf r}j_{l+1}(\omega_n
r)
      \end{array}\right)|j,m_j\rangle_{\kappa=+1},
\end{eqnarray}
and
\begin{eqnarray}
({\rm ii})\ {\rm for}\ \kappa=-1&&\bigg(j=l-\frac{1}{2}\bigg)\nonumber\\
&&\varphi^s_{\overline{m}}=N_2\left(\begin{array}{c}
      j_l(\omega_n r)\\ -i\mbox{\boldmath
$\sigma$}\cdot\hat{\bf r}j_{l-1}(\omega_n r)
      \end{array}\right)|j,m_j\rangle_{\kappa=-1},
\end{eqnarray}
where $N_1$ and $N_2$ have the same form as those for the hedgehog
states replacing $\Omega_n$ and $K$ by $X_n\ (=\omega_n R)$ and
$l$, respectively. Using the boundary condition eq.~(\ref{sbc}),
the energy eigenvalues are obtained from
\begin{equation}
j_{l+1}(X_n)=+j_l(X_n),
\end{equation}
for $\kappa=+1$, and
\begin{equation}
j_{l-1}(X_n)=-j_l(X_n),
\end{equation}
for $\kappa=-1$. The energy spectrum is degenerate with respect to
the quantum number $m_j$ and has the property
\begin{equation}
X_{\overline{m}}=-X_{\overline{n}}
\end{equation}
due to the invariance of the Dirac equation and the boundary
condition eq.~(\ref{sbc}) under the CP-operation. Here
$\overline{m}=\{j,m_j,P,n,\kappa,\epsilon\}$ and
$\overline{n}=\{j,m_j,-P,n,-\kappa,-\epsilon\}$.

\section{The Baryon Number Fractionization} As can be seen in
Fig.~\ref{helevel}, the energy level dives into the Dirac sea at
$\theta_s=-\pi/2$. This means, that even if one has unit baryon
number at $\theta_s=0$ by putting $N_c$ quarks into the bag, there
can be a leakage of baryon number as $\theta_s$ varies. This fact
brings to baryon number fractionization when only the quark phase
is considered.

By filling up the valence quark states and all negative energy
eigenstates with quarks of different colors, the $K=0$  ground
state  hedgehog baryon can be constructed as
\begin{eqnarray}
|{\rm H}_0\rangle=\left\{\begin{array}{ll}
\frac{\epsilon^{\alpha_1\cdots\alpha_{N_c}}}{\sqrt{N_c!}}
a_{0\alpha_1}^{\dagger}\cdots
a_{0\alpha_{N_c}}^{\dagger}|0\rangle,&
{\rm for}\ -\frac{\pi}{2}\leq\theta_s<0\\
|0\rangle,& {\rm for}\ -\pi\leq\theta_s<-\frac{\pi}{2},
\end{array}\right.
\label{H_0}
\end{eqnarray}
where $|0\rangle$ is the vacuum state defined by
$a_n^{\alpha}|0\rangle
=b_n^{\alpha}|0\rangle=c_n^{\alpha}|0\rangle=d_n^{\alpha}|0\rangle=0$
and the subscript ``0" in the creation operator indicates that $K
= 0$. The baryon number is given by
\begin{equation}
{\rm B}_B=\frac{1}{N_c}\langle {\rm H}_0|\int_B d^3r\
\psi^{\dagger}\psi\ |{\rm H}_0\rangle.
\end{equation}
Applying Wick's contraction leads to
\begin{eqnarray}
{\rm B}_B=\left\{\begin{array}{ll}
1+{\rm B}_{{\rm vac}} & {\rm for}\ -\frac{\pi}{2}\leq\theta_s<0,\\
{\rm B}_{{\rm vac}} & {\rm for}\ -\pi\leq\theta_s<-\frac{\pi}{2}.
\end{array}\right.
\label{B_V}
\end{eqnarray}
While the usual vacuum cannot carry any baryon number, the quark
vacuum of the hedgehog state can get an induced baryon number
through a non-trivial polarization by the interaction with the
meson phase outside the bag as first pointed out by Vento et
al~\cite{ventoetal}. The induced baryon number of the hedgehog
quark vacuum can be obtained by evaluating the regularized
spectral asymmetry \cite{GoldstoneJaffe}
\begin{eqnarray}
{\rm B}_{{\rm vac}}&=&\lim_{\tau\rightarrow 0+}\bigg(-\frac{1}{2}
\sum_n {\rm sign}(\varepsilon_n)e^{-\tau|\varepsilon_n|}\bigg)\nonumber\\
&=&\left\{\begin{array}{ll}
\frac{1}{\pi}(\theta_s-\sin\theta_s\cos\theta_s) &
-\frac{\pi}{2}\leq\theta_s\leq
0,\\
1+\frac{1}{\pi}(\theta_s-\sin\theta_s\cos\theta_s)&-\pi\leq\theta_s<-\frac{\pi}{2},
\end{array}\right.
\end{eqnarray}
where $\tau$ in the first line is introduced for regularization.
The non-trivial vacuum polarization and the non-vanishing baryon
number of the hedgehog quark vacuum result from the CP-symmetry
breaking in the energy spectrum.

As discussed in the previous section, the baryon current,
eq.~(\ref{baryoncurrent}), gets a contribution from the meson
phase outside the bag through the Wess-Zumino-Witten
term~\footnote{That is for $N_f=3$. For $N_f<3$, the argument is
indirect as discussed in the previous section.}. Substituting the
classical configuration (hedgehog solution), eq.~(\ref{ansatz}),
into the meson part in eq.~(\ref{baryoncurrent}) yields the baryon
number
\begin{eqnarray}
{\rm B}_{\bar{B}}&=&-\frac{1}{24\pi^2}\int_{\bar{B}}d^3r
\epsilon_{0ijk}{\rm Tr}
(U^{\dagger}\partial^{i}UU^{\dagger}\partial^{j}U
U^{\dagger}\partial^{k}U)\nonumber\\
&=&-\frac{1}{\pi}(\theta_s - \sin\theta_s\cos\theta_s),
\label{B_Vbar}\end{eqnarray} where the condition
$\theta(r\rightarrow\infty)=0$ has been used. Therefore, to get
baryon number one, two quantities, $B_B$ and $B_{\bar{B}}$, should
be added. In other words, although the quark is confined
classically, the quantum fluctuations due to the hedgehog solution
of the meson induce  a leakage at the bag surface of baryon number
so that there is a contribution to the baryon number from the
outside region. That is
\begin{equation}
{\rm B}={\rm B}_B+{\rm B}_{\bar{B}}=1.
\end{equation}
This mechanism  is known as the baryon number fractionization,
and, as just seen, in the $SU(3)$ case it is identical to that
previously studied for the $SU(2)$ model \cite{pr}.

\section{The Cheshire Cat Principle (CCP)}

We have seen that the baryon number in the chiral bag model (CBM)
arises from summing the contributions arising from the quark and
meson phases. This is a particular case of a general statement,
namely that for any observable ${\cal O}$ in the CBM its value
arises from adding the contribution of both phases,

\begin{equation}
{\cal O}={\cal O}_{B}+{\cal O}_{\bar{B}},
\end{equation}
where ${\cal O}_{B}$ and ${\cal O}_{\bar{B}}$ are the
contributions from the quark and meson phases, respectively. It
may be recalled, that in the case of the baryon number, although
each contribution independently depended on the bag radius $R$,
its sum did not. This is because baryon charge is a topological
quantity. It has been observed that this is a general trend, i.e.,
when a correct calculation for {\it any observable} is performed
within the CBM, the result tends to be almost radius-independent
over a sizeable range of $R$ \cite{pv} \cite{ht}. This statement
has become known as the approximate {\it Cheshire Cat Principle}
(CCP) \cite{ccp} \cite{ccpph}.

In order to understand the profound meaning of the CCP we have to
recall some results from quantum  field theories in 1+1
dimensions. In this case fermionic theories are exactly
bosonizable, i.e., one can write for any fermionic theory a
bosonic theory which leads exactly to the same S-matrix. Thus in
1+1 dimensions the Cheshire Cat Principle is an exact statement
and its meaning very clear \cite{pm}. Let us divide space into two
arbitrary regions. In one of them we describe the physics by means
of a certain theory of fermions. In the other by its equivalent
bosonic theory. The boundary conditions, which couple the two
theories, arise from the bosonization rules associated with given
symmetries. Any observable one calculates arises from the addition
of the contribution of the two sectors and naturally it is
independent on the position of the boundary. Thus the Cheshire Cat
Principle is a corollary of exact bosonization and the proper
definition of the boundary conditions. One can phrase this freedom
in terms of a gauge symmetry

In four dimensions there is no exact bosonization technique known
up to date. This is because one would in principle need infinitely
many mesonic degrees of freedom to write a theory equivalent to a
fermionic theory. Thus the CCP can, in general, be only an
approximate statement. The exact CCP for the baryon number is a
special case because of its topological character, i.e., from all
possible mesons fields only the hedgehog carries baryon number.
Therefore the CCP transforms from a corollary of exact
bosonization in 1+1 dimensions to a predictive statement in 3+1
dimensions. It basically asserts the quality and indicates the
limitation of our effective theories and calculations. The closer
our theories represent the true theory in their corresponding
regime and the better we perform our calculations, the larger will
be the range of radius independence of our observables.

Many calculations have been performed for different observables
and in all of them a certain degree of radius independence has
been observed \cite{vento}. We will show in this thesis the
realization of the CCP in a very complex physical scenario.


\section{Collective Coordinate Quantization} We have studied the
hedgehog solution for the ground state of the baryon in the
previous section. This solution does not carry spin nor isospin
and therefore does not correspond to any baryon of the spectrum.
It can be regarded as a superposition of physical $B=1$ baryons
with various spins and isospins constrained by the relation ${\bf
K}={\bf J}+{\bf I}=0$ \cite{zb} \cite{ccpph}. In other words, the
independent spin and isospin symmetries of the baryons are mixed
up in the ${\bf K}$-symmetry of the hedgehog solution. By using
the collective coordinate quantization method, this problem can be
overcome and baryons with the appropriate quantum numbers can be
obtained.

We shall consider firstly the $SU(2)$ case and then its  extension
to $SU(3)$ which contains the Wess-Zumino-Wess term .

The hedgehog solution is degenerate in energy with respect to an
arbitrary constant rotation in $SU(2)$ space $A$, which transforms
the fields as
\begin{eqnarray}
\psi&\rightarrow& \psi'=A\psi, \nonumber\\
U&\rightarrow& U'=UAU^{\dagger}, \label{field'}
\end{eqnarray}
and which can be parameterized as
\begin{equation}
A=a_0+i{\bf a}\cdot\mbox{\boldmath $\tau$},
\end{equation}
where the parameters are constrained by $a_0^2+{\bf a}^2=1$.
Allowing  $A$ to be time-dependent introduces three independent
collective coordinates. Substituting these new fields into the
Lagrangian leads to
\begin{equation}
L'=L_0+L_{{\rm rot}}^{\rm quark}+L_{\rm rot}^{\rm meson},
\end{equation}
where $L_0$ is the original Lagrangian and
\begin{eqnarray}
L_{\rm rot}^{\rm quark}&=&-\frac{1}{2}\int_B d^3r\ \psi^{\dagger}
\mbox{\boldmath $\tau$}\cdot\mbox{\boldmath $\omega$}\psi,
\nonumber\\
L_{\rm rot}^{\rm meson}&=&\frac{1}{2}{\cal I}_{\rm meson}
\mbox{\boldmath $\omega$}^2. \label{perL}
\end{eqnarray}
Here $\mbox{\boldmath $\omega$}$ represents  a rotational
velocity, which is defined by
\begin{equation}
\mbox{\boldmath $\omega$}=-i{\rm tr}\ [\mbox{\boldmath $\tau$}
A^{\dagger}\partial_0 A]=a_0\dot{{\bf a}}-{\bf a}\dot{a}_0 +{\bf
a}\times\dot{{\bf a}}
\end{equation}
and ${\cal I}_{\rm meson}$ a moment of inertia arising from the
meson phase due to the collective rotation, which is given by
\begin{equation}
{\cal I}_{\rm meson} =
\left\{\frac{8\pi}{3}\int_R^{\infty}r^2dr\sin^2\theta\left[f_{\pi}^2
+\frac{1}{e^2}\bigg(\frac{d\theta}{dr}\bigg)^2+\frac{\sin^2\theta}{r^2}\right]
\right\}. \label{I_meson}
\end{equation}
Note in  $L_{\rm rot}^{\rm meson}$, eq.~(\ref{perL}), that the
rotational effects associated with the hedgehog  appear in the
mesonic sector to second order in $\mbox{\boldmath $\omega$}$,
while in the quark phase in first order in $\mbox{\boldmath
$\omega$}$. The equation motion for the quark changes to
\begin{equation}
\bigg(i\dpa-\frac{\gamma^0}{2}\mbox{\boldmath $\tau$}\cdot
\mbox{\boldmath $\omega$}\bigg)\psi=0
\end{equation}
with the boundary condition given by eq.~(\ref{bcforquark}). With
the adiabatic assumption, i.e. slow rotation, the additional terms
in the equation of motion can be treated perturbatively.  The
single quark eigenstate obtained by means of standard time
independent perturbation theory is given by
\begin{equation}
\varphi_n(\vec{r})=\varphi_n^{(0)}+\frac{1}{2}\sum_{m}
\frac{\langle m|\mbox{\boldmath $\tau$}\cdot\mbox{\boldmath
$\omega$}
|n\rangle}{\varepsilon_m^0-\varepsilon_n^0}\varphi_m^{(0)}(\vec{r})+\cdots,
\end{equation}
where $\varphi_m^{(0)}$ represents the $n$-th eigenstate of the
unperturbed equation with eigenenergy $\varepsilon_m^0$. The
baryon state is also modified by the well known Thouless formula
\cite{thou};
\begin{equation}
|{\rm H}\rangle=\exp\bigg(\sum_{p\not\in {\rm H}_0, h\in {\rm
H}_0}\frac{1}{2} \cdot \frac{\langle p|\mbox{\boldmath
$\tau$}\cdot \mbox{\boldmath $\omega$}|h\rangle}
{\varepsilon_p^0-\varepsilon_h^0}\cdot a_p^{\dagger}a_h|{\rm
H}_0\rangle\bigg), \label{H}
\end{equation}
where $|{\rm H}_0\rangle$ is the unperturbed baryon state given by
eq.~(\ref{H_0}) and $a_p^{\dagger}\ (a_h)$ is the creation
(annihilation) operator for a particle (hole) state. The energy of
the system arises from both phases. The contribution from the
quark phase can be calculated by taking the expectation value of
the Hamiltonian operator with respect to the baryon state
eq.~(\ref{H});
\begin{eqnarray}
E_{\rm quark}&=&\left\langle\ \int_Bd^3r\ \psi^{\dagger}
\mbox{\boldmath $\alpha$}
\cdot\frac{1}{i}\mbox{\boldmath $\nabla$}\psi\ \right\rangle \nonumber\\
&=&\frac{\langle {\rm H}|\ \int_Bd^3r\ \psi^{\dagger}
\mbox{\boldmath $\alpha$}\cdot\frac{1}{i}\mbox{\boldmath $\nabla$}
\psi\ |{\rm H}\rangle}{\langle {\rm H}|{\rm H}\rangle}\nonumber\\
&=&E^0_{\rm quark}+\frac{1}{2}{\cal I}_{\rm quark} \mbox{\boldmath
$\omega$}^2+\cdots,
\end{eqnarray}
where the moment of inertia from the quark phase, ${\cal I}_{\rm
quark}$, is defined by
\begin{equation}
{\cal I}_{\rm quark}=\frac{1}{2}\sum_{p\not\in {\rm H}_0,h\in {\rm
H}_0} \frac{|\langle
p|\tau_z|h\rangle|^2}{\varepsilon^0_p-\varepsilon^0_h}.
\end{equation}
Here, only $\tau_z$ appears because we are choosing the axis of
rotation along the $z$-direction. The contribution from the meson
phase is obtained by substituting $U'$ into the meson part of the
Hamiltonian eq.~(\ref{hamiltonian});
\begin{equation}
E_{\rm meson}=E^0_{\rm meson}+\frac{1}{2}{\cal I}_{\rm meson}
\mbox{\boldmath $\omega$}^2,
\end{equation}
where $E^0_{\rm meson}$ and ${\cal I}_{\rm meson}$ are given in
eq.~(\ref{E^0_meson}) and eq.~(\ref{I_meson}), respectively.
Including the volume energy, the energy of the baryon becomes up
to second order in $\mbox{\boldmath $\omega$}$
\begin{eqnarray}
E_{\rm baryon}&=&\bigg(E^0_{\rm quark}+E^0_{\rm meson}
+\frac{4}{3}\pi R^3 B\bigg)+\frac{1}{2}({\cal I}_{\rm quark}
+{\cal I}_{\rm meson})\mbox{\boldmath $\omega$}^2\nonumber\\
&\equiv& E_0+\frac{1}{2}{\cal I}\mbox{\boldmath $\omega$}^2.
\end{eqnarray}

Let us proceed to  describe the Isospin and Spin in this formalism
and see how they enter into the energy expression. Substituting
the fields given in eq.~(\ref{field'}) into the expressions,
eq.~(\ref{LRcurrent}) and eq.~(\ref{VAcurrent}), of the vector
current, replacing $\lambda_a$ by $\mbox{\boldmath $\tau$}$ in the
$SU(2)$ case, and integrating the time component of the current
over space, we obtain the isospin in terms of the collective
variables
\begin{eqnarray}
{\bf T}_i&=&\frac{1}{2}\left\langle\ \int_B d^3r\
\psi^{\dagger}\tau_i\psi\ \right\rangle
+{\cal I}_{\rm meson}\omega_i\nonumber\\
&=&({\cal I}_{\rm quark}+{\cal I}_{\rm meson})\omega_i={\cal
I}\omega_i,
\end{eqnarray}
where the expectation value for the quark fields has been taken
with respect to $|{\rm H}\rangle$.

The conjugate momenta $\Pi_{\mu}\ (\mu=0,1,2,3)$ associated with
the collective coordinates $a_{\mu}\ (\mu=0,1,2,3)$ can be derived
from the Lagrangian, eq.~(\ref{perL}), and yield
\begin{eqnarray}
\Pi_0&=&\frac{\partial L}{\partial \dot{a}_0}=2{\bf T}\cdot{\bf
a},
\nonumber\\
\mbox{\boldmath $\Pi$}&=&\frac{\partial L}{\partial \dot{\vec{\bf
a}}} =2(-{\bf T}a_0+{\bf a}\times{\bf T}).
\end{eqnarray}
These relations lead to an expression for the isospin in the form
\begin{equation}
{\bf T}=\frac{1}{2}\bigg({\bf a}\Pi_0-a_0\mbox{\boldmath $\Pi$}
+{\bf a}\times\mbox{\boldmath $\Pi$}\bigg).
\end{equation}
Requiring the commutation relation
\begin{equation}
[a_{\mu}, \Pi_{\nu}]=i\delta_{\mu\nu},
\end{equation}
the quantum mechanical isospin operator can be represented as
\begin{equation}
{\bf T}_i=\frac{i}{2}\bigg(a_0\frac{\partial}{\partial a_i}
-a_i\frac{\partial}{\partial a_0}
-\epsilon_{ijk}a_j\frac{\partial}{\partial a_k}\bigg).
\end{equation}
We now proceed with the description of spin. For ${\bf K}=0$ the
space rotation turns opposite to the isospace rotation, therefore,
the quantum mechanical spin operator can be built by replacing
$\vec{\bf a}$ with its negative value
\begin{equation}
{\bf J}_i=\frac{i}{2}\bigg(a_i\frac{\partial}{\partial a_0}
-a_0\frac{\partial}{\partial
a_i}-\epsilon_{ijk}a_j\frac{\partial}{\partial a_k} \bigg).
\end{equation}
With these results and the fact, $[{\bf J},{\bf T}]=0$, the energy
of the baryon can be written, keeping in mind that ${\bf K}=0$, as
\begin{equation}
E_{\rm baryon}=E_0+\frac{{\bf T}^2}{2{\cal I}} =E_0+\frac{{\bf
J}^2}{2{\cal I}},
\end{equation}

This expression may be interpreted as the energy of the rotating
top with the moment of inertia ${\cal I}$. Since we are concerned
with the static case, this energy can be regarded as the mass of
the baryon. In terms of the eigenvalues of the isospin and the
spin, the mass becomes
\begin{equation}
M_{\rm baryon}=M_0+\frac{I(I+1)}{2{\cal
I}}=M_0+\frac{J(J+1)}{2{\cal I}},
\end{equation}
so that the corresponding masses of the nucleons and the $\Delta$s
are
\begin{eqnarray}
M_N(I=J=1/2)&=&M_0+\frac{1}{2{\cal I}}\frac{3}{4},\nonumber\\
M_{\Delta}(I=J=3/2)&=&M_0+\frac{1}{2{\cal I}}\frac{15}{4}.
\end{eqnarray}

The extension to $SU(3)$ case is more complex. The hedgehog is
sitting in $SU(2)$ and only the collective coordinates are
extended. By replacing $A$ in eq.~(\ref{field'}) by a $SU(3)$
matrix, eight collective variables $q_{\alpha}\
(\alpha=1,\cdots,8)$ are defined through the relation
\begin{equation}
\frac{i}{2}\lambda_{\alpha}\dot{q}_{\alpha}=-A^{\dagger}\partial_0
A.
\end{equation}
Substituting the fields $A\psi$ and $AUA^{\dagger}$ into the
Lagrangian eq.~(\ref{Lag}), it becomes in terms of the collective
variables
\begin{equation}
L=L_{Q}(A,\dot{A})+L_{M}(A,\dot{A})+L_{QM}-\frac{4}{3}\pi R^3 B,
\label{LSU(3)}
\end{equation}
with
\begin{eqnarray}
L_Q&=&\int_Bd^3r\ \bar{\psi}\bigg(i\dpa+\frac{\gamma_0}{2}
\lambda_{\alpha}\dot{q}_{\alpha}\bigg)\psi,\nonumber\\
L_M&=&\frac{1}{2}{\cal I}_{\rm meson}\dot{q}_i^2 +\frac{1}{2}{\cal
I'}_{\rm meson}\dot{q}_M^2+\frac{N_c}{2\sqrt{3}}B_{\bar{B}}
\dot{q}_8 -M_{\rm meson}^0,\nonumber\\
L_{QM}&=&-\frac{1}{2}\int_{\partial B} d^3r\ \bar{\psi}U_5\psi
\Delta_B.
\end{eqnarray}
Here the static meson field $U({\bf
r})=\exp(i\lambda_i\hat{r}_i\theta(r))$ has been integrated out to
get the $L_M$. The index $i\ (M)$ in the $L_M$ runs $1,2,3\
(4,\cdots,7)$. The quantities ${\cal I}_{\rm meson}, {\cal
I'}_{\rm meson}$, $B_{\bar{B}}$, and $M^0_{\rm meson}$ are given
by
\begin{eqnarray}
{\cal I}_{\rm meson}&=&\frac{8\pi}{3}\int_R^{\infty}r^2dr\
\sin^2\theta
\left\{f_{\pi}^2+\frac{1}{e^2}\left[\bigg(\frac{d\theta}{dr}\bigg)^2
+\frac{\sin^2\theta}{r^2}
\right]\right\},\nonumber\\
{\cal I'}_{\rm meson}&=&2\pi\int_R^{\infty}r^2dr(1-\cos\theta)
\left\{f_{\pi}^2+\frac{1}{4e^2}\left[\bigg(\frac{d\theta}{dr}\bigg)^2
+\frac{2\sin^2\theta}{r^2}\right]\right\},\\
{\rm B}_{\bar{B}}&=&-\frac{1}{\pi}[\theta_s-\sin\theta_s\cos\theta_s],\nonumber\\
M_{\rm
meson}^0&=&4\pi\int_R^{\infty}r^2dr\left\{\frac{f_{\pi}^2}{2}
\left[\bigg(\frac{d\theta}{dr}\bigg)^2+\frac{2\sin^2\theta}{r^2}\right]
+\frac{\sin^2\theta}{2e^2r^2}
\left[2\bigg(\frac{d\theta}{dr}\bigg)^2+\frac{\sin^2\theta}{r^2}\right]\right\}.
\nonumber\label{abBM}
\end{eqnarray}
The ${\cal I}_{\rm meson}$ and ${\cal I'}_{\rm meson}$ are the
meson contributions to the moments of inertia. Note that in case
of $R=0$, they are just the moments of inertia of the standard
$SU(3)$ Skyrmion \cite{Skyrme} \cite{pra}. The $B_{\bar{B}}$ is
the baryon number carried by the meson arising from the
Wess-Zumino-Witten term. Since the change of the meson field by
the collective rotation occurs to second order in the time
derivative of $A$, $\theta(r)$ has been substituted by
$\theta_0(r)$, the value which minimizes the energy(mass) $M_{\rm
meson}^0$ in eq.~(\ref{abBM}).

For the quark field, the effect of the collective rotation appears
in first order, as did in the $SU(2)$ case. Replacing
$\mbox{\boldmath $\tau$}\cdot\mbox{\boldmath $\omega$}$ in the
$SU(2)$ case by $\lambda_{\alpha}\dot{q}_{\alpha}$ gives the
equation of motion in the form of
\begin{equation}
\bigg(i\dpa+\frac{\gamma_0}{2}\lambda_{\alpha}\dot{q}_{\alpha}\bigg)\psi=0
\end{equation}
and the boundary condition
\begin{equation}
-i\mbox{\boldmath $\gamma$}\cdot\hat{\bf r}\psi=U_5\psi.
\end{equation}
Assuming an adiabatic collective rotation, the change in the
single quark eigenstate can be calculated by  standard
time-independent perturbation theory. Taking the wave functions of
eq.~(\ref{phi^0}) as the unperturbed solutions, the single quark
eigenstates are given by
\begin{eqnarray}
\phi_n^h(\vec{r})&=&\phi_n^{0h}(\vec{r})+\sum_{m\neq n}
\frac{\left\langle \underline{m}|\frac{\lambda_i}{2}\dot{q}_i
|\underline{n}\right\rangle}
{\varepsilon_m^0-\varepsilon_n^0}\phi_m^{0h}(\vec{r}) +\sum_{m\neq
n}\frac{\left\langle \overline{m}|\frac{\lambda_M}{2}\dot{q}_M
|\underline{n}\right\rangle}
{\omega_m^0-\varepsilon_n^0}+{\cal O}(\dot{q}^2),\nonumber\\
\phi_n^s(\vec{r})&=&\phi_n^{0s}(\vec{r})+\sum_{m\neq n}
\frac{\left\langle \underline{m}|\frac{\lambda_M}{2}\dot{q}_M
|\overline{n}\right\rangle} {\varepsilon_m^0-\omega_n^0},
\label{hedgehogstrange}
\end{eqnarray}
where $\underline{m}$ denotes the hedgehog state and
$\overline{n}$ the strange state, respectively. As mentioned,
$i=1, 2, 3$ and $M=4, 5, 6, 7.$ Note that $\lambda_8$ does not
contribute to these perturbations of the wave functions because
$\phi_m^{0h}$ and $\phi_m^{0s}$ are the eigenstates of $\lambda_8$
with eigenvalues $\frac{1}{\sqrt{3}}$ and $-\frac{2}{\sqrt{3}}$,
respectively. The effect of $\lambda_8$ is only to shift the
eigenvalues of the quark fields. The perturbation modifies the
ground state in a form analogous to eq.~(\ref{H})
\begin{eqnarray}
|{\rm
H}\rangle&=&\exp\bigg(\sum_{\varepsilon_m^0>0,\varepsilon_n^0<0}\frac{1}{2}\cdot
\frac{\langle\underline{m}|\lambda_i\dot{q}_i|\underline{n}\rangle}
{\varepsilon_m^0-\varepsilon_n^0}a_m^{\dagger}b_n^{\dagger}
+\sum_{\varepsilon_m^0>0,\omega_n^0<0}\frac{1}{2}\cdot
\frac{\langle\underline{m}|\lambda_M\dot{q}_M|\overline{n}\rangle}
{\varepsilon_m^0-\omega_n^0}a_m^{\dagger}d_n^{\dagger}
\nonumber\\
&&\hspace{1.5cm}+\sum_{\omega_m^0>0,\varepsilon_n^0<0}\frac{1}{2}\cdot
\frac{\langle\overline{m}|\lambda_M\dot{q}_M|\underline{n}\rangle}
{\omega_m^0-\varepsilon_n^0}c_m^{\dagger}b_n^{\dagger}
+\sum_{\varepsilon_m^0>0}\frac{1}{2}\cdot
\frac{\langle\underline{m}|\lambda_i\dot{q}_i|v\rangle}
{\varepsilon_m^0-\varepsilon_v^0}a_m^{\dagger}a_v
\nonumber\\
&&\hspace{1.5cm}+\sum_{\omega_m^0>0}\frac{1}{2}\cdot
\frac{\langle\underline{m}|\lambda_M\dot{q}_M|v\rangle}
{\omega_m^0-\varepsilon_v^0}c_m^{\dagger} a_v\bigg)|{\rm
H}_0\rangle, \label{HSU(3)}
\end{eqnarray}
where $|{\rm H}_0\rangle$ is defined by eq.~(\ref{H_0}) and
$|v\rangle$ stands for the valence quark state.

The Hamiltonian in terms of the collective variables can be
obtained from the Lagrangian eq.~(\ref{LSU(3)}) as follows. The
canonical momenta $\mbox{\boldmath $\Pi$}_{\alpha}\
(\alpha=1,\cdots,8)$ conjugate to $q_{\alpha}$ are
\begin{eqnarray}
\mbox{\boldmath $\Pi$}_{\alpha}=\frac{\partial L}{\partial
\dot{q}_{\alpha}}&=&\frac{1}{2}\int_B d^3r\
\psi^{\dagger}\lambda_{\alpha}\psi +{\cal I}_{\rm
meson}\dot{q}_i\delta_{i\alpha}\nonumber\\
&&+{\cal I'}_{\rm meson}\dot{q}_M \delta_{M\alpha}
+\frac{N_c}{2\sqrt{3}}B_{\bar{B}}\delta_{8\alpha}.
\end{eqnarray}
Here the quark field operator should be taken as the expectation
value with respect to the rotated hedgehog ground state $|{\rm
H}\rangle$ of eq.~(\ref{HSU(3)}) by consistency with the classical
meson sector. Then, the canonical momenta are written as
\begin{equation}
\mbox{\boldmath $\Pi$}_{\alpha}= ({\cal I}_{\rm quark}+{\cal
I}_{\rm meson})\dot{q}_i\delta_{i\alpha} +({\cal I'}_{\rm
quark}+{\cal I'}_{\rm meson})
\dot{q}_M\delta_{M\alpha}+\frac{\sqrt{3}}{2}\delta_{8\alpha},
\end{equation}
where the following expectation value for the quark field operator
has been used
\begin{equation}
\frac{1}{2}\langle {\rm H}|\int_B d^3r\
\psi^{\dagger}\lambda_{\alpha}\psi\ |{\rm H}\rangle ={\cal I}_{\rm
quark}\dot{q}_i\delta_{i\alpha} +{\cal I'}_{\rm
quark}\dot{q}_M\delta_{M\alpha}
+\frac{\sqrt{3}}{2}Y_R\delta_{8\alpha}
\end{equation}
where the following definitions apply
\begin{eqnarray}
{\cal I}_{\rm quark}&=&
\frac{3}{2}\cdot\sum_{\varepsilon_m^0>0,\varepsilon_n^0<0}
\frac{|\langle\underline{m}|\lambda_3|\underline{n}\rangle|^2}
{\varepsilon_m^0-\varepsilon_n^0}
+\frac{3}{2}\cdot\sum_{m,\varepsilon_m^0>\varepsilon_v}
\frac{|\langle\underline{m}|\lambda_3|v\rangle|^2}
{\varepsilon_m^0-\varepsilon_v^0}, \nonumber\\
{\cal I'}_{\rm quark}&=&
\frac{3}{2}\cdot\sum_{\varepsilon_m^0>0,\omega_n^0<0}
\frac{|\langle\underline{m}|\lambda_4|\overline{n}\rangle|^2}
{\varepsilon_m^0-\omega_n^0}
+\frac{3}{2}\cdot\sum_{\omega_m^0>0,\varepsilon_n^0<0}
\frac{|\langle\overline{m}|\lambda_4|\underline{n}\rangle|^2}
{\omega_m^0-\varepsilon_n^0}
\nonumber\\
&&+\frac{3}{2}\cdot\sum_{m,\omega_m^0>0}
\frac{|\langle\underline{m}|\lambda_4|v\rangle|^2}
{\omega_m^0-\varepsilon_v^0},\nonumber\\
Y_R&=&1-B_{\bar{B}}.
\end{eqnarray}
We have summed over $N_c=3$ colors. By taking the expectation
value of the quark operator with respect to $|{\rm H}\rangle$, the
classical Hamiltonian is obtained as
\begin{eqnarray}
H&=&(M_{\rm quark}^0+M_{\rm meson}^0)+\frac{1}{2}({\cal I}_{\rm
quark} +{\cal I}_{\rm meson})\dot{q}_i^2+\frac{1}{2}({\cal
I'}_{\rm quark}
+{\cal I'}_{\rm meson}) \dot{q}_M^2\nonumber\\
&=&M_0+\frac{\mbox{\boldmath $\Pi$}_i^2}{2{\cal I}}
+\frac{\mbox{\boldmath $\Pi$}_M^2}{2{\cal I'}} \label{1.92}
\end{eqnarray}
with $M^0=M_{\rm quark}^0+M_{\rm meson}^0,\ {\cal I}={\cal I}_{\rm
quark} +{\cal I}_{\rm meson},$ and ${\cal I'}={\cal I'}_{\rm
quark}+{\cal I'}_{\rm meson}.$ Quantization of the Hamiltonian can
be done by promoting the canonical momenta to a quantum mechanical
operator,
\begin{equation}
\tilde{H}=M_0+\frac{1}{2}\bigg(\frac{1}{{\cal I}}-\frac{1}{{\cal
I'}}\bigg) \tilde{\bf J}^2+\frac{1}{2{\cal I'}}(\tilde{\bf
C}_2^2-\tilde{Y}_R^2), \label{HamSU(3)}\end{equation} where
$\tilde{\bf C}_2^2$ is the quadratic Casimir operator for flavor
$SU(3)$, $\tilde{\bf J}^2$ the corresponding one for the spin of
$SU(2)$, and $\tilde{Y}_R$ the ``right" hypercharge operator,
needed to represent the Wess-Zumino-Witten constraint, namely that
physical states obey
\begin{equation}
\tilde{Y}_R|{\rm phys}\rangle=|{\rm phys}\rangle.
\end{equation}
The eigenstates of the Hamiltonian can be written in terms of the
Wigner$-D$ functions \cite{nmp} as
\begin{eqnarray}
\Phi_{a,b}^{(p,q)}&=&\sqrt{{\rm dim}(p,q)}\langle
a|D^{(p,q)}(A)|b\rangle, \nonumber\\
&&|a\rangle=|II_3;Y\rangle,\nonumber\\&&|b\rangle=|I'I'_3;Y'\rangle,
\label{Phi}\end{eqnarray} where $(p,q)$ label the irreducible
representation of $SU(3)$, $D^{(p,q)}(A)$ the corresponding
element, $|a\rangle$ and $|b\rangle$ the basis on which $D(A)$
act, $I$ the isospin of the baryon, $I_3$ the third component, $Y$
the hypercharge and the primed quantities are the right isospin,
right hypercharge etc. With this collective-coordinate wave
function and $|{\rm H}\rangle$ eq.~(\ref{HSU(3)}), the baryon is
described by the wave function of the form
\begin{equation}
|{\rm B}\rangle=\Phi_{a,b}^{(p,q)}\otimes|{\rm
H}\rangle.\label{baryon}
\end{equation}

The mass formula from eq.~(\ref{HamSU(3)}) and eq.~(\ref{Phi}) is
\begin{eqnarray}
M(p,q;II_3Y:JJ_3)&=&M_0+\frac{1}{2}\bigg(\frac{1}{{\cal I}}
-\frac{1}{{\cal I'}}\bigg)J(J+1)\nonumber\\
&+&\frac{1}{2{\cal
I'}}\bigg(\frac{1}{3}\left\{p^2+pq+q^2+3(p+q)\right\}
-\frac{3}{4}\bigg),
\end{eqnarray}
which yields the mass formulas for the baryon octet and decuplet
\begin{eqnarray}
M_8&=&M_0+\frac{3}{8{\cal I}}+\frac{3}{4{\cal I'}},\nonumber\\
M_{10}&=&M_0+\frac{15}{8{\cal I}}+\frac{3}{4{\cal I'}}.
\end{eqnarray}
Since the quark masses are ignored, all the particles in the
baryon octet (decuplet) have the same mass.

\section{The Gluons} The treatment of the non-perturbative
interaction between the pseudoscalar mesons and the quarks has
been discussed in the previous sections. The chiral bag model
contains besides the quarks and the octet pseudoscalar mesons
other degrees of freedom, namely gluons and $\eta'$ meson, which
we will incorporate in a perturbative fashion.

The gluons appear in two ways, as produced by the quark sources
and through the vacuum properties of the cavity, i.e. the so
called vacuum fluctuation.

Let ${\bf E}^a$ and ${\bf B}^a$ be the color electric and magnetic
fields, respectively. The index $a$ denotes the gluon color and
runs from 1 to 8. They satisfy generalized Maxwell equations for
$r<R$,
\begin{eqnarray}
\mbox{\boldmath $\nabla$}\cdot{\bf E}^a&=&J^{0,a},\label{E1}\\
\mbox{\boldmath $\nabla$}\times{\bf E}^a&=&0,\label{E2}\\
\mbox{\boldmath $\nabla$}\cdot{\bf B}^a&=&0,\label{B2}\\
\mbox{\boldmath $\nabla$}\times{\bf B}^a&=&{\bf J}^a,\label{B1}
\end{eqnarray}
with the boundary conditions due to confinement at the bag surface
$r=R$,
\begin{eqnarray}
\hat{\bf r}\cdot{\bf E}^a&=&0,\label{bcforE}\\
\hat{\bf r}\times{\bf B}^a&=&0,\label{bcforM}
\end{eqnarray}
where $J^{\mu,a}$ is the color charge current given by
\begin{equation}
J^{\mu,a}=g_s\bar\psi\gamma^{\mu}\frac{\lambda^a_c}{2}\psi.
\end{equation}
The boundary conditions resemble the case of the perfect conductor
in electrodynamics, but the roles of ${\bf E}^a$ and ${\bf B}^a$
are interchanged due to the structure of the QCD vacuum
\cite{TDLEE}.

The solution of Maxwell equations, for example, eqs.~(\ref{E1})
and (\ref{E2}), can be written as
\begin{equation}
{\bf E}^a({\bf r})=\mbox{\boldmath $\nabla$}\int_B d^3r\ G({\bf
r},{\bf r'}) J^{0,a}({\bf r'}),
\end{equation}
with a proper static cavity propagator which satisfies the
boundary condition \cite{MaxwellVento} \cite{TDLEE2}. Since all
the valence quarks
\footnote{In case of $-\pi\leq\theta_s<-\frac{\pi}{2}$, there is
no valence quark. In this case, the quarks which occupy the lowest
energy $K^P=0^+$ state are taken as valence quarks even though
they are in the negative energy sea.}
have the same quantum numbers except for color, the color charge
density operator, $J^{0,a}_{\rm val}$, and the current operator,
${\bf J}^{a}_{\rm val}$, can be written in the form
\begin{eqnarray}
J_{\rm val}^{0,a}&=&g_s\phi^{\dagger}_v({\bf r})\phi_v({\bf r})
\sum_{\alpha\beta}\langle\alpha|\frac{\lambda^a_c}{2}|\beta\rangle
a^{\dagger}_{\alpha}a_{\beta},
\nonumber\\
{\bf J}_{\rm val}^{a}&=&g_s\phi^{\dagger}_v({\bf
r})\mbox{\boldmath $\alpha$} \phi_v({\bf r})
\sum_{\alpha\beta}\langle\alpha|\frac{\lambda^a_c}{2}
|\beta\rangle a^{\dagger}_{\alpha}a_{\beta},
\end{eqnarray}
where $|\alpha\rangle$ and $|\beta\rangle\
(\alpha,\beta=1,\cdots,N_c)$ denote the color states and
$\phi_v({\bf r})$ is the spatial, spin, and flavor wave function.

With the help of eq.~(\ref{hedgehogstrange}), $\phi_v({\bf r})$
becomes
\begin{equation}
\phi_v({\bf r})=\phi_v^{0h}({\bf r})+\frac{\dot{q}_i}{2}
\cdot\sum_{n,\varepsilon_n^0>\varepsilon_v^0}\frac{\langle
n|\lambda_i|v\rangle} {\varepsilon_n^0-\varepsilon_v^0}\phi_n^{0h}
+\frac{\dot{q}_M}{2}
\cdot\sum_{m,\omega_m^0>\varepsilon_v^0}\frac{\langle
n|\lambda_M|v\rangle} {\omega_m^0-\varepsilon_v^0}\phi_m^{0s},
\end{equation}
up to the lowest order in the collective variables, $\dot{q}_a$.
Here $\phi_v^{0h}$ is the unperturbed hedgehog quark state with
$K^P=0^+$ and $M_K=0$ in the lowest energy level. Because of the
matrix element $\langle n|\lambda_i |v\rangle\ (\langle
m|\lambda_M|v\rangle)$, the summation over $n\ (m)$ is restricted
to the hedgehog quark states with $K^P=1^+$ and $M_K=0,\pm 1$
(strange quark states with $K^P=j^P=\frac{1}{2}^+$ and
$m_j=\pm\frac{1}{2}$). The conditions in the summation are
necessary to be consistent with the particle-hole picture.

Substituting the explicit wave functions to $\phi_v^{0h},
\phi_n^{0h},$ and $\phi_m^{0s}$ leads to
\begin{eqnarray}
J_{\rm val}^{0,a}({\bf r})&=&g_s\frac{\rho'(r)}{4\pi
r^2}\sum_{\alpha,\beta}
\langle\alpha|\frac{\lambda^a_c}{2}|\beta\rangle
a^{\dagger}_{\alpha}a_{\beta},
\nonumber\\
{\bf J}_{\rm val}^{a}({\bf r})&=&-g_s\frac{3}{4\pi}(\hat{\bf
r}\times{\bf S}) \frac{\mu'(r)}{r^3}\sum_{\alpha,\beta}
\langle\alpha|\frac{\lambda^a_c}{2}|\beta\rangle
a^{\dagger}_{\alpha}a_{\beta},
\end{eqnarray}
where ${\bf S}$ is the spin for ${\bf K}=0$ defined by
\begin{equation}
{\bf S}_i=-{\cal I}\dot{q}_i
\end{equation}
in terms of collective variables $q_i\ (i=1,2,3)$ and of the
moment of inertia ${\cal I}$ given in eq.~(\ref{1.92}). The
quantities $\rho'(r)$ and $\mu'(r)$ are given by
\begin{eqnarray}
\rho'(r)&=&N^2r^2(j_0^2(\ \varepsilon_v^0r)+j_1(\varepsilon_v^0r)\ ),\\
\mu'(r)&=&\frac{r^3}{3{\cal
I}}\sum_{n,\varepsilon_n^0>0}\frac{\tilde{\mu}_n}
{\varepsilon_n^0-\varepsilon_v^0}\left\{\begin{array}{l}
\sqrt{\frac{1}{2}}\alpha N_1N(\
j_0(\varepsilon_v^0r)j_1(\varepsilon_n^0r)
-j_1(\varepsilon_v^0r)j_2(\varepsilon_n^0r)\ )\\
-\beta N_2N(\ j_0(\varepsilon_v^0r)j_1(\varepsilon_n^0r)+
j_1(\varepsilon_v^0r)j_0(\varepsilon_n^0r)\ )
\end{array}\right\},\nonumber
\end{eqnarray}
where the sum in $\mu'(r)$ runs over all positive energy
eigenstates of $K^P=1^+$, $\varepsilon_n^0$ is the energy of the
$n$-th eigenstate, and
\begin{equation}
\tilde{\mu}_n=N\beta N_2\int_0^R r^2dr (\
j_0(\varepsilon_v^0r)j_0(\varepsilon_n^0r)
+j_1(\varepsilon_v^0r)j_2(\varepsilon_n^0r)\ ).
\end{equation}
Here $N,\ \alpha N_1$ and $\beta N_2$ are the normalization
constants for the wave functions as appeared in eq.~(\ref{1.28});
\begin{eqnarray}
N^{-2}&=&\int_0^R r^2 dr(\
j_0^2(\varepsilon_v^0r)+j_1^2(\varepsilon_v^0r)\ ),
\nonumber\\
N^{-2}_1&=&\int_0^R r^2 dr(\
j_1^2(\varepsilon_n^0r)+j_2^2(\varepsilon_n^0r)\ ),
\nonumber\\
N^{-2}_2&=&\int_0^R r^2 dr(\
j_0^2(\varepsilon_n^0r)+j_1^2(\varepsilon_n^0r)\ ).
\end{eqnarray}
Note that the color charge current can have a non-vanishing value
due to the perturbation in the valence quark by the collective
rotation. The color charge current does not get any contribution
from the strange quark.

Following the refs.~\cite{MIT} and \cite{Myhrer}, the color
electric and magnetic fields have the form
\begin{eqnarray}
{\bf E}^a_{\rm val}&=&g_s\frac{\rho(r)}{4\pi r^2}\hat{\bf
r}\sum_{\alpha,\beta}
\langle\alpha|\frac{\lambda^a_c}{2}|\beta\rangle
a^{\dagger}_{\alpha}a_{\beta},
\\
{\bf B}^a_{\rm val}&=&g_s\left\{\frac{{\bf S}}{4\pi}\bigg(2M(r)
+\frac{\mu(R)}{R^3}-\frac{\mu(r)}{r^3}\bigg)+\frac{3\hat{\bf
r}}{4\pi} (\hat{\bf r}\cdot{\bf
S})\frac{\mu(r)}{r^3}\right\}\sum_{\alpha,\beta}
\langle\alpha|\frac{\lambda^a_c}{2}|\beta\rangle
a^{\dagger}_{\alpha}a_{\beta},\nonumber \label{EB}
\end{eqnarray}
with
\begin{eqnarray}
\rho(r)&=&\int_0^rdr'\rho'(r'),\nonumber\\
\mu(r)&=&\int_0^rdr'\mu'(r'),\nonumber\\
M(r)&=&\int_r^Rdr'\frac{\mu'(r')}{r^{\prime 3}}.
\end{eqnarray}
These fields are of the same form as those of the MIT bag model
\cite{MIT} except for the numerical details due to the
modification of the valence quark wave function by the chiral
boundary condition.

It is well known that the color electric field given by
eq.~(\ref{EB}) does not satisfy the boundary condition
eq.~(\ref{bcforE}) dynamically. One can have it satisfied by
imposing that hadrons are color singlet states only at the level
of expectation values \cite{MIT}. However, it is quite unnatural
that while the color magnetic field in eq.~(\ref{EB}) and all the
other multi-pole electric fields automatically satisfy the
boundary condition \cite{Myhrer}, the monopole part requires an
additional prescription. Here an alternative choice is proposed to
make the monopole electric field satisfy the boundary condition.
Suppose that a sphere of radius $\varepsilon\ll R$ around the
origin is excluded so that the $\delta$ function term associated
with the equation for the electric field, eq.~(\ref{E1}), is not
present. Then, the most general solution for this field is given
by eq.~(\ref{EB}) where now
\begin{equation}
\rho(r)=\int_{\lambda}^r dr'\rho'(r'),
\end{equation}
with an arbitrary $\lambda$. The confinement boundary condition,
eq.~(\ref{bcforE}), can be satisfied if $\lambda=R $ \cite{pv}.
Because of the singularity of the field at the origin which
introduces an additional $\delta$ source in eq.~(\ref{E1}), this
function is not a solution to the initial problem. However, if one
is willing to accept that the electric field is discontinuous,
{\it i.e.}, zero at the origin, and assumes this function away
from the origin, then all dynamical requirements will be
satisfied. This solution satisfies the boundary conditions at the
price of relaxing the continuity of the electric field inside the
cavity. We classify the solutions of the electric field as;
(solution I) if $\lambda=0$ and (solution II) if $\lambda=R$.

We now enter the description for the Casimir effect. The vacuum
fluctuation of the abelianized gluon fields is described by the
time dependent Maxwell equations without any sources
\begin{eqnarray}
&&\hspace{0.7cm}\mbox{\boldmath $\nabla$}\cdot{\bf E}^a=0,
\hspace{1cm}
\mbox{\boldmath $\nabla$}\cdot{\bf B}^a=0,\nonumber\\
&&\mbox{\boldmath $\nabla$}\times{\bf E}^a= -\frac{\partial {\bf
B}^a}{\partial t},\hspace{1cm} \mbox{\boldmath $\nabla$}\times{\bf
B}^a= \frac{\partial {\bf E}^a}{\partial t} \label{tmeq}
\end{eqnarray}
and satisfy the MIT confinement boundary conditions
eq.~(\ref{bcforE}) and eq.~(\ref{bcforM}). The classical
eigenmodes of the abelianized gluons can be classified by the
total spin quantum number $(J,M)$ given by the vector sum of the
orbital angular momentum ${\bf L}$ and the spin {\bf S},
\begin{equation}
{\bf J}={\bf L}+{\bf S},
\end{equation}
and the radial quantum number $n$. There are two kinds of
classical eigenmodes according to their relations between the
parity and the total spin; (i) M-mode with the parity $P=-(-1)^J$
and (ii) E-mode with the parity $P=-(-1)^{J+1}$. Here the extra
minus sign is due to the negative intrinsic parity of the gluon.

It is convenient to introduce the vector potentials, $G_{\mu}^a$,
and choose the Coulomb gauge condition;
\begin{equation}
G_0^a=0\hspace{1cm}{\rm and}\hspace{1cm}\mbox{\boldmath $\nabla$}
\cdot{\bf G}^a=0.
\end{equation}
Then, the electric field and the magnetic field are obtained in
terms of the vector potential through the relations
\begin{eqnarray}
{\bf E}^a&=&-\frac{\partial {\bf G}^a}{\partial t},\nonumber\\
{\bf B}^a&=&\mbox{\boldmath $\nabla$}\times{\bf G}^a.
\end{eqnarray}
{}Omitting the color index, from  Maxwell equations
eq.~(\ref{tmeq}), the solutions become
\begin{eqnarray}
{\rm (i)\ M-modes}:&&{\bf G}^M_{n,J,M}={\cal N}_Mj_J(\omega_n r)
{\bf Y}_{J,J,M}(\hat{\bf r}),\label{Mmode}\\
{\rm (ii)\ E-modes}:&&{\bf G}^E_{n,J,M}={\cal
N}_E\bigg[-\sqrt{\frac{J}{2J+1}}
j_{J+1}(\omega_nr){\bf Y}_{J,J+1,M}(\hat{\bf r})\nonumber\\
&&\hspace{2cm}+\sqrt{\frac{J+1}{2J+1}}j_{J-1} (\omega_nr){\bf
Y}_{J,J-1,M}(\hat{\bf r})\bigg],
\end{eqnarray}
where ${\bf Y}_{J,l,M}(\hat{\bf r})$ are the vector spherical
harmonics of the total spin $J$ carrying the orbital angular
momentum $l$. The energy eigenvalues are determined by the MIT
boundary conditions eq.~(\ref{bcforE}) and eq.~(\ref{bcforM}) as
\begin{eqnarray}
{\rm (i)\ M-modes}:\ && X_nj'_J(X_n)+j_J(X_n)=0,\label{energyM}\\
{\rm (ii)\ E-modes}:\ && j_J(X_n)=0,\label{energyE}
\end{eqnarray}
where $X_n=\omega_nR$. The normalization constants ${\cal
N}_{M,E}$ will be specified below.

The field operator ${\bf G}({\bf r},t)$ is expanded in terms of
the classical eigenmodes in the form of
\begin{equation}
{\bf G}({\bf r},t)=\sum_{\{\nu\}}\bigg(a_{\{\nu\}}{\bf
G}_{\{\nu\}}({\bf r}) e^{-i\omega_n t}+a^{\dagger}_{\{\nu\}}{\bf
G}^*_{\{\nu\}}({\bf r})e^{i\omega_n t} \bigg), \label{G}
\end{equation}
where $\{\nu\}$ denotes the quantum number set $(n,J,M,\lambda=E\
{\rm or}\ M).$

The normalization constants ${\cal N}_{M,E}$ are determined in
such way that the free gluon Hamiltonian operator
\begin{equation}
H=\frac{1}{2}\int_Bd^3r({\bf E}\cdot{\bf E}+{\bf B}\cdot{\bf B})
\label{HofG}
\end{equation}
becomes
\begin{equation}
H=\sum_{\{\nu\}}\omega_{\{\nu\}}a^{\dagger}_{\{\nu\}}a_{\{\nu\}},
\end{equation}
when eq.~(\ref{G}) is substituted into eq.~(\ref{HofG}). It leads
to a normalization condition for the classical eigenmodes given by
\begin{equation}
\int_Bd^3r\ {\bf G}^*_{\{\nu\}}\cdot{\bf
G}_{\{\mu\}}=\frac{1}{2\omega_{\{\nu\}}} \delta_{\{\nu\}\{\mu\}}.
\end{equation}
Then the normalization constants are determined explicitly
\begin{eqnarray}
{\cal N}_M&=&\{\ X_nR^2[\ j_J^2(X_n)-j_{J-1}(X_n)j_{J+1}(X_n)\ ]\ \}^{-1/2},\\
{\cal N}_E&=&\{\ X_nR^2j_{J-1}^2(X_n)\ \}^{-1/2}.
\end{eqnarray}

\section{The $\eta'$ Meson} The $\eta'$ field is incorporated in
the Lagrangian eq.~(\ref{Lag}) by allowing the $U$ field to be
$U(3)$ valued. Since the $\eta'$ cannot have any topological
structure, it satisfies the usual Klein-Gordon equation of motion
\begin{equation}
(\partial_0^2-\nabla^2+m_{\eta'}^2)\eta'=0.
\end{equation}
Moreover, the $\eta'$ field decouples from the pseudoscalar octet
meson fields. However, there is some secondary coupling between
them via the quark-$\eta'$ interaction on the bag surface. Notice
that the introduction of $\eta'$ field modifies the quark boundary
condition to
\begin{equation}
-i\mbox{\boldmath $\gamma$}\cdot\hat{\bf
r}\psi=\exp\bigg(i\gamma_5
(\eta'/f_0+\lambda_i\hat{r}_i\theta_s)\bigg)\psi,
\end{equation}
which shows how the $\eta'$ field can affect the quark fields {\it
directly} and the hedgehog solution {\it indirectly}. Assuming its
effect to be small, however, the possible modification of the
hedgehog solution by the $\eta'$ field will be not considered. As
in the gluon case the $\eta'$ field will be treated
perturbatively.

The boundary condition for $\eta'$ field arises from the
continuity of the flavor singlet axial current on the bag surface,
\begin{equation}
\langle {\rm H}|\ \hat{\bf r}\cdot\bar{\psi}\mbox{\boldmath
$\gamma$}\gamma_5\psi\ |{\rm H}\rangle=\langle {\rm H}|\ \hat{\bf
r}\cdot(2f_{\pi}\mbox{\boldmath $\nabla$}\eta') \ |{\rm H}\rangle.
\label{bcforeta}
\end{equation}
Before collective coordinate quantization, the hedgehog solution
cannot have a flavor singlet axial current, so that the $\eta'$
field is identically zero. When the hedgehog solution is rotated
by the collective rotation, the matrix element of the flavor
singlet axial current is linear in the spin operator. Thus, in
order to satisfy the boundary condition eq.~(\ref{bcforeta}), the
$\eta'$ field should be linear in the spin operator. One possible
static solution of the Klein-Gordon equation with this constraint
is
\begin{equation}
\eta'=C{\bf S}\cdot\mbox{\boldmath
$\nabla$}\bigg(\frac{e^{-m_{\eta'}r}}{r}\bigg) =-C({\bf
S}\cdot\hat{\bf
r})\bigg(\frac{1+m_{\eta'}r}{r^2}\bigg)e^{-m_{\eta'}r},
\label{eta}
\end{equation}
where the constant $C$ can be determined by the boundary condition
eq.~(\ref{bcforeta}).


\chapter{Anomalies}
\section{Preliminary Remarks} Symmetries and their corresponding
conservation laws play an important role in describing the
fundamental forces of nature. However, it might turn out that a
certain conservation law or symmetry, which is valid in the
classical level, is violated at the quantum level. This phenomenon
is known as the anomaly. If the symmetry so violated is a local
gauge symmetry, then such an anomaly must be cancelled at the
quantum level. This is the so-called anomaly cancellation for
gauge theories. We will see below that this is relevant in our
development. If the symmetry is however global, then the anomaly
can and does manifest itself in observables. A well known example
is the $U_A(1)$ anomaly. To see a role of $U_A(1)$ anomaly in QCD,
let's consider the QCD Lagrangian with the three light quarks
\begin{equation}
{\cal L}=i\bar{\psi}\gamma^{\mu}D_{\mu}\psi-\bar{\psi}m\psi
-\frac{1}{2}G_{\mu\nu}G^{\mu\nu},
\end{equation}
where $D_{\mu}=\partial_{\mu}-ig_s\frac{\lambda^a_c}{2}G_{\mu}^a$
is the covariant derivative and $\lambda^a_c$'s are the Gell-Mann
matrices for the color structure. If we take the current quark
masses to be equal, the QCD Lagrangian is invariant under the
global transformation $\exp(i\lambda^a\theta^a/2)$ in flavor
space. Besides this symmetry, when the current quarks are
massless, the QCD Lagrangian is also invariant under the global
axial transformation $\exp(i\gamma^5\lambda^a\theta^a/2)$ in
flavor space. In other words, the QCD Lagrangian has the chiral
symmetry $SU_L(N_F)\times SU_R(N_F)$ with $N_F$ flavors. Through
the spontaneously symmetry breaking mechanism, the real symmetry
of  QCD becomes $SU_V(N_F)$ and a pseudoscalar octet of Goldstone
bosons appear.

There are two additional global symmetries in the massless QCD
Lagrangian. One is the global $U(1)$ symmetry which corresponds to
the conservation of the baryon number. The other is a global axial
transformation $U_A(1)$. The $U_A(1)$ symmetry requires parity
doublets in the hadron spectrum which are never seen. Therefore,
it should be broken and there be the accompanying Goldstone boson.
The only known candidate for this particle is $\eta'$. However
$\eta'$ is too heavy to be regarded as the Goldstone boson.  This
is known as the $U_A(1)$ problem. The resolution of this problem
may be in the fact that the $U_A(1)$ is not a physical symmetry,
i.e. the $U_A(1)$ symmetry is broken explicitly due to  a quantum
effect, the so called $U_A(1)$ anomaly. In addition to the
$U_A(1)$ problem, without the existence of the $U_A(1)$ anomaly,
the process $\pi_0\rightarrow 2\gamma$ cannot be
understood~\cite{Itz}.\vskip 1ex

We have been discussing, in the previous chapters,  two phase
scenarios in which the theory is described differently in each
phase. We next show that in them the realization of the symmetries
is more complex than in conventional field theory. The presence of
two phases generates two sorts of anomalies, one global and the
other local. The global symmetry involved is the $U_V(1)$
corresponding to the baryon number and the local one is the local
QCD color anomaly. Both should be conserved in a realistic theory.
We described above how the baryon number is conserved in the
two-phase picture. When considered on its own, the bag boundary
induces the baryon charge to leak from the interior which can be
interpreted as an effect of an induced axial current on the
surface which leads to an anomaly in the $U_V(1)$ current. This
baryon charge leaked from the bag interior is picked up by the
hedgehog pion that lives outside of the bag in such a way that the
total baryon charge is preserved~\cite{ccpph}.

As explained below, the bag boundary induces the color charge to
leak out also. In contrast to the baryon charge case, there is no
topological field outside to absorb the color charge accumulated
on the surface, this charge must be cancelled by a boundary
condition. We will find that this requires the presence of a
surface term that violates the local color symmetry. This means
that that the classical action violates the gauge invariance which
is rectified only at the quantum level. The way anomaly figures in
this case is opposite to the global case mentioned above where a
symmetry which is manifest at the classical level gets broken at
the quantum level. \vskip 1ex In this chapter, the first section
is devoted to a derivation of the $U_A(1)$ anomaly in QED, to
understand how it is generated by using the Schwinger's model in
(1+1) dimension, and to extension of the $U_A(1)$ anomaly to the
non-abelian anomaly. In the next section, the relation between the
$U_A(1)$ anomaly and the flavor singlet axial charge, introduced
previously, is described. The discussion on the color anomaly
follows and completes this chapter.

\section{Axial anomaly in QED} There are many methods to obtain
the $U_A(1)$ anomaly~\cite{ChFZ}. Here, the method of taking the
divergence of the axial vector current of QED in the position
space is considered since this method will be used again in the
discussion that follows. It is well known that operator products
at the same space-time point are singular. So the axial vector
current consisting of two fermion field operators
\begin{equation}
j_5^{\mu}(x)=\bar{\psi}(x)\gamma^{\mu}\gamma_5\psi(x)
\end{equation}
may be singular. The regularization of this current leads to the
anomaly~\cite{ChFZ}. The point splitting method may be used to
regularize the current operator. In this regularization the
operators are separated by a small vector $\epsilon^{\mu}$ in the
following way:
\begin{equation}
j_5^{\mu}(x,\epsilon)=\bar{\psi}(x+\epsilon/2)\gamma^{\mu}\gamma_5
\psi(x-\epsilon/2)\exp\bigg(ie\int_{x-\epsilon/2}^{x+\epsilon/2}
dy^{\nu} A_{\nu}(y)\bigg)
\end{equation}
and the regularized axial vector current is defined as
\begin{equation}
j_{5{\rm reg}}^{\mu}(x)=\lim_{\epsilon\rightarrow
0}j_5^{\mu}(x,\epsilon).
\end{equation}
Here, the exponential of the gauge field is introduced to keep
gauge invariance. $\epsilon$ should be sent to zero after all the
calculations have been performed. The Dirac equations
\begin{eqnarray}
(i\dpa -m+e\dda)\psi=0,\nonumber\\
\bar{\psi}(i\dpa +m-e\dda)=0
\end{eqnarray}
yield the divergence of the point-split axial vector current
\begin{equation}
\partial_{\mu}j_5^{\mu}=2imP(x,\epsilon)
-iej_5^{\mu}(x,\epsilon)\epsilon^{\nu}(\partial_{\nu}A_{\mu}(x)
-\partial_{\mu}A_{\nu}(x)),
\end{equation}
where the definition of the point-split pseudoscalar density
\begin{equation}
P(x,\epsilon)=\bar{\psi}(x+\epsilon/2)\gamma_5\psi(x-\epsilon/2)
\exp\bigg(ie\int_{x-\epsilon/2}^{x+\epsilon}dy^{\nu}A_{\nu}(y)\bigg)
\end{equation}
has been introduced. Naively taking the limit $\epsilon\rightarrow
0$ would yield the classical (partial) conservation law,
\begin{equation}
\partial_{\mu}j_5^{\mu}=2im\bar{\psi}\gamma_5\psi
\end{equation}
leading to the exact conservation in the massless limit. However,
this procedure is incorrect since the operator
$j_5^{\mu}(x,\epsilon)$ is singular. Let's consider the vacuum
expectation value of the second term with non-vanishing
$\epsilon$:
\begin{eqnarray}
&&\epsilon^{\nu}\langle0|j_5^{\mu}(x,\epsilon)|0\rangle\nonumber\\
&=&\epsilon^{\nu}{\rm tr}
\bigg(\gamma_5\gamma^{\mu}S(x-\epsilon/2,x+\epsilon/2)\bigg)
\exp\bigg(ie\int_{x-\epsilon/2}^{x+\epsilon/2}
dy^{\rho}A_{\rho}(y)\bigg), \label{epj5}
\end{eqnarray}
where the fermion propagator in the external field $A_{\rho}$ has
been introduced as
\begin{equation}
S(x-\epsilon/2,x+\epsilon/2) =\langle0|{\rm
T}\psi(x-\epsilon/2)\bar{\psi(x+\epsilon/2)}|0\rangle.
\end{equation}
The fermion propagator can be expanded in powers of $A_{\rho}$
\begin{equation}
S(x-\epsilon/2,x+\epsilon/2) =S_0(-\epsilon)+ie\int d^4z\
S_0(x-\epsilon/2-z) \dda(z)S_0(z-x-\epsilon/2)+\cdots,
\end{equation}
where $S_0(x)$ is the free fermion propagator. The first free term
vanishes because of the properties of the Dirac trace. Since the
degree of divergence in the expansion decreases, only the linear
term in $A_{\rho}$ survives. In momentum space the fermion
propagator has the following representation
\begin{equation}
S(x-\epsilon/2,x+\epsilon/2)\simeq
-ie\int(dq)e^{-iqx}\int(dp)e^{ip\epsilon}
\frac{1}{\ddp-m}\dda(q)\frac{1}{\ddp-\ddq-m},
\end{equation}
where the abbreviation, $(dp)=\frac{d^4p}{(2\pi)^4}$, has been
used. Substituting this representation into eq.~(\ref{epj5}) gives
\begin{eqnarray}
&&\epsilon^{\nu}\langle0|j_5^{\mu}(x,\epsilon)|0\rangle =-ie\int
(dq) e^{-iqx}A_{\lambda}(q)\int (dp)
\epsilon^{\nu}e^{ip\epsilon}\nonumber\\
&&\hspace{1cm}\cdot\ {\rm
tr}\frac{\gamma_5\gamma^{\mu}(\dP+m)\gamma^{\lambda}
(\dP-\dq+m)}{(p^2-m^2)(\ (p-q)^2-m^2 \ )}
\exp\bigg(ie\int_{x-\epsilon/2}^{x+\epsilon/2}dy^{\rho}A_{\rho}(y)
\bigg).
\end{eqnarray}
Because of $\gamma_5$, the trace generates a linear divergent
term,
\begin{eqnarray}
&&\epsilon^{\nu}\langle0|j_5^{\mu}(x,\epsilon)|0\rangle=
4e\varepsilon^{\mu\alpha\lambda\beta}\int (dq)e^{-iqx}
q_{\beta}A_{\lambda}(q)
\\
&&\cdot\epsilon^{\nu}\int (dp) e^{ip\epsilon}
\frac{p_{\alpha}}{(p^2-m^2)(\ (p-q)^2-m^2\ )}
\exp\bigg(ie\int_{x-\epsilon/2}^{x+\epsilon/2} dy^{\rho}
A_{\rho}(y)\bigg)\nonumber,
\end{eqnarray}
where the trace of $\gamma$ matrices, ${\rm
tr}\gamma_5\gamma^{\mu}\gamma^{\alpha}\gamma^{\lambda}
\gamma^{\beta}=-4i\varepsilon^{\mu\alpha\lambda\beta}$, has been
performed with the convention $\varepsilon^{0123}=-\varepsilon_{0123}=1$.
For the limit $\epsilon\rightarrow0$, the integral over
$p$ becomes
\begin{eqnarray}
\int (dp)e^{ip\epsilon}\frac{p_{\alpha}}{(p^2-m^2)(\ (p-q)^2-m^2\
)}&\rightarrow&\int(dp)\frac{p_{\alpha}}{p^4}e^{ip\epsilon}\nonumber\\
&=&\frac{\partial}{i\partial\epsilon^{\alpha}}
\int(dp)\frac{e^{ip\epsilon}}{p^4}\nonumber\\
&=&-\frac{1}{8\pi^2}\frac{\epsilon_{\alpha}}{\epsilon^2},
\end{eqnarray}
with the help of the following integral
\begin{equation}
\int(dp)\frac{e^{ip\epsilon}}{p^4}=-\frac{i}{16\pi^2}\ln
\epsilon^2.
\end{equation}
The symmetrization
\begin{equation}
\lim_{\epsilon\rightarrow0}\frac{\epsilon^{\mu}\epsilon^{\nu}}{\epsilon^2}
=\frac{g^{\mu\nu}}{4}\label{2ep}
\end{equation}
leads to
\begin{eqnarray}
&&-ie\epsilon^{\nu}\langle0|j_5^{\mu}(x,\epsilon)
|0\rangle|_{\epsilon\rightarrow0}(\partial_{\nu}A_{\mu}(x)
-\partial_{\mu}A_{\nu}(x))\nonumber\\
&&=-\frac{e^2}{2\pi^2}\varepsilon^{\mu\alpha\lambda\beta}
\partial_{\beta}A_{\lambda}(x) F_{\nu\mu}(x)
\lim_{\epsilon\rightarrow0}
\frac{\epsilon^{\nu}\epsilon_{\alpha}}{\epsilon^2}\nonumber\\
&&=-\frac{e^2}{16\pi^2}\varepsilon^{\alpha\beta\mu\nu}
F_{\alpha\beta}F_{\mu\nu},
\end{eqnarray}
Collecting the results and using the fact that the pseudoscalar
$P(x,\epsilon)$ is regular, the divergence of the regularized
axial vector current has the form of
\begin{equation}
\partial_{\mu}j_{5{\rm reg}}^{\mu}(x)=2imP(x)
-\frac{e^2}{16\pi^2}\varepsilon^{\alpha\beta\mu\nu}
F_{\alpha\beta}F_{\mu\nu}.
\end{equation}
This equation, which expresses the non-conservation of the axial
vector current even for case of the massless fermion, is known as
Adler-Bell-Jackiw anomaly.
\footnote{When we apply the point-split regularization to the
vector current, the term $\epsilon^{\nu}{\rm tr}(\gamma^{\mu}
S(x-\epsilon/2,x+\epsilon/2))$ appears in the divergence of the
vector current. Although ${\rm
tr}(\gamma^{\mu}S(x-\epsilon/2,x+\epsilon/2))$ is quadratically
divergent, due to the symmetric structure as in eq.(\ref{2ep}),
there remains only a logarithmic divergence so that the divergence
of the vector current vanishes.}
It was proved that this anomaly is correct to all orders in
perturbation theory for QED~\cite{AdBa}.

As we have seen in deriving the axial anomaly, the Dirac vacuum (or
sea) plays a crucial role. To see how the axial anomaly is
generated from the Dirac sea~\cite{Shifman91}, let's consider
2-dimensional QED, the Schwinger's model, for simplicity. The
Schwinger's model is composed of one massless fermion coupled to
an abelian gauge field. The corresponding Lagrangian reads
\begin{equation}
{\cal
L}=\bar{\psi}(i\dpa+\dda)\psi-\frac{1}{4}F_{\mu\nu}F^{\mu\nu}.
\end{equation}
We assume that the fermion has unit charge. In two dimensions, we
choose the Dirac matrices as
\begin{equation}
\gamma^0=\sigma_2,\ \ \ \ \ \gamma^1=i\sigma_1,\ \ \ \ \
\gamma^5=\gamma^0\gamma^1=\sigma_3,
\end{equation}
where the $\sigma_i$'s denote the usual Pauli matrices. From the
fact that there is no mass term in the Lagrangian,  chirality is a
good quantum number. Therefore, the  two component Dirac spinor
has the form of
\begin{equation}
\psi=\left(
           \begin{array}{c}
            \psi_L\\
            \psi_R\\
           \end{array}
     \right),
\end{equation}
where $\psi_{L,R}$ are the eigenstates of $\gamma_5$, i.e.
$\gamma_5\psi_{L,R}=\pm\psi_{L,R}$. Classically, there are two
conserved currents:
\begin{equation}
j^{\mu}=\bar{\psi}\gamma^{\mu}\psi,\ \ \ \ \ \ \ \ \
j_5^{\mu}=\bar{\psi}\gamma^{\mu}\gamma_5\psi,
\end{equation}
as in four dimensions. To see the quantum effect on these
currents, let's assume that the system has a finite length $L$ and
satisfies the boundary conditions
\begin{eqnarray}
A_{\mu}(t,x=0)&=&A_{\mu}(t,x=L),\nonumber\\
\psi(t,x=0)&=&-\psi(t,x=L).
\end{eqnarray}
In addition to these assumptions, we choose the Coulomb gauge so
that $A_0$ can be neglected and $A_1$ is independent of $x$. Then,
with the above $\gamma$ matrices, the Dirac equation becomes
\begin{equation}
\left[i\frac{\partial}{\partial
t}+\sigma_3\bigg(\frac{\partial}{\partial x}
-A_1\bigg)\right]\psi=0.
\end{equation}
According to the boundary conditions the fermion wave function can
be expanded into the Fourier series
\begin{equation}
\psi(t,x)=\frac{1}{\sqrt{L}}\sum_{k}u(k)e^{-iE_kt}
\exp\left[i\frac{2\pi}{L}\bigg(k+\frac{1}{2}\bigg)x\right]
\end{equation}
which yields the following energy eigenvalue for the  $L-$ and
$R-$fermion eigenstates
\begin{eqnarray}
E_k^L&=&\frac{2\pi}{L}\bigg(k+\frac{1}{2}\bigg)+A_1, \nonumber\\
E_k^R&=&-\frac{2\pi}{L}\bigg(k+\frac{1}{2}\bigg)-A_1,
\end{eqnarray}
with $k=0, \pm 1, \pm 2, \cdots$. Each type of the fermion has an
infinite tower of energy levels. For $A_1=0$ the energy levels for
$L-$ and $R-$fermions are degenerate. If $A_1$ is not zero, the
levels split; the energy of the $L-$levels increases whereas that
of the $R-$levels decreases. For $A_1=2\pi/L$, the original level
structure is reproduced exactly as it should be because of gauge
invariance.
\footnote{If $A_1$ changes by the finite value, $\frac{2\pi}{L}$,
from $A_1=0$, the Wilson loop,
$$\exp\bigg(i\int_0^L dx A_1(x)\bigg),$$
has the same value as that of $A_1=0$. Therefore,
$A_1=\frac{2\pi}{L}$ is equivalent to $A_1=0$ under a gauge
transformation.}

\begin{figure}
\centerline{\epsfig{file=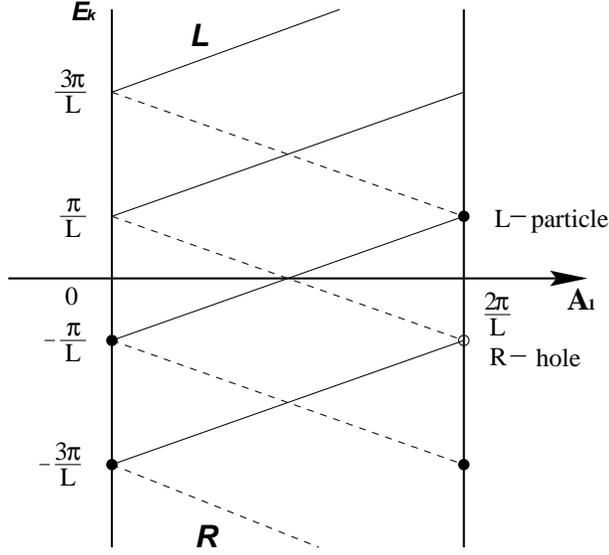,width=8cm}} \caption{Shift
of the energy levels of the fermion due to a change of
$A_1$.}\label{elevel}
\end{figure}

Now suppose that the system is in the vacuum state in which all
negative energy levels are filled up and all positive levels empty
with $A_1=0$. Increasing the value of $A_1$ from $0$ to
$\frac{2\pi}{L}$ produces a $L-$particle and a $R-$hole. This
situation is shown in Fig. \ref{elevel}.
Because the electric charges of the particle and the hole are
opposite, the total electric charge vanishes under the change of
$A_1$:
\begin{equation}
Q(t)=\int dx \ j^0(t,x)=Q_L+Q_R=0,
\end{equation}
where $Q_{L,R}$ are defined by
\begin{equation}
Q_{L,R}=\int dx \ \bar{\psi}_{L,R}\gamma_0\psi_{L,R} =\int dx \
\psi^{\dagger}_{L,R}\psi_{L,R}.
\end{equation}
Consequently the vector current is conserved. On the contrary, the
axial charges can have a non-vanishing value
\footnote{In Fujikawa's method~\cite{Ramond}, a nontrivial Jacobian
appears in the path integral measure of the fermion field when  a
chiral transformation is performed. Using the eigenstates of the
Dirac equation as basis, we can construct the Jacobian which
contains the term
$$\sum_n\psi_n^{\dagger}(x)\gamma_5\psi_n(x),$$ which is equal to
the difference of the zero modes of each chirality once the volume
integration is performed. The Atiyah-Singer's index theorem gives
the following result:
$$\sum_{n,{\rm zero}}\int d^4x\ \psi_n^{\dagger}(x)\gamma_5\psi_n(x)
=\frac{e^2}{16\pi^2}\int d^4x \ F_{\mu\nu}\tilde{F}^{\mu\nu}$$ in
the QED of the (3+1) dimensions.}
according to its definition:
\begin{equation}
Q_5=\int dx  \ j_5^0(t,x)=Q_L-Q_R=2.
\end{equation}
We can rewrite this expression as follows:
\begin{equation}
\Delta Q_5=\frac{L}{\pi}\Delta A_1
\end{equation}
and per unit time
\begin{equation}
\frac{\Delta Q_5}{\Delta t}=\frac{L}{\pi}\frac{\Delta A_1}{\Delta
t}.
\end{equation}
Using its definition, we obtain the relation for the axial current
\begin{equation}
\partial_0j_5^0=\frac{1}{\pi}\partial_0A_1
\end{equation}
and finally we can write the anomaly for Schwinger's model in a
Lorentz invariant form
\begin{equation}
\partial_{\mu}j_5^{\mu}=\frac{1}{\pi}\varepsilon_{\mu\nu}\partial^{\mu}A^{\nu}
=\frac{1}{2\pi}\varepsilon_{\mu\nu}F^{\mu\nu}.
\end{equation}
\vskip 1ex For the explicit calculation that we will perform in
the next chapter, we consider the flavor singlet axial current in
QCD. It has the same anomaly equation except for the appropriate
group theoretic factor. Applying the same regularization as in the
preceding discussion and with gluon fields,
$G_{\mu}=\frac{\lambda^a}{2}G^a_{\mu}$, where
$\frac{\lambda^a}{2}$'s are the Gell-Mann matrices, the flavor
singlet axial current, $A^0_{\mu}$, satisfies the following
anomaly equation:
\footnote{From now on, the notation, $G^a_{\mu\nu}$ for the gauge
field strength will be used for non-abelian case.}
\begin{eqnarray}
\partial^{\mu}A_\mu^0&=&-\frac{N_F\alpha_s}{2\pi}\
{\rm Tr}{\bf G}_{\mu\nu}\tilde{\bf G}^{\mu\nu}\nonumber\\
&=&-\frac{N_F\alpha_s}{4\pi}\ \sum_a\
G_{\mu\nu}^a\tilde{G}^{\mu\nu}_a,\label{singletanomaly}
\end{eqnarray}
where the trace of the Gell-Mann matrices, ${\rm
Tr}\bigg(\frac{\lambda}{2}\frac{\lambda^b}{2}\bigg)=\frac{\delta^{ab}}{2},$
has been used in the last line and $N_F$ is the flavor number of
quark.

\section{Application of the $U_A(1)$ anomaly to the flavor
singlet axial charge} As discussed in Chapter 2, the flavor
singlet axial anomaly supplies the key to understand the proton
spin problem. In this section we give the relation between the
flavor singlet axial charge and the singlet axial anomaly. From
eq.~(\ref{singletanomaly}) and the fact that
$G_{\mu\nu}^a\tilde{G}^{\mu\nu}_a$ can be written in terms of the
Chern-Simons current, $K^{\mu}$ as
\begin{equation}
\frac{1}{2}G_{\mu\nu}^a\tilde{G}^{\mu\nu}_a=\partial_{\mu}K^{\mu},
\end{equation}
with the definition of $K^{\mu}$
\begin{equation}
K^{\mu}=\epsilon^{\mu\nu\rho\sigma}G_{\nu}^a\bigg(G^a_{\rho\sigma}
-\frac{g_s}{3}f_{abc}G_{\rho}^bG_{\sigma}^c\bigg),
\end{equation}
where $f_{abc}$ is the structure constant of QCD, we can define
the following conserved axial vector current for massless
quarks (in the chiral limit):
\begin{equation}
\tilde{A}_{\mu}^0=A_{\mu}^0-N_F\frac{\alpha_s}{2\pi}K_{\mu}.
\label{tildea}
\end{equation}

In the gauge,
\footnote{The gluon spin and orbital angular are not separately
gauge invariant and hence a choice of gauge is necessary. More
detail is given in the next chapter}
$G^a_0(x)=0$, the gluon spin spin operator, $\hat{\bf S}^g$,
becomes
\begin{equation}
\hat{\bf S}^g_i=-\epsilon_{ijk}G^j_a\partial^0G_a^k.
\end{equation}
Moreover, in this gauge, the cubic term  vanishes for the spatial
components of $K_{\mu}$. Thus, one finds the relation,
\begin{equation}
K_i=-\hat{\bf S}^g_i,
\end{equation}
and its forward proton matrix element,
\begin{equation}
\langle p,S|K^{\mu}|p,S\rangle=-S^{\mu}\Delta g, \label{deltag}
\end{equation} where $S^{\mu}$ is the proton spin
and $\Delta g$ is the net helicity of the gluon along the proton
spin.

Due to the conservation of $\tilde{A}_{\mu}^0$ in the chiral
limit, its forward proton matrix elements are independent of the
renormalization scale and the form factor, $\tilde{a}^0$, defined
through the matrix element
\begin{equation}
\langle p, S|\tilde{A}_{\mu}^0|p,S\rangle=\tilde{a}^0S_{\mu}
\end{equation}
should correspond with the value in the quark-model, i.e.,
$\tilde{a}^0=\Delta\Sigma.$ From eq.~(\ref{tildea}) and
eq.~(\ref{deltag}), therefore, one has the scale dependent flavor
singlet axial charge in terms of $\Delta\Sigma$ and $\Delta g$ as:
\begin{equation}
a^0(Q^2)=\Delta\Sigma-N_F\frac{\alpha_s}{2\pi}\Delta g(Q^2),
\label{a^0}
\end{equation}
as given in Chapter 2.

Finally, let's consider the gauge dependence of the Chern-Simons
current. For an abelian case like QED, the gauge transformation,
\begin{equation}
A_{\mu}(x)\rightarrow A_{\mu}(x)+\partial_{\mu}\Lambda(x),
\end{equation}
induces the change in the Chern-Simon current
\begin{equation}
K_{\mu}(x)\rightarrow
K_{\mu}(x)-\frac{1}{2}[\partial_{\mu}\Lambda(x)]
\epsilon_{\mu\nu\rho\sigma}F^{\rho\sigma}(x).
\end{equation}
While the Chern-Simons current changes, its forward proton matrix
element (or expectation value) does not since the expectation value
of $F^{\rho\sigma}$ vanishes due to the derivatives in the
definition of the field strength. Thus, the flavor singlet axial
charge is gauge invariant for the abelian case.

On the other hand, for the non-abelian case as in QCD, the
situation is more subtle. Under
\begin{equation}
G_{\mu}(x)\rightarrow
U(x)G_{\mu}(x)U^{-1}(x)+\frac{i}{g_s}(\partial_{\mu}U(x))
U^{-1}(x),
\end{equation}
with the definition, $G_{\mu}=G_{\mu}^a\frac{\lambda_c^a}{2}$, the
change of the Chern-Simons current becomes
\begin{eqnarray}
K_{\mu}&\rightarrow&
K_{\mu}+\frac{2i}{g_s}\epsilon_{\mu\nu\rho\sigma}
\partial^{\nu} {\rm Tr}(G^{\alpha}U^{-1}\partial^{\beta}U)\nonumber\\
&&+\frac{2}{3g_s^2}\epsilon_{\mu\nu\rho\sigma}{\rm
Tr}\{U^{-1}(\partial^{\nu}U)
U^{-1}(\partial^{\alpha}U)U^{-1}(\partial^{\sigma}U)\}.
\end{eqnarray}
The second term is a total divergence so that does not contribute
to the forward proton matrix element. Although the third term can
also be shown to be a divergence~\cite{Jac85}, but it cannot be
discarded because of the non-trivial topological
structure~\cite{'tHoot76} of QCD. As a result, its forward proton
matrix element is not gauge invariant. To avoid these problems in
our discussion, we will treat gluons as abelianized fields in the
next discussions.

\section{Color anomaly in the chiral bag
model~\cite{nrwz} \cite{ccpph}} In the previous chapter, the
boundary conditions for the confined gluons in the chiral bag
model have been given as
\begin{equation}
\hat{\bf r}\cdot{\bf E}^a=0,\hspace{1cm}\hat{\bf r}\times{\bf
B}^a=0 \label{MIT}
\end{equation} with $\hat{\bf r}$ the outward
unit vector normal to the bag surface and $a$ the color index.
These conditions mean that the color electric fields, ${\bf E}^a$,
point along the surface, while the color magnetic fields, ${\bf
B}^a$, are orthogonal to the surface. These are the usual MIT bag
boundary conditions. We will show that the leakage of the color
charge resulting from the $\eta'$ field at the bag surface, as
discussed at the beginning of this chapter, makes these boundary
conditions change. As a result, the color electric field has a
component normal to the bag surface and the color magnetic field a
component along the surface.

The boundary condition for the quarks due to $\eta'$ field at the
bag surface is given by
\begin{equation}
(i\mbox{\boldmath $\gamma$}\cdot\hat{\bf
r}+e^{i\gamma_5\eta(\beta)})\psi=0, \label{bcforeta'}
\end{equation}
where $\eta=\eta'/f_{\eta'}\equiv\eta'/f_0$ and $\beta$ is a point
on the surface. At the classical level this boundary condition
makes the color current of the quark, confined inside the bag, to
have no leakage outside. As a result, the color charge of the bag
is a constant in time,
\begin{eqnarray}
\langle0|\dot{Q}^a|0\rangle&=&-\int_{\Sigma} d\beta\ \langle0|{\bf
j}^a
\cdot\hat{\bf r}|0\rangle \nonumber\\
&=&-g_s\oint_{\Sigma} d\beta\ \langle0|\ \bar{\psi}
\mbox{\boldmath $\gamma$}\cdot\hat{\bf r} \frac{\lambda^a}{2}\psi\
|0\rangle =0,
\end{eqnarray}
where we have made use of the quasi-abelian approximation for
simplicity and $\Sigma=\partial B$. Once the calculation is
completed, we will extract the non-abelian structure by
inspection.

The quantum correction to $\dot{Q}^a$ due to the $\eta$ can be
obtained by introducing the gauge invariant point-splitting
regularization to the color current operator as before. If we use
a point-splitting in time direction and choose the temporal gauge
condition, $G_0^a=0$, the regularized color current operator in
the quasi-abelian approximation at the surface becomes
\begin{eqnarray}
{\bf j}^a_{\rm reg}(\beta)&=&g_s\bar{\psi}(\beta+\epsilon/2)
\mbox{\boldmath $\gamma$}\
\frac{\lambda^a}{2}\exp\bigg(ig_s\int_{\beta-\frac{1}{2}\epsilon}
^{\beta+\frac{1}{2}\epsilon}dz^{\mu}G_{\mu}(z)\bigg)
\psi(\beta-\epsilon/2)
\nonumber\\
&=&g_s\bar{\psi}(\beta+\epsilon/2)\mbox{\boldmath $\gamma$}\
\frac{\lambda^a}{2} \psi(\beta-\epsilon/2).
\end{eqnarray}
Then, we have
\begin{equation}
\dot{Q}^a\equiv\langle0|\dot{Q}^a|0\rangle
=-g_s\lim_{\epsilon\rightarrow 0} \oint_{\Sigma} d\beta\
\langle0|\ \bar{\psi}(\beta+\epsilon/2)\mbox{\boldmath $\gamma$}
\cdot\hat{\bf r}\frac{\lambda^a}{2}\psi(\beta-\epsilon/2) \
|0\rangle
\end{equation}
as the regularized expression. Using the boundary condition,
eq.~(\ref{bcforeta'}), the effect of the $\eta$ appears explicitly
as
\begin{eqnarray}
\dot{Q}^a &=&g_s\lim_{\epsilon\rightarrow 0} \oint_{\Sigma}
d\beta\ \langle0|\ \bar{\psi}(\beta+\epsilon/2)\mbox{\boldmath
$\gamma$} \cdot\hat{\bf r}\frac{\lambda^a}{2}
e^{-i\gamma_5\eta(\beta+\epsilon/2)}
e^{i\gamma_5\eta(\beta-\epsilon/2)} \psi(\beta-\epsilon/2) \
|0\rangle
\nonumber\\
&=&-i\frac{g_s}{2} \lim_{\epsilon\rightarrow 0}\epsilon
\oint_{\Sigma} d\beta\ \langle0|\
\bar{\psi}(\beta+\epsilon/2)\mbox{\boldmath $\gamma$}
\cdot\hat{\bf r}\frac{\lambda^a}{2} \gamma_5
\psi(\beta-\epsilon/2)\ |0\rangle \dot{\eta}(\beta),
\end{eqnarray}
where the relation
\begin{eqnarray}
e^{-i\gamma_5\eta(\beta+\epsilon/2)}
e^{i\gamma_5\eta(\beta-\epsilon/2)} &=&
e^{-i\gamma_5(\eta(\beta+\epsilon/2)-\eta(\beta-\epsilon/2))}
e^{-\frac{1}{2}[\eta(\beta+\epsilon/2),\ \eta(\beta-\epsilon/2)]}\nonumber\\
&\simeq&1-i\gamma_5\dot\eta(\beta)\epsilon
\end{eqnarray}
up to $O(\epsilon)$, and  point splitting in the time direction,
have been used. Rewriting the vacuum expectation values in terms
of the components gives
\begin{equation}
\dot{Q}^a=-i\frac{g_s}{2}\lim_{\epsilon\rightarrow
0}\epsilon\oint_{\Sigma} d\beta\ {\rm
Tr}\bigg(\gamma_5\vec{\gamma}\cdot\hat{\bf r}\frac{\lambda^a}{2}
S^+(\beta-\epsilon/2;\beta+\epsilon/2)\bigg)\dot{\eta}(\beta),
\end{equation}
where $S^+(\beta,\beta')$ is defined by
\begin{equation}
S^+(\beta,\beta')=\lim_{(x,x')\rightarrow(\beta,\beta')}S(x,x')
\end{equation}
from the confined fermion propagator $S(x,x')$. The multiple
reflection expansion method~\cite{HanssonJaffe83} produces the
relation
\begin{equation}
S^+(\beta-\epsilon/2;\beta+\epsilon/2)=\frac{1}{4}
(1+\mbox{\boldmath $\gamma$}_E\cdot\hat{\bf
r}U_5)S(\beta-\epsilon/2;\beta+\epsilon/2) (3-U_5\mbox{\boldmath $
\gamma$}_E\cdot\hat{\bf r})\label{S+}
\end{equation}
in Euclidean space.
\footnote{The $\gamma$ matrices of  Euclidean space are related to
those of  Minkowski space according to the rules: ${\bf x}_E={\bf
x}, x_4=ix_0, \mbox{\boldmath $\gamma$}_E=-i\mbox{\boldmath
$\gamma$}, \gamma_4=\gamma_0, \ {\rm and}\
\gamma_5=-\gamma^1_E\gamma^2_E\gamma^3_E\gamma^4$ so that all
$\gamma_{\mu}$ become the hermitian matrices. The boundary
conditions for the quark due to $\eta$ becomes
$$\mbox{\boldmath $\gamma$}_E\cdot\hat{\bf r}\psi=U_5\psi \ \ {\rm and}\ \
\bar{\psi}\mbox{\boldmath $\gamma$}_E\cdot\hat{\bf
r}=-\bar{\psi}U_5.$$ From these, the confined fermion propagator
satisfies the boundary conditions
$$U_5(\alpha)S(\alpha,x')=\mbox{\boldmath $\gamma$}_E\cdot\hat{\bf r}S(\alpha,x')
\ \ {\rm and}\ \
S(x,\alpha)U_5(\alpha)=-S(x,\alpha)\mbox{\boldmath $
\gamma$}_E\cdot\hat{\bf r}$$ for any vector $\alpha$ on the
boundary.}
Here $U_5=e^{i\gamma_5\eta}$. Assuming that $\eta$ on the bag
surface does not depend on the location, but only on time,
\footnote{One can generalize the result to a variation with a
space-time dependent $\eta(x,t)$ by replacing the ordinary
derivative by a functional derivative.}
we have
\begin{eqnarray}
\frac{d Q^a}{d\eta}&=& -i\frac{g_s}{2}\lim_{\epsilon\rightarrow
0}\epsilon\oint_{\Sigma} d\beta\ {\rm
Tr}\bigg(\gamma_5\mbox{\boldmath $\gamma$}\cdot\hat{\bf
r}\frac{\lambda^a}{2}
S^+(\beta-\epsilon/2;\beta+\epsilon/2)\bigg)\nonumber\\
&=&\frac{g_s}{2}\lim_{\epsilon\rightarrow 0}\epsilon\oint_{\Sigma}
d\beta_E\ {\rm
Tr}\bigg(\gamma_5\frac{\lambda^a}{2}(\mbox{\boldmath $
\gamma$}_E\cdot\hat{\bf r}+U_5)
S(\beta-\epsilon/2;\beta+\epsilon/2)\bigg)\nonumber\\
&=&g_s\lim_{\epsilon\rightarrow 0}\epsilon\int_{\Sigma} d\beta_E\
{\rm Tr}\bigg(\gamma_5\frac{\lambda^a}{2} \mbox{\boldmath $
\gamma$}_E\cdot\hat{\bf r}
S(\beta-\epsilon/2;\beta+\epsilon/2)\bigg).\label{Q/eta}
\end{eqnarray}
Here the eq.~(\ref{S+}) and the boundary condition of the confined
propagator in the footnote have been used in order to get the last
result. Because of the structure of the $\gamma$ matrices in
eq.~(\ref{Q/eta}) the singular contribution arises from the first
order term in the gluon interaction in the perturbative expansion
of the confined fermion propagator. In the Minkowski space the
confined fermion propagator is
\begin{equation}
S(\beta-\epsilon/2;\beta+\epsilon/2)\sim\int_B d^4x\
S_0(\beta-\epsilon/2-x) (-ig_s\gamma\cdot
G^b(x)\lambda^b/2)S_0(x-\beta-\epsilon/2),
\end{equation}
where $S_0(x,x')$ is the free fermion propagator. Substituting
this result into the eq.~(\ref{Q/eta}) leads to the integral in
momentum space,
\begin{eqnarray}
\frac{d Q^a}{d\eta}&=&-ig_s\lim_{\epsilon\rightarrow 0}\epsilon
\oint_{\Sigma} d\beta\ {\rm Tr}\bigg(\gamma_5\mbox{\boldmath $
\gamma$}\cdot\hat{\bf r}
\frac{\lambda^a}{2}S(\beta-\epsilon/2;\beta+\epsilon/2)\bigg)\nonumber\\
&=&g_s^2\lim_{\epsilon\rightarrow 0}\epsilon\oint_{\Sigma} d\beta\
{\rm Tr}\bigg(\gamma_5\mbox{\boldmath $\gamma$}\cdot\hat{\bf
r}\frac{\lambda^a}{2} \frac{\lambda^b}{2}\int (dq)\
G^b_{\alpha}(q)e^{iq\cdot(\beta+\epsilon/2)}\nonumber\\
&&\hspace{3.5cm}\cdot\
\int(dp)\frac{1}{\ddp}\gamma^{\alpha}\frac{1}{\ddp-\ddq}
e^{ip\cdot\epsilon} \bigg).
\end{eqnarray}
From the relations
\begin{equation}
{\rm
Tr}(\gamma^5\gamma^i\gamma^{\beta}\gamma^{\alpha}\gamma^{\sigma})
=-4i \epsilon^{i\beta\alpha\sigma},\hspace{1cm} {\rm
Tr}\bigg(\frac{\lambda^a}{2}\frac{\lambda^b}{2}\bigg)
=\frac{\delta^{ab}}{2}
\end{equation}
and the fact that for the limit $\epsilon\rightarrow 0$ the
integral over $p$ becomes
\begin{equation}
\int (dp)\frac{p_{\beta}}{p^4}e^{ip\cdot\epsilon}
=-\frac{1}{8\pi^2}\frac{\epsilon_\beta}{\epsilon^2},
\end{equation}
we arrive at the final result in the form of a surface integral
\begin{equation}
\frac{d Q^a}{d\eta}=
-\frac{g_s^2}{4\pi^2}\epsilon^{0i\alpha\sigma}\oint_{\Sigma}
d\beta\ \hat{\bf r}_i\partial_{\alpha}G^a_{\sigma}(\beta).
\end{equation}
We can write this result in terms of the color magnetic fields in
the quasi--abelian case as:
\begin{equation}
\frac{d Q^a}{d\eta}=N_F\frac{g_s^2}{4\pi^2}\oint_{\partial B}
d\beta\ {\bf B}^a\cdot\hat{\bf r}, \label{colorano.}
\end{equation}
when there are $N_F$ massless quarks. Note that
eq.~(\ref{colorano.}) has been obtained with unconstraint radiated
gluon. Since we consider a confine theory, there should be the
additional factor of 1/2 in the final result by making gluon
confined only inside the bag. Then, the result becomes
\begin{equation}
\frac{d Q^a}{d\eta}=N_F\frac{g_s^2}{8\pi^2}\oint_{\partial B}
d\beta\ {\bf B}^a\cdot\hat{\bf r}. \label{Ncolorano.}
\end{equation}

Eq.~(\ref{Ncolorano.}) means that color charge disappears from the
bag due to the time variation of the $\eta$ field which is the
artifact of the effective approach as ours. In other words, the
color charge confined at the classical level leaks out at the
quantum level. A simple way to remedy this problem is to introduce
a surface term of the following type as a counter term in the
action to remove the artifact:
\begin{equation}
S_{\rm CT}\sim-N_F\frac{g_s^2}{8\pi^2}\oint_{\partial
B}d\beta\ G_0^a {\bf B}^a\cdot\hat{\bf r}\eta(\beta). \label{LCT0}
\end{equation}
Chiral invariance, covariance, and general gauge transformation
properties allow us to rewrite eq.~(\ref{LCT0}) in a general form
\begin{equation}
S_{\rm CT}=i\frac{g_s^2}{32\pi^2}\oint_{\partial B}d\beta\
K^{\mu}n_{\mu} {\rm Tr}(\ln U^{\dagger}-\ln U), \label{LCT}
\end{equation}
where $K^{\mu}$ is the Chern-Simons current defined by
\begin{equation}
K^{\mu}=\epsilon^{\mu\nu\alpha\beta}\bigg(G^a_{\nu}G^a_{\alpha\beta}-\frac{1}{3}
g_sf^{abc}G^a_{\nu}G^b_{\alpha}G^c_{\beta}\bigg)
\end{equation}
and $n_{\mu}$ is the outward unit four vector normal to the bag
surface. Here, $U=e^{i\eta'/f_0}e^{i\pi/f}$ as given in the
previous chapter. This counter term describes that the colorless
$\eta$ field outside the bag interacts with gluons at the bag
surface by the help of quarks. We expect that this term may be
generated naturally in the effective action of the chiral bag
model as the interaction term between the neutral pion and photons
in the effective action of QED~\cite{effectiveaction}.

Note that eq.~(\ref{LCT}) is not gauge invariant on the bag
surface because of the Chern-Simons current. Thus at the classical
level, the Lagrangian is not gauge invariant. However, at the
quantum level, the invariance is restored by the cancellation
between the anomalous term, eq.~(\ref{Ncolorano.}), and the
surface term, eq.~(\ref{LCT}).

Including this color anomaly phenomenon, the action for the chiral
bag given in the previous chapter is generalized to
\begin{eqnarray}
S&=&S_B+S_{\bar{B}}+S_{\partial B}\nonumber\\
S_B&=&\int_B d^4x\bigg(\bar{\psi}i\dd\psi-\frac{1}{2}{\rm
Tr}G_{\mu\nu}G^{\mu\nu}\bigg),\nonumber\\
S_{\bar{B}}&=&\frac{f_\pi^2}{4}\int_{\bar{B}}\bigg({\rm
Tr}\partial_{\mu}U^{\dagger}\partial^{\mu}U+\frac{1}{4N_F}m_{\eta'}^2[{\rm
Tr}(\ln U^{\dagger}-\ln U)]^2\bigg)+\cdots+S_{WZW},\nonumber\\
S_{\partial B}&=&\frac{1}{2}\oint_{\partial B} d\Sigma^{\mu}
\left\{n_{\mu}\bar{\psi}U_5\psi+i\frac{g_s^2}{16\pi^2}K_{\mu} {\rm
Tr}(\ln U^{\dagger}-\ln U)\right\}.
\end{eqnarray}

From the presence of the Chern-Simons term in the surface action,
the boundary conditions for gluon fields at classical level are
affected. In place of the MIT conditions eq.~(\ref{MIT}), we have
instead
\begin{eqnarray}
\hat{\bf r}\cdot{\bf E}^a&=&-\frac{N_Fg_s^2}{8\pi^2f_0}\hat{\bf
r}\cdot{\bf B}^a\eta',\nonumber\\
\hat{\bf r}\times{\bf B}^a&=&\frac{N_Fg_s^2}{8\pi^2f_0}\hat{\bf
r}\times{\bf E}^a\eta'. \label{colorBC}
\end{eqnarray}
The fact that the color electric field can have a radial component
contrary to the MIT conditions plays an important role on the
proton spin problem. Besides, the $\eta'$ mass can be estimated by
using these boundary conditions and the axial anomaly~\cite{NRWZ2}.


\chapter{Flavor singlet axial charge in the chiral bag model}

In this chapter, we present the calculation of the Flavor Singlet
Axial Charge (FSAC) in the chiral bag model scenario and
demonstrate how the Cheshire Cat
Principle~\cite{ccp,ccpph} operates for the observables that are
not topological as baryon charge discussed in a previous chapter.
In order to do so we need a specific formulation of the model
through its equations of motion and boundary conditions. We should
stress that although we are truncating the model, the model itself
is quite general and represents low-energy dynamics of QCD.

We recall that the equations of motion have been shown in Chapter
3 and the color boundary conditions have been introduced in
Chapter 4. Our calculation will be carried out in the static
spherical cavity approximation, that is, our bag, polarized along
$z-$direction, is a static sphere of radius $R$ dividing two
regions of space in which the theory is implemented by QCD for $r<
R$, and by an effective meson theory for $r > R $. Besides, we
treat, throughout the calculation, the gluons as abelian fields,
i.e., we work in the lowest order approximation to QCD in
perturbation theory.

First we calculate the various static contributions to this
observable, and then, after considering the gluon spin, we proceed
to include the Casimir contribution which arises due to the change
in the boundary conditions of gluons associated to the color
anomaly.

\section{The formalism} To obtain the FSAC, we need to calculate
the matrix elements of the flavor singlet axial current. According
to the Cheshire Cat Principle (CCP) discussed in Chapter 3, the
flavor singlet current appears in the chiral bag model as the sum
of two terms, one coming from the interior of the bag and the
other from the outside,  populated among others by the
$\eta^\prime-$meson \footnote{From now on we will omit the prime
and write simply $\eta$ for this meson.}
\begin{equation}
A^0_{\mu} =A^0_{\mu,B} \Theta_B + A^0_{\mu,\bar{B}}
\bar{\Theta}_B.\label{current}
\end{equation}
Here, $\Theta_B$ and $\bar{\Theta}_B$ are defined,  as before, as
$\Theta_B=\theta(R-r)$ and $\bar{\Theta}_B=1-\Theta_B$, with $R$
being the radius of the bag. We demand that the $U_A(1)$ anomaly
is given in the model by
\begin{equation}
\partial_\mu A^0_{\mu} =
-\frac{\alpha_s N_F}{4\pi}\sum_a G_{\mu\nu}^a \tilde{G}^{\mu\nu,a}
\Theta_{B}+ 2f_\pi m_\eta^2 \eta \bar{\Theta}_{B}.\label{abj}
\end{equation}
Our task is to construct a FSAC in the chiral bag model that is
gauge-invariant and consistent with this anomaly equation. Our
basic assumption is that in the nonperturbative sector outside of
the bag, the only relevant $U_A(1)$ degree of freedom is the
massive $\eta$ field. This assumption allows us to write
\begin{equation}
A^0_{\mu,\bar{B}} = A^0_{\mu,\eta} = 2f_\pi\partial^\mu \eta
\end{equation}
with its divergence given by
\begin{equation} \partial^\mu A^0_{\mu,\eta} =
2f_\pi m_\eta^2 \eta.
\end{equation}
The immediate question to ask is: what is the gauge-invariant and
regularized $A^0_{\mu,B}$ such that the anomaly equation,
eq.~(\ref{abj}), is satisfied? To address this question, we
rewrite the current, eq.~(\ref{current}), absorbing the theta
functions, as
\begin{equation}
A^0_{\mu}=A^0_{\mu,B_{Q}}+ A^0_{\mu,B_{G}} + A^0_{\mu,\eta}
\label{sep}
\end{equation}
such that
\begin{eqnarray}
\partial^\mu (A^0_{\mu,B_{Q}} + A^0_{\mu,\eta}) &=& 2f_\pi m_\eta^2 \eta
\bar{\Theta}_{B},\label{Dbag}\\
\partial^\mu A^0_{\mu,B_G}&=&
\frac{\alpha_s N_F}{\pi}\sum_a {\bf E}^a \cdot {\bf B}^a
\Theta_{B}, \label{Dmeson}
\end{eqnarray}
where the relation $G_{\mu\nu}^a\tilde{G}^{\mu\nu,a}=-4{\bf
E}^a\cdot{\bf B}^a$ has been used. The sub-indices $Q$ and $G$
signal that these currents are written in terms of quark and gluon
fields respectively. In writing eq.~(\ref{Dbag}), we have ignored
the up and down quark masses. Going to the static situation,
appropriate to our discussion, eq.~(\ref{Dmeson}) can be written
as
\begin{equation}
\mbox{\boldmath $\nabla$}\cdot{\bf A}^0_{B_G}=\frac{\alpha_s
N_F}{\pi}\sum_a {\bf E}^a \cdot {\bf B}^a
\Theta_{B}\label{SDmeson}.
\end{equation}

We should stress that since we are dealing with an interacting
theory, there is no unique way to separate the different
contributions from the gluon, quark and $\eta$ components. In
particular, the separation we adopt, (\ref{Dbag}) and
(\ref{Dmeson}), is non-unique although the sum is free of
ambiguities. We find, however, that this separation leads to a
natural partition of the contributions in the framework of the bag
description for confinement we use.

\section{The quark contribution} The quark current is given by
\bq A^\mu_{B_Q} = \bar{\psi} \gamma^\mu \gamma_5 \psi \eq where
$\psi$ should be understood to be the {\it bagged} quark field,
which means that the quark field is a cavity mode before turning
on the quark-gluon interaction. The quark current contribution to
the FSAC is given by \bq a^0_{B_Q} = \langle {\rm P}|\int_B d^3r
\bar{\psi} \gamma_3 \gamma_5 \psi\ |{\rm P}\rangle \la{aq}\eq with
the proton state, $|{\rm P}\rangle$, given by eq.~(\ref{baryon}).



\begin{figure}
\centerline{\epsfig{file=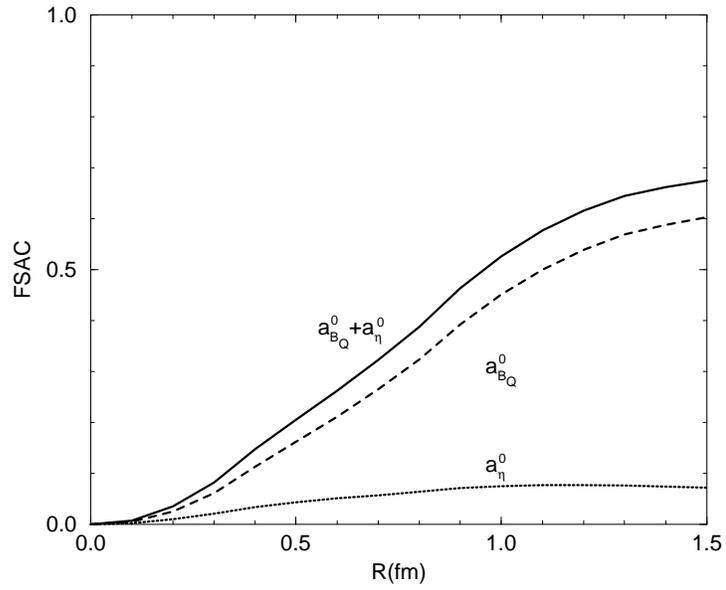, width=11cm}} \caption{Various
contributions to the flavor singlet axial current of the proton as
a function of bag radius : (a) quark  contribution $a^0_{B_Q}$;
(b) $\eta^\prime$ contribution $a^0_\eta$ and (c) the
sum.}\label{fig1}
\end{figure}

The calculation of this type of matrix elements in the chiral bag
model is nontrivial due to the baryon charge leakage between the
interior and the exterior through the Dirac sea. But we know how
to do this in an unambiguous way. A complete account of such
calculations can be found in \ct{pv,pr,p}. The leakage produces an
$R$ dependence, as shown in Fig. \ref{fig1},  which would
otherwise not be there in the matrix element, eq.~(\ref{aq}). It
is worth stressing that, as seen in the Fig. \ref{fig1},  for zero
radius, that is, in the pure skyrmion scenario for the  proton
this matrix element vanishes. The contribution grows as a function
of $R$ towards the pure MIT result that technically is reached for
infinite radius. The result of this calculation was presented in
refs. \ct{pv,pvrb}. No new ingredient has been added.

\section{The {\it meson} current $A^\mu_\eta$} We can get the
$\eta$ field contribution in terms of the quark contribution. From
the boundary condition eq.~(\ref{bcforeta}) for the  $\eta$ field,
\begin{equation}
\langle {\rm P}|\ \hat{\bf r}\cdot(\bar{\psi}\mbox{\boldmath
$\gamma$}\gamma_5\psi)\ |{\rm P}\rangle=\langle {\rm P}|\ \hat{\bf
r}\cdot(2f_{\pi}\mbox{\boldmath $\nabla$}\eta)\ |{\rm
P}\rangle,\hspace{0.5cm}{\rm at}\ r=R,
\label{bcetaQ}\end{equation} and the fact that the bagged current,
$\bar{\psi}\mbox{\boldmath $\gamma$}\gamma_5\psi$, is
divergenceless for massless quarks, we have the following
identity:
\begin{equation}
\langle {\rm P}| \int_{\partial B}d^2s\ \hat{\bf r}\cdot(r_3
2f_\pi\mbox{\boldmath $\nabla$}\eta)\ |{\rm P}\rangle=\langle {\rm
P}| \int_B d^3r(\bar{\psi}\gamma_3\gamma_5\psi)\ |{\rm P}\rangle.
\end{equation}
Substituting $\eta$ field given by
\begin{equation}
\eta=C{\bf S}\cdot\mbox{\boldmath $\nabla$}\bigg(\frac{e^{-m_\eta
r}}{r}\bigg),
\end{equation}
where ${\bf S}$ is the proton spin operator, into the left hand
side leads to
\begin{equation}
\langle {\rm P}| \int_{\partial B}d^2s\ \hat{\bf r}\cdot(r_3
2f_\pi\mbox{\boldmath $\nabla$}\eta)\ |{\rm
P}\rangle=Cf_\pi\frac{8\pi}{3}S_3[2(1+y_\eta)+y_\eta^2]e^{-y_\eta}.
\end{equation}
Here \begin{equation} y_\eta\equiv m_\eta R. \end{equation}
Therefore $C$ is determined by
\begin{equation}
C=\frac{3}{4\pi f_\pi}[2(1+y_\eta)+y_\eta^2]e^{y_\eta}\langle {\rm
P}| \int_B d^3r(\bar{\psi}\gamma_3\gamma_5\psi)\ |{\rm P}\rangle.
\label{Cineta}\end{equation} Consequently, the $\eta$ contribution
becomes \bq a^0_\eta = \langle {\rm P}|\int_{\bar{B}}d^3r
2f_\pi\mbox{\boldmath $\nabla$}_3\eta\ |{\rm P}\rangle= \frac{1
+y_\eta}{2(1+y_\eta) +y_\eta^2} \langle {\bf P}|\int_B d^3r
\bar{\psi} \gamma_3 \gamma_5 \psi\ |{\bf P}\rangle. \la{aeta} \eq
Thus we have the result that the $\eta$ contribution outside is
entirely given in terms of the quark contribution inside and the
$\eta$ mass.

In Fig.~\ref{fig1} we show the radial dependence of several
contributions. We show  the results of \ct{pv,pvrb}, which arise
from the charge leakage mechanism, and follow the quark
distribution. We show also the contribution of the $\eta$ just
calculated, by taking the quark current matrix element also from
\ct{pv,pvrb}. Since the $\eta$ field has no topological structure,
its contribution vanishes in the skyrmion limit. Our calculation
illustrates how the dynamics of the exterior can be mapped onto
that of the interior by the boundary conditions. We may summarize
the analysis of these two contributions by stating that no trace
of the CCP is apparent in Fig.~\ref{fig1}. Thus if the CCP is to
emerge, the only possibility one foresees, is that the gluons do
the miracle!

\section{The {\it gluon } current $A^\mu_{B_G}$} Understanding
the FSAC and its implications in the present framework involves
crucially the role of the gluon contribution, in particular its
static properties and vacuum fluctuations, i.e., the Casimir
effects. The calculation of the Casimir effects in the next
section constitutes the principal aim of this work.

Since we have assigned the anomaly to the gluon fields,
eq.~(\ref{Dmeson}), the gluonic axial current has the form
\begin{equation}
A^0_{\mu,B_G}=\frac{N_F\alpha_s}{4\pi}\epsilon_{\mu\nu\rho\lambda}A^{\nu
a} (G^{\rho\lambda a}-\frac{1}{3}g_sf^{abc}A^{\rho b}A^{\lambda
c}).
\end{equation}
Note that this expression is not locally gauge invariant. In fact
there is no gauge invariant dimension-3 vector operator with the
gauge field alone.
%
%

With this gluonic axial current, the boundary condition,
eq.~(\ref{bcetaQ}), changes to
\begin{equation}
\langle{\rm P}|\ \hat{\bf r}\cdot({\bf A}^0_{B_Q}+{\bf
A}^0_{B_G})\ |{\rm P}\rangle=\langle{\rm P}|\ \hat{\bf
r}\cdot(2f_\pi\mbox{\boldmath $\nabla$}\eta)\ |{\rm
P}\rangle,\hspace{0.5cm}{\rm at}\ r=R. \label{modbc}
\end{equation}
Using this boundary condition and the anomaly,
eq.~(\ref{SDmeson}), the total FSAC of the proton can be
constructed. Since the quark is a divergenceless field, we rewrite
eq.~(\ref{SDmeson}) in the form
\begin{equation}
\mbox{\boldmath $\nabla$}\cdot({\bf A}^0_{B_Q}+{\bf
A}^0_{B_G})=N_F\frac{\alpha_s}{\pi}\sum_a{\bf E}^a\cdot{\bf
B}^a\Theta_B.
\end{equation}
Integrating this equation with respect to the bag volume after
multiplying ${\bf r}$, the matrix element for proton becomes
\begin{eqnarray}
&&\langle{\rm P}|\int_Bd^3r({\bf A}^0_{B_Q}+{\bf A}^0_{B_G})\
|{\rm P}\rangle=-\langle{\rm P}|N_F\frac{\alpha_s}{\pi}\int_Bd^3r\
\sum_a{\bf E}^a\cdot{\bf B}^a{\bf r}\ |{\rm
P}\rangle\nonumber\\
&&\hspace{5.3cm}+\langle{\rm P}|\int_{\partial B}d\mbox{\boldmath
$\Sigma$}\cdot({\bf A}^0_{B_Q}+{\bf A}^0_{B_G}){\bf r}\ |{\rm
P}\rangle.
\end{eqnarray}
Using the boundary condition, eq.~(\ref{modbc}), we see that the
second term yields the contribution from $\eta$ outside the bag.
Therefore, we have
\begin{eqnarray}
&&\langle{\rm P}|\int_Bd^3r({\bf A}^0_{B_Q}+{\bf
A}^0_{B_G})+\int_{\bar{B}}d^3r{\bf A}^0_\eta\ |{\rm
P}\rangle=-\langle{\rm P}|N_F\frac{\alpha_s}{\pi}\int_Bd^3r\
\sum_a{\bf E}^a\cdot{\bf
B}^a{\bf r}\ |{\rm P}\rangle\nonumber\\
&&\hspace{6.3cm}-\langle{\rm P}|\int_{\bar{B}}d^3r(2f_\pi
m_\eta^2\eta){\bf r}\ |{\rm P}\rangle,
\end{eqnarray}
and, since there is no explicit coupling between gluons and the
$\eta$ field at the tree level in the model Lagrangian, the first
term on the right-hand side of the above equation  may be
considered as the {\it gluonic} contribution to the FSAC, namely
\begin{equation}
a^0_G=\langle{\rm P}|-\frac{N_F\alpha_s}{\pi}\int_Bd^3r\ r_3\
\sum_a{\bf E}^a\cdot{\bf B}^a\  |{\rm P}\rangle.
\end{equation}
This corresponds to the second term in eq.~(\ref{a^0}) in Chapter 4.

Let us proceed to calculate the gluonic contribution. We begin by
dividing the gluon current into two terms according to their
origin \bq A^\mu_{B_G} = A^\mu_{G,stat}  + A^\mu_{G,vac}. \eq The
first term arises from the quark and $\eta$ sources, while the
second is associated with the properties of the vacuum of the
model. One might worry that this contribution cannot be split in
these two terms without double counting. That there is no cause
for worry can be seen in  several ways. Technically, it is easy to
show it in terms of mode creation and annihilation operators. One
can also show this intuitively by making the analogy to the
condensate expansion in QCD \ct{svz}, where the perturbative terms
and the vacuum condensates enter additively to the lowest order.
%
%
%

Let us first describe the static term which effectively accounts
for the mixing of the light quarks with the ${\bf E}^a\cdot{\bf
B}^a$ of the anomaly. The boundary conditions for the gluon field
corresponds in our approaximation to the original MIT one
\ct{MIT}. The source for it is the quark current that appears in
the equations of motion after performing a perturbative expansion
in the QCD coupling constant, i.e., the quark color current \bq
g_s{\bar \psi} \gamma_\mu \lambda^a \psi \eq where the $\psi$
fields represent the lowest cavity modes. In this lowest mode
approximation, the color electric and magnetic fields are given by
\bq {\bf E}^a = g_s \frac{\lambda^a}{4\pi} \frac{\hat{\bf r}}{r^2}
\rho (r) \label{ef} \eq \bq {\bf B}^a = g_s
\frac{\lambda^a}{4\pi}\left\{ \frac{\mu (r)}{r^3}(3 \hat{\bf r}
\mbox{\boldmath $\sigma$} \cdot \hat{\bf r} - \mbox{\boldmath $
\sigma$}) + \bigg(\frac{\mu (R)}{R^3} + 2 M(r)\bigg)
\mbox{\boldmath $\sigma$}\right\} \label{bf} \eq where $\rho$ is
related to the quark density $\rho^\prime$ as\footnote{Note that
the quark density that appears here is associated with the color
charge, {\it not} with the quark number (or rather the baryon
charge) that leaks due to the hedgehog pion.} \bq \rho
(r,\Gamma)=\int_\Gamma^r ds \rho^\prime
(s)\label{density}\nonumber \eq and $\mu, M$ to the vector current
density \ba \mu (r) &=& \int_0^r ds \mu^\prime (s),\nonumber\\ M
(r)&=& \int_r^R ds \frac{\mu^\prime (s)}{s^3}.\nonumber \ea As
considered in Chapter 3, the lower limit $\Gamma$ is taken to be
zero in the MIT bag model $-$ in which case the boundary condition
is satisfied only {\it globally}, that is, after averaging $-$ and
$\Gamma=R$ in the so called {\it monopole solution} \ct{pv,rv} $-$
in which case, the boundary condition is satisfied {\it locally}.

Now we  introduce the $\eta$ field. We perform the same
calculation however with the color anomaly boundary conditions given in
Chapter 4,
\begin{equation}\hat{\bf r}\cdot{\bf
E}^a=-\frac{N_Fg_s^2}{8\pi^2f_0}\hat{\bf r}\cdot{\bf
B}^a\eta(R),\hspace{1cm}\hat{\bf r}\times{\bf
B}^a=\frac{N_Fg_s^2}{8\pi^2f_0}\hat{\bf r}\times{\bf E}^a\eta(R).
\label{Cbc}\end{equation} In the approximation of keeping the
lowest non-trivial term, the boundary conditions become \bq
\hat{\bf r}\cdot {\bf E}_{stat}^a=-\frac{N_F g_s^2}{8\pi^2 f_0}
\hat{\bf r}\cdot {\bf B}^a_g \eta (R)\label{Eg} \eq \bq \hat{\bf
r}\times {\bf B}_{stat}^a=\frac{N_F g_s^2}{8\pi^2 f_0} \hat{\bf
r}\times {\bf E}^a_g \eta(R).\label{Bg} \eq Here ${\bf E}^a_g$ and
${\bf B}^a_g$ are the lowest order fields \ct{pv,rv} given by
(\ref{ef}) and (\ref{bf}) and $\eta (R)$ is the $\eta$-meson field
at the boundary. The $\eta$ field is given by \bq \eta ({\bf r}) =
-\frac{g_{NN\eta}}{4\pi M} {\bf S} \cdot \hat{\bf r}
 \frac{1+m_\eta r}{r^2} e^{-m_\eta r}
\la{eta}\eq where the coupling constant is determined from the
surface conditions, eq.~(\ref{Cineta}), the results of
refs.~\ct{pv,rv} and this expression for the $\eta$.

Note that the magnetic field is not affected by the new boundary
conditions, since ${\bf E}^a_g$ points into the radial direction.
The effect on the electric field is just a change in the charge,
i.e., \bq \rho_{stat}(r) = \rho(r,\Gamma) + \rho_\eta(R) \eq where
\bq \rho_\eta (R) = \frac{N_F g_s^2}{64 \pi^3 M}
\frac{g_{NN\eta}}{f} (1+y_\eta) e^{-y_\eta}. \eq

\begin{figure}
\centerline{\epsfig{file=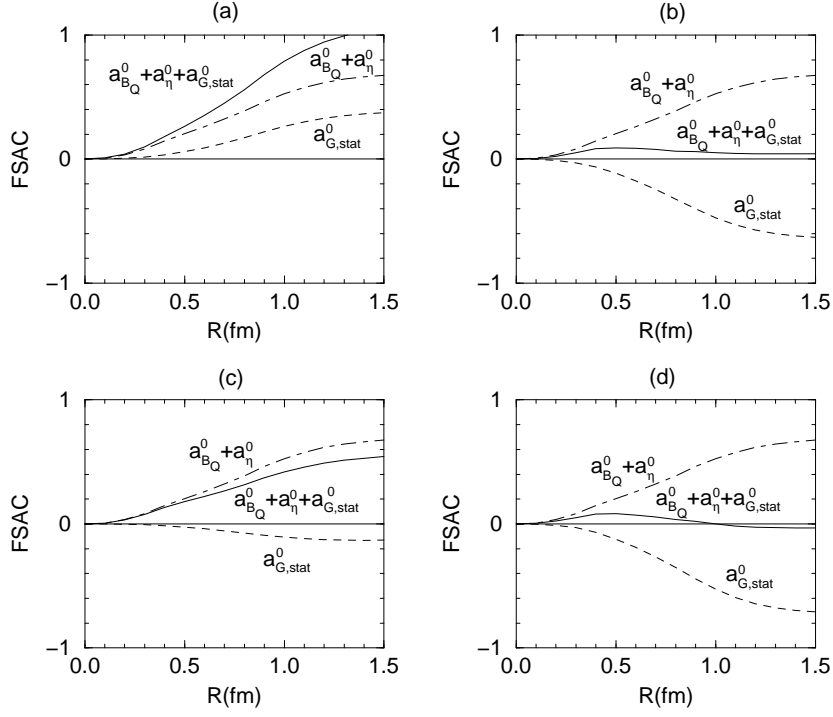, width=11cm}}
\caption{Dependence of $a_{G,stat}^0$ on the choice of $\Gamma$
and the boundary conditions as a function of bag radius : (a) with
an MIT-like electric field without $\eta$ coupling, (b) with a
{\it monopole}-like electric field without $\eta$ coupling, (c)
with an MIT-like electric field {\em with} $\eta$ coupling, and
(d) with a {\it monopole}-like electric field {\em with} $\eta$
coupling.}\label{fig2}
\end{figure}

The contribution to the FSAC arising from these fields is
determined from the expectation value of the anomaly \bq
a^0_{G,stat} = \langle {\rm P}|-\frac{N_F\alpha_s}{\pi} \int_B
d^3r\ r_3\ {\bf E}^a_{stat} \cdot {\bf B}^a_{stat}|{\rm P}\rangle
. \label{astat}\eq The result of this contribution appears in
Fig.~\ref{fig2} , where we show the MIT solution, the {\it
monopole} one and the correction associated to both due to the
color coupling\footnote{We have also investigated electric fields
of the form $(\frac{A}{r^2} + Br)\hat{r}$, but the results do not
change much with respect to the ones shown since the $B$ term
tends to be small.}. One sees that including the $\eta$
contribution in $\rho_{stat} (r)$ produces a non-negligible
modification of the FSAC but does not modify the result
qualitatively. What is most striking is the drastic difference
between the effect of the MIT-like electric field and that of the
monopole-like electric field: The former is totally incompatible
with the Cheshire Cat property whereas the latter remains
consistent independently of whether or not the $\eta$ contribution
is included in $\rho_{stat}$.

Before going to the $A_{G,vac}$, let's consider the gluon spin to
see another observable where the {\it monopole} solution seems to
be favored by experiment.

\section{The gluon spin in the chiral bag model \cite{HDBVR2}} In
this section we study the gluon polarization contribution to the
proton spin $\Gamma$, which is identical to $\Delta g$ in previous
chapters, at the quark model renormalization scale in
the chiral bag model. It is evaluated, as we will show later in
detail,  by taking the expectation value of the forward matrix
element of a local gluon operator in the axial gauge ${\bf A}^+=0$
in the light cone frame. We show that the confining boundary
condition for the color electric field plays an important role in
the outcome. When a solution satisfying the boundary condition for
the color electric field, the so called monopole solution, which
is not the conventional one, but which we favor, is used, $\Gamma$
has a positive value for {\it all} bag radii and its magnitude is
comparable to the quark spin polarization. This results in a
significant reduction in the relative fraction of the proton spin
carried by the quark spin, which is consistent with the small
flavor singlet axial current measured in the EMC experiments.
\vskip 0.25cm

As we have discussed in Chapter 2, the EMC experiment \cite{Ashman}
revealed the surprising fact that less than 30\% of the proton
spin may be carried by the quark spin. This is at variance with
what one expects from non-relativistic or relativistic constituent
quark models. This discrepancy $-$ so called ``the proton spin
crisis" $-$ can be understood as an effect associated with the
axial anomaly \cite{abj}. If we follow the
argument \cite{Anselmino95} given in Chapter 2, the experimentally
measured quantity is not merely the quark spin polarization
$\Delta\Sigma$ but rather the flavor singlet axial charge, to
which the gluons contribute through the axial anomaly. Therefore,
to understand the proton spin crisis, the sign of the gluon
contribution is crucial.

Although not directly observable, an equally interesting quantity
related to the proton spin is the fraction of spin in the proton
that is carried by the gluons. In Ref.~\cite{Jaffe96}, the gluon
spin $\Gamma$ is introduced as
\begin{equation}
\textstyle \frac12 = \frac12 \Sigma +  L_Q + \Gamma + L_G,
\label{QG}
\end{equation}
where $L_{Q,G}$ is the orbital angular momentum of the
corresponding constituent and $\Gamma$ is defined as the integral
of the polarized gluon distribution in analogy to $\Sigma$. The
spin of course is gauge-invariant but the individual components in
eq.~(\ref{QG}) may not be. $\Gamma$ can be expressed as a matrix
element of products of the gluon vector potentials and field
strengths in the nucleon rest frame and in the ${\bf A}^+=0$
gauge. When evaluated with the gluon fields responsible for the
$N-\Delta$ mass splitting,  $\Gamma$ turns out to be negative,
$\Gamma\sim -0.1 \alpha_{\mbox{\scriptsize{bag}}}$, in the MIT bag
model and even more so in the non-relativistic quark model.

By contrast, there are several other calculations that give
results with opposite sign. For example, the QCD sum rule
calculation \cite{MPS97} yields a positive value $2\Gamma \sim 2.1
\pm 1.0$ at 1 $\mbox{GeV}^2$. In Ref.~\cite{BCD98}, it is
suggested that the negative $\Gamma$ of Ref.~\cite{Jaffe96} could
be due to neglecting ``self-angular momentum."  The authors of
\cite{BCD98} show that when self-interaction contributions are
included, one obtains a positive value $\Gamma\sim +0.12$ in the
Isgur-Karl quark model at the scale $\mu^2_0 \approx 0.25\mbox{
GeV}^2$. Once the gluons contribute a significant fraction to the
proton spin, due to the normalization eq.~(\ref{QG}), the relative
fraction of the proton spin lodged in the quark spin changes.
Thus, the positive gluon spin seems to be consistent with the EMC
experiment.

We address this issue in the chiral bag model and pay special
attention to the confining boundary condition for the gluon
fields.

Let us start by briefly reviewing how the gluon spin operator was
derived in Ref.~\cite{Jaffe96,JM90}. From the Lagrangian
\begin{equation}
{\cal L}=-\frac{1}{2}{\rm Tr}\bigg({\bf G}^{\mu\nu}{\bf
G}_{\mu\nu}\bigg)
\end{equation}
with ${\bf G}^{\mu\nu}=\frac{\lambda^a}{2}G^{a\mu\nu}$, one gets
the gluon angular momentum tensor
\begin{equation}
M^{\mu\nu\lambda}=2{\rm Tr}\bigg( x^{\nu}{\bf G}^{\mu\alpha}{{\bf
G}_{\alpha}}^{\lambda} -x^{\lambda}{{\bf
G}_{\mu\alpha}}^{\nu}\bigg)
-(x^{\nu}g^{\mu\lambda}-x^{\lambda}g^{\mu\nu}){\cal L}.
\end{equation}
Integrating by parts, we have
\begin{eqnarray}
M^{\mu\nu\lambda}&=&2{\rm Tr}\bigg( -{\bf G}
^{\mu\alpha}(x^{\nu}\partial^{\lambda}
-x^{\lambda}\partial^{\nu}){\bf A}_{\alpha}
+{\bf G}^{\mu\lambda}{\bf A}^{\nu}+{\bf G}^{\nu\mu}{\bf A}^{\lambda} \nonumber\\
&&+\partial_{\alpha} (x^{\nu}{\bf G} _{\mu\alpha}{\bf A}
^{\lambda}-x^{\lambda}{\bf G}_{\mu\alpha}{\bf A}^{\nu})
+\frac{1}{4}{\bf G}^{\mu\nu}{\bf G}_{\mu\nu}
(x^{\nu}g^{\mu\lambda}-x^{\lambda}g^{\mu\nu})\bigg)
\label{M}\end{eqnarray} with ${\bf
A}_{\mu}=\frac{\lambda^a}{2}A^a_{\mu}$. It seems reasonable to
interpret the first term as the gluon orbital angular momentum
contribution and the second as that of the gluon spin, while
recalling that this is a gauge dependent statement. We will not
consider the fourth term hereafter, since it contributes only to
boosts. In Ref.~\cite{Jaffe96,JM90}, the third term is also
dropped as is done in the open space field theory. When finite
space is involved, as in the bag model, dropping this term
requires that the gluon fields satisfy boundary conditions on the
surface of the region, as we next show. Let us express the gluon
angular momentum operator in terms of the Poynting vector, i.e.,
\begin{equation}
{\bf J}_G = 2{\rm Tr}\int_V d^3r[{\bf r}\times({\bf E}\times{\bf
B})]. \label{Jg}
\end{equation}
Now doing the partial integration for ${\bf B}=\mbox{\boldmath $
\nabla$}\times{\bf A}$, we have
\begin{equation}
J^k_G = 2{\rm Tr}\left\{\int_B d^3r\bigg({\bf E}^l({\bf
r}\times\mbox{\boldmath $\nabla$})^k {\bf A}_l + ({\bf
E}\times{\bf A})^k\bigg) - \int_{\partial B}d^2r({\bf r}\cdot{\bf
E})({\bf r}\times{\bf A})^k \right\}. \label{J^k_g}
\end{equation}
The surface term is essential to make the whole angular momentum
operator gauge-invariant, but the surface term only vanishes, if
the electric field satisfies the boundary condition on the
surface,
\begin{equation}
{\bf r}\cdot{\bf E}=0. \label{bcE}\end{equation} This is just the
MIT boundary condition for gluon confinement. However, the static
electric field traditionally used \cite{MIT} does not satisfy
this condition. Instead the color singlet nature of the hadron
states is imposed to assure confinement globally.

We next show that the negative $\Gamma$ of Ref. \cite{Jaffe96}
results if this procedure to confine color is imposed. To proceed,
we choose the ${\rm A}^+=0$ gauge and write the gluon spin in a
local form as
\begin{equation}
\Gamma=\langle {\rm P}, \uparrow |2{\rm Tr}\int_V d^3x \bigg(
({\bf E}\times{\bf A})^3+{\bf B}_{\perp}\cdot{\bf A}_{\perp}
\bigg) |{\rm P}, \uparrow\rangle,
\end{equation}
where $\perp$ denotes the direction perpendicular to the proton
spin polarization and the superscript $+$ indicates the light cone
coordinates defined as $x^\pm = \frac{1}{\sqrt 2}(x^0 \pm x^3)$.
We shall evaluate this expression by incorporating the exchange of
the static gluon fields between $i-$th and $j-$th quarks ($i\neq
j$) which are responsible for the $N-\Delta$ mass splitting in the
bag model.

As discussed in Chapter 3, in the chiral bag model, the static
gluon fields are generated by the color charge and current
distributions of the $i-$th valence quark given by \cite{pv}
\begin{eqnarray}
J^{0a}_i({\bf r}) &=& \frac{g_s}{4\pi} {\rho(r)}
\frac{\lambda^a_i}{2},
\label{J0}\\
{\bf J}^a_i({\bf r}) &=& \frac{g_s}{4\pi} 3 (\hat{\bf r} \times
{\bf S}) \frac{\mu^\prime (r)}{r^3} \frac{\lambda^a_i}{2},
\label{vecJ}
\end{eqnarray}
where $\rho(r)$ and $\mu^\prime (r)$ are, respectively, the quark
number  and current densities determined by the valence quark wave
functions given in Chapter 3. They are very similar in form to
those of the MIT bag model. There is, however, an essential
difference, namely, that the spin in the chiral bag model is given
by the collective rotation of the whole system while in the MIT
bag it is given by a individual contribution of each constituent,
i.e., there is no index $i$ in the spin operator in
eq.~(\ref{vecJ}).

The charge and current densities yield the color electric and
magnetic fields as
\begin{eqnarray}
{\bf E}^a_i &=& \frac{g_s}{4\pi}
\frac{Q(r)}{r^2}\frac{\lambda^a_i}{2} \hat{\bf r},
\label{vecE}\\
{\bf B}^a_i &=& \frac{g_s}{4\pi} \left\{ {\bf S} \left( 2M(r) +
\frac{\mu(R)}{R^3} - \frac{\mu(r)}{r^3} \right) + 3 \hat{\bf r}
(\hat{\bf r}\cdot {\bf S}) \frac{\mu(r)}{r^3} \right\}
\frac{\lambda^a_i}{2}, \label{vecB}
\end{eqnarray}
where
\begin{eqnarray}
\mu(r) &=& \int^r_0 d r^\prime \mu^\prime (r^\prime), \\
M(r) &=& \int^R_r dr^\prime \frac{\mu^\prime (r^\prime)}{r^{\prime
3}}.
\end{eqnarray}
The quantity $Q(r)$ can be determined from Maxwell's equations.
The most general solution can be written as
\begin{equation}
Q_\lambda(r) = 4\pi\int^r_\lambda dr^\prime r^{\prime 2}
\rho(r^\prime),
\end{equation}
with an arbitrary $\lambda$. The standard procedure is to choose
$\lambda=0$ so that the electric field is regular at the origin.
This has been the adopted convention in the early days \cite{MIT}.
We will refer to this solution as $Q_0(r)$. However, $Q_0(r)$ does
not satisfy the {\it local} boundary condition eq.~(\ref{bcE}),
since it is normalized as $Q_0(R)=1$. In Ref.~\cite{MIT}, the fact
that hadrons are color singlet states, had to be imposed in order
to justify the use of this solution.

Another solution, namely {\it monopole} solution presented in
previous discussion, is obtained by setting $\lambda=R$ \cite{pv}
and we will look for its consequences here. This choice satisfies
the local boundary condition but requires the relaxation of the
continuity of the electric fields inside the bag. It has been
shown in \cite{pv}, and will be shown here again, that these two
solutions, $Q_0(r)$ and $Q_R(r)$, lead to dramatic differences for
certain observables.

By using the static Green functions and the Coulomb gauge
condition, one can obtain time-independent scalar and vector
potentials from the charge and current densities, eqs.~(\ref{J0})
and (\ref{vecJ}),
\begin{eqnarray}
\Phi^a_i({\bf r})&=&\frac{g_s}{4\pi} \frac{Q_\lambda(r)}{r}
\frac{\lambda^a_i}{2},
\label{Phi} \\
{\bf U}^a_i({\bf r})&=&\frac{g_s}{4\pi} h(r) ({\bf S}\times{\bf
r}) \frac{\lambda^a_i}{2}, \label{U}
\end{eqnarray}
where
\begin{equation}
h(r) = \bigg( \frac{\mu(R)}{2R^3} + \frac{\mu(r)}{r^3} + M(r)
\bigg).
\end{equation}
{}From these, the appropriate scalar and vector potentials
satisfying the ${\bf A}^+=0$ gauge condition can be constructed:
\begin{eqnarray}
A^{a0}_i({\bf r}) &=& \Phi^a_i ({\bf r}),
\nonumber\\
{\bf A}^a_i({\bf r}) &=& {\bf U}^a_i({\bf r}) -\nabla \int_0^z
d\zeta\  \Phi^a_i(x,y,\zeta),
\end{eqnarray}
where the direction of the proton polarization is taken as that of
the $z-$axis. Finally, we obtain
\begin{eqnarray}
\Gamma_\lambda &=& \sum_{i\neq j}\sum_{a=1}^8 \langle {\rm P}
,\uparrow|\bigg(2\int_V d^3x ({\bf E}^a_i \times {\bf U}^a_j )^3
\nonumber\\
&& + \int_{\partial V} d^2s \hat{\bf z} \cdot \hat{\bf r} \bigg(
U_i^{a1}({\bf x}) \int_0^z d \zeta \ E_i^{a2}(x,y,\zeta)
\nonumber\\
&& - U_i^{a2}({\bf x})\int_0^z d\zeta\
E_i^{a1}(x,y,\zeta)\bigg)\bigg) |{\rm P},\uparrow\rangle
\nonumber\\
&=&\frac{4}{3} \alpha_s \int_0^Rrdr\ Q_\lambda(r) (h(R)-2h(r)),
\end{eqnarray}
where $\alpha_s=g^2_s/4\pi$. The numerical factor in front of the
final formula comes from the fact that $\sum_a \langle \lambda^a_i
\lambda^a_j \rangle_{\mbox{\scriptsize baryon}}= -8/3$ for $i \neq
j$ so that
\begin{equation}
\sum_{i\neq j}\sum_{a=1}^8 \langle {\rm P}, \uparrow | S_3
\frac{\lambda_i^a}{2} \frac{\lambda_j^a}{2} |{\rm P},\uparrow
\rangle = -2,
\end{equation}
and the integration over angles yields $1/3$. It is different from
8/9 of the MIT bag model \cite{Jaffe96}, which comes from the
expectation value
\begin{equation}
\sum_{i\neq j}\sum_{a=1}^8 \langle {\rm P}, \uparrow | \sigma_i^3
\frac{\lambda_i^a}{2} \frac{\lambda_j^a}{2} |{\rm P},\uparrow
\rangle = - 4/3.
\end{equation}

It is interesting to note that, if we naively substitute the
static gluon fields $\Phi^a_i$ and ${\bf U}^a_i$ of
eqs.~(\ref{Phi}) and (\ref{U}) satisfying the Coulomb gauge
condition into the second term of eq.~(\ref{J^k_g}), we get
\begin{equation}
\Gamma^\prime = -\frac{4}{3}\alpha_s \int_0^R r dr \
Q_\lambda(r)h(r),
\end{equation}
which is the same expression \cite{pv,Zahed} that was used in the
previous section to evaluate the anomalous gluon contribution to
the flavor singlet axial current $a_0$ with the extra factor
$(-N_F\alpha_s/2\pi)$, i.e., $a_0 = \Sigma - (N_F \alpha_s/2\pi)
\Gamma^\prime_\lambda$. On the other hand, in Ref.~\cite{BCD98},
the gluon spin $\Gamma$ instead of $\Gamma^\prime$ is used for the
anomaly correction term because the calculation is performed in
the ${\rm A}^+=0$ gauge.

If the gluons can contribute to the proton spin, then the
collective coordinate quantization scheme of the chiral bag model
has to be modified to incorporate their contribution. That is
because there is a natural sum rule namely that the total proton
spin must come out to be $\frac12$, whatever the various
contributions are. In the chiral bag model, where the mesonic
degrees of freedom also play an important role, the proton spin is
described by the following contributions
\begin{equation}
\textstyle \frac12 = \frac12 \Sigma +  L_Q + \Gamma + L_G + L_M,
\end{equation}
where $L_M$, the orbital angular momentum  of the mesons, has to
be added to eq.~(\ref{QG}). The proton spin is generated by
quantizing the collective rotation associated with the zero modes
of the classical soliton solution of the model Lagrangian. To the
collective rotation, each constituent responds with the
corresponding moment of inertia. The moments of inertia of the
quarks and mesons, ${\cal I}_Q$ and ${\cal I}_M$, have been
extensively studied in the literature \cite{pr}. Substitution of
the color electric and magnetic fields, given by eqs.~(\ref{vecE})
and (\ref{vecB}) respectively, into eq.~(\ref{Jg}) defines a new
moment of inertia of the static gluon fields with respect to the
collective rotation as
\begin{equation}
\langle {\bf J}_G \rangle = - {\cal I}_G {\bf \omega}, \label{IG}
\end{equation}
where the expectation value is taken keeping only the exchange
terms, and ${\bf\omega}$ is the classical angular velocity of the
collective rotation.

We show in Figs.~\ref{fig7} (a) and (b) the gluon moment of
inertia evaluated by using the color electric fields with $Q_R(r)$
and $Q_0(r)$. In the case of $Q_R(r)$, ${\cal I}_G$ is positive
for all bag radii and comparable in size to ${\cal I}_Q$, the
quark moment of inertia. On the other hand, $Q_0(r)$ results in a
negative ${\cal I}_G$. This ``negative" moment of inertia may
appear to be bizarre but it may not be a problem from the
conceptual point of view. The ${\cal I}_G$ defined by
eq.~(\ref{IG}) can be interpreted as the one-gluon exchange
correction to the corresponding quantity of the quark phase, which
is still positive anyway. The point is that the spin fractionizes
in the same way as the moment of inertia does. This means that we
have
\begin{eqnarray}
L_Q+\textstyle\frac12 \Sigma &=& \frac{{\cal I}_Q}{2({\cal
I}_Q+{\cal I}_G+{\cal I}_M)},
\nonumber\\
L_G+\Gamma &=& \frac{{\cal I}_G}{2({\cal I}_Q+{\cal I}_G+{\cal
I}_M)},
\\
L_M &=& \frac{{\cal I}_M}{2({\cal I}_Q+{\cal I}_G+{\cal I}_M)}.
\nonumber
\end{eqnarray}
Each fraction as a function of the bag radius is presented in the
small boxes inside each figure. Note in the case of adopting
$Q_R(r)$ that at the large bag limit the proton spin is equally
carried by quarks and gluons somewhat like the momentum of the
proton. The negative ${\cal I}_G$ obtained with $Q_0(r)$, thus,
yields a scenario where the gluons are anti-aligned with the
proton spin.

The dashed and dash-dotted curves in Figs.~\ref{fig8} (a) and (b)
show the values for $\Gamma_\lambda$ and $\Gamma^\prime_\lambda$.
For comparison, we draw $\frac12\Sigma$ by a solid curve. Note
that, because of the difference in ${\cal I}_G$, even
$\frac12\Sigma$ is different according to which $Q_\lambda$ is
used. Again, both $\Gamma_0$ and $\Gamma^\prime_0$ are
anti-aligned with the proton spin. Note of course that the
negative $\Gamma_0^\prime$ is apparently at variance with the
general belief that the anomaly is to cure the proton spin
problem.

To conclude, we show in Figs.~\ref{fig9} (a) and (b) the
flavor-singlet axial current including the $U_A(1)$ anomaly given
by
\begin{equation}
a_0 = \Sigma - \frac{N_F\alpha_s}{2\pi}\Gamma^\prime_\lambda.
\end{equation}
For simplicity, we neglect other contributions to $a_0$ studied in
previous discussion. They show that the positive $\Gamma$ is
consistent with the small $a_0$ measured in the EMC experiments.
The radius dependence of each component may be viewed as gauge
dependence both in color gauge symmetry and in the ``Cheshire Cat"
gauge symmetry discussed by Damgaard, Nielsen and
Sollacher~\cite{CC}.

\section{The Casimir effect on the FSAC due to the color
anomaly}Finally we proceed to study the term $A_{G,vac}$, which
arises from the so called Casimir effect associated with the
anomaly. The vacuum in the cavity and the perturbative vacuum in
free space are different due to the geometry of the cavity. This
difference might lead to observable consequences and it has been
considered for many observables and also for the quarks in our
calculation \ct{pv,pr,p}, but never for the gluons. We proceed
next to describe the Casimir effect for the FSAC.

The calculation of Casimir effects is in general complex and is
plagued by divergences, which have to be properly taken care of.
In order to clarify these issues we consider first the Casimir
energy \cite{Cas}, which is well studied, and shows the structure
of divergences.

In the canonical quantization formalism of field theories in
infinite (free) space-time, vacuum energies are divergent. By the
Wick's normal ordering procedure, which is based on the fact that
the physical measured energy is the difference between the state
energy and the vacuum energy, which acts as the energy origin,
these divergences disappear. However, when one considers a theory
in a finite space region with a boundary, the vacuum is changed
with respect to the free one, the change depending on the geometry
of the region. The Casimir energy is defined as the difference
between the vacuum energy in the presence of a boundary and that
of free space \cite{Plunien85}.

For example, let us consider the case of the free massless scalar
field $\phi$ in the region between two plates normal to the
$z-$axis. The plates are separated by the distance $L$ with the
left plate at $z=0.$ If one imposes Dirichlet (Neumann) boundary
conditions, namely $\phi=0\ (\partial_z\phi=0)\ \mbox{\rm at the
plates}$, one obtains from the Hamiltonian density, by direct
summation of modes, the vacuum energy density as
\begin{equation}
{\cal
E}_0^{D,N}=\lim_{\Lambda\rightarrow\infty}\bigg(\frac{3\Lambda^4}{2\pi^2}
\mp\frac{\Lambda^3}{4\pi L}-\frac{\pi^2}{1440L^4}\bigg),
\end{equation}
where superscripts $D(N)$ and the negative (plus) sign in the
second term denote the Dirichlet (Neumann) boundary condition.
Here the cut-off $\Lambda$ has been introduced to treat the
divergent summation. The first term is the divergent vacuum energy
for infinite space-time corresponding to the limit
$\Lambda\rightarrow\infty.$ Therefore, the Casimir energy has the
form of
\begin{equation}
{\cal E}_{\rm Cas}={\cal E}_0^{D,N}-{\cal
E}_0=\mp\frac{\Lambda^3}{4\pi L} -\frac{\pi^2}{1440L^4}.
\end{equation}
Although there is an ambiguity due to the first divergent term,
the plates feel a well defined attractive force, which can be
obtained from this energy density. On the other hand, if one uses
dimensional regularization instead of the cut-off, the vacuum
energy density does not have any divergent terms~\footnote{By fiat
power divergences are ``killed" in dimensional regularization.
This is the power of dimensional regularization for renormalizable
field theories but one should be aware of that power divergences
play a physical role in effective field theories where the
standard renormalizability requirement is not applicable.
\cite{weinbergII}} and is given by,
\begin{equation}
{\cal E}_0^{D,N}=-\frac{1}{12\pi
L}\bigg(\frac{\pi}{L}\bigg)^3\zeta(-3) =-\frac{\pi^2}{1440L^4},
\end{equation}
where the Riemann's zeta function, defined by
$\zeta(a)=\sum_{n=1}^\infty n^{-a}$, has been used. Thus, in this
regularization scheme, the Casimir energy can be obtained without
any ambiguity. With this result, it was pointed out
~\cite{Ravndal00} that when electromagnetic fields confined
between plates are considered, which can be regarded as two scalar
fields, one with Dirichlet and one with Neumann boundary
conditions, the vacuum energy density becomes
\begin{equation}
{\cal E}_0^{D+N}=\frac{3\Lambda^4}{\pi^2}-\frac{\pi^2}{720L^4}
\end{equation}
so that the Casimir energy has the same form as that obtained by
dimensional regularization. However, this result is a special
case. In general, it is known that divergences  appear in the
Casimir energy and depend on the regularization used
\cite{DivCas}.

Similarly, the Casimir effect on any physical observable, ${\cal
O}$, may be defined by the difference between the vacuum
expectation value of ${\cal O}$ with a boundary and that without
boundary, i.e.,
\begin{equation}
{\cal O}_{\rm Cas}=\langle 0|{\cal O}|0\rangle_{\rm boundary}
-\langle 0|{\cal O}|0\rangle_{\rm free},
\end{equation}
and has a similar structure of divergences to those of the Casimir
energy.

The quantity that we wish to calculate is the gluonic vacuum
contribution to the FSAC of the proton. It can be done by
evaluating the expectation value
\begin{equation}
\langle 0_B| -\frac{N_F\alpha_s}{\pi} \int_V d^3r\ r_3 ( {\bf E}^a
\cdot {\bf B}^a) |0_B\rangle \label{f1}\end{equation} where
$|0_B\rangle$ denotes the vacuum in the bag. Note that since the
electric and magnetic fields are orthogonal to each other, there
is no contribution from the MIT boundary conditions and free
space. The standard way to evaluate this expectation value would
be to expand the field operators in terms of the classical
eigenmodes that satisfy the equations of motion and the color
anomaly boundary conditions, eq.~(\ref{Cbc}). Although
well-defined, this approach is technically involved. We have not
yet obtained any quantitative results to report. In here, we shall
proceed in the opposite direction. Instead of arriving at the CCP
as in the standard approach, we shall {\it assume} the CCP and
evaluate the Casimir contribution with the expression that follows
from the assumption. The idea goes as follows.

The CCP states that at low energy, hadronic phenomena do not
discriminate between QCD degrees of freedom (quarks and gluons) on
the one hand and meson degrees of freedom (pions, $\eta$,...) on
the other, provided that all necessary quantum effects (e.g.,
quantum anomalies) are properly taken into account. If we consider
the limit where the $\eta$ excitation is a long wavelength
oscillation of zero frequency, the CCP asserts that it does not
matter whether we choose to describe the $\eta$, in the interior
of the infinitesimal bag, in terms of quarks and gluons or in
terms of mesonic degrees of freedom. This statement, together with
the color boundary conditions, leads to an extremely and useful
{\it local} formula \ct{NRWZ2},
 \begin{equation}
{\bf E}^a \cdot {\bf B}^a \approx -\frac{N_F g_s^2}{8\pi^2}
\frac{\eta}{f_0} \frac12 G^2, \label{f2}\end{equation} where only
the term up to the first order in $\eta$ is retained in the
right-hand side. Here we adapt this formula to the chiral bag
model. This means that the couplings are to be understood as the
average bag couplings and the gluon fields are to be expressed in
the cavity vacuum through a mode expansion. In fact, by comparing
the expression for the $\eta$ mass derived in \ct{NRWZ2}
using eq.~(\ref{f2}) with that obtained by Novikov et al \ct{NSVZ}
in QCD sum-rule method, we note that the matrix element of $G^2$
in (\ref{f2}) should be evaluated in the {\it absence} of light
quarks. This means, in the bag model, the cavity vacuum. That the
surface boundary condition can be interpreted as a local operator
is a rather strong CCP assumption which while justifiable for
small bag radius, can only be validated a posteriori by the
consistency of the result. This procedure is the substitute to the
condensates in the conventional discussion.

Substituting eq.~(\ref{f2}) into eq.~(\ref{f1}) we obtain
\begin{eqnarray}&& \hskip -2em
\langle 0_B| -\frac{N_F\alpha_s}{\pi} \int_V d^3r r_3 ({\bf E}^a
\cdot {\bf B}^a) |0_B\rangle \nonumber\\ && \approx
\left(-\frac{N_F\alpha_s}{\pi} \right) \left(-\frac{N_F
g^2}{8\pi^2} \right) \frac{y(R)}{f_0} \langle p|S_3|p\rangle
\langle 0_B |\int_V d^3r \frac12 G^2 r_3 \hat{r}_3 |0_B\rangle \nonumber \\
&& \approx  \left(-\frac{N_F\alpha_s}{\pi} \right)
\left(-\frac{N_F g^2}{8\pi^2} \right) \frac{y(R)}{f_0}
\langle p|S_3|p\rangle (N_c^2-1) \nonumber\\
&& \hskip 3em \times\sum_n \int_V d^3r ({\bf B}_n^* \cdot {\bf
B}_n - {\bf E}_n^* \cdot {\bf E}_n) r_3 \hat{r}_3 ,
\label{ps1}\end{eqnarray} where we have used that $\eta$ has a
structure like $({\bf S}\cdot\hat{\bf r}) y(R)$. Since we are
interested only in the first order perturbation, the field
operator can be expanded by using MIT bag eigenmodes (the zeroth
order solution). Thus, the summation runs over all the classical
MIT bag eigenmodes. The factor $(N_c^2-1)$ comes from the sum over
the abelianized gluons.

The next steps are the  numerical calculations to evaluate the
mode sum appearing in eq.~(\ref{ps1}): (i) introduction of the
heat kernel regularization factor to classify the divergences
appearing in the sum and (ii) subtraction of the ultraviolet
divergences. \vskip 1ex

As given in Chapter 3, the classical eigenmodes of the
(abelianized) gluons confined in the MIT bag can be classified by
the total spin quantum numbers $(J,M)$ given by the vector sum of
the orbital angular momentum ${\bf L}$ and the spin ${\bf S}$,
\begin{equation}
{\bf J}\equiv {\bf L} + {\bf S},
\end{equation}
and the radial quantum number $n$. In the Coulomb gauge there are
two kinds of classical eigenmodes according to the relations
between the parity and the total spin $J$:
\begin{eqnarray}
\mbox{(i) M-modes : }&& \pi=-(-1)^J\nonumber\\
&&{\bf G}^{(M)}_{(n,J,M)}({\bf r}) = {\cal N}_M j_{J}(\omega_n r)
{\bf Y}_{J,J,M}(\hat{\bf r}),
\label{A_M} \\
 \mbox{(ii) E-modes : }&& \pi=-(-1)^{J+1}\nonumber\\
&&{\bf G}^{(E)}_{(n,J,M)}({\bf r}) = {\cal N}_E \left[
-\sqrt{\frac{J}{2J+1}} j_{J+1}(\omega_n r) {\bf
Y}_{J,J+1,M}(\hat{\bf r})
\right.\label{A_E} \\
&& \hskip 9em \left. +\sqrt{\frac{J+1}{2J+1}} j_{J-1}(\omega_n r)
{\bf Y}_{J,J-1,M}(\hat{\bf r}) \right], \nonumber
\end{eqnarray}
where ${\bf Y}_{J,\ell,M}$ is the vector spherical harmonics of
the total spin $J$ composed of the angular momentum $\ell$ and
$j_{\ell}(x)$ is the spherical Bessel functions. The energy
eigenvalues are determined to satisfy the MIT boundary conditions
as
\begin{eqnarray}
 \mbox{(i) M-modes : }&&\nonumber\\
&& X_n j^\prime_J(X_n) + j_J(X_n) = 0, \\
\mbox{(ii) E-modes : }&&\nonumber\\
&& j_J(X_n) = 0.
\end{eqnarray}
{}From the results in Chapter 3, the normalization constants
${\cal N}_{M,E}$ are specified as:
\begin{eqnarray}
{\cal N}_M &=& \left[ X_n R^2 \left( j^2_J(X_n) - j_{J-1}(X_n)
j_{J+1}(X_n) \right) \right]^{-1/2}, \\
{\cal N}_E &=& \left[ X_n R^2 j_{J-1}^2(X_n) \right]^{-1/2}.
\end{eqnarray}
\vskip 1ex

The first step is to calculate the matrix elements
\begin{equation}
Q_{\{\nu\}} \equiv  2\int_B d^3 r ( {\bf B}^*_{\{\nu\}} \cdot {\bf
B}_{\{\nu\}} - {\bf E}^*_{\{\nu\}} \cdot {\bf E}_{\{\nu\}} ) x_3
\hat{x}_3.
\end{equation}
{}From eq.~(\ref{A_M}), we obtain
\begin{eqnarray}
{\bf E}_{\{\nu\}}({\bf r}) & = & (+i\omega_n) {\cal N}_M
j_J(\omega_n r)
{\bf Y}_{J,J,M}(\hat{\bf r}), \\
{\bf B}_{\{\nu\}}({\bf r}) & = & (+i\omega_n) {\cal N}_M \left[
-\sqrt{\frac{J}{2J+1}} j_{J+1}(\omega_n r) {\bf
Y}_{J,J+1,M}(\hat{\bf r})
\right.\nonumber \\
&& \hskip 8em \left. +\sqrt{\frac{J+1}{2J+1}} j_{J-1}(\omega_n r)
{\bf Y}_{J,J-1,M}(\hat{\bf r}) \right],
\end{eqnarray}
for the M-modes and the similar equations with ${\bf E}$ and ${\bf
B}$ being interchanged for the E-modes.

We encounter in the calculation the following angular integrals
\begin{equation}
\int d\Omega {\bf Y}_{J,\ell,M}^* \cdot {\bf Y}_{J,\ell,M}
\hat{r}_3^2.
\end{equation}
By using that $\hat{x}_3^2 = (4/3)\sqrt{\pi/5} Y_{20} + 1/3$ and
the Wigner-Eckart theorem, we obtain
\begin{equation}
\int d\Omega {\bf Y}_{J,\ell,M}^* \cdot {\bf Y}_{J,\ell,M}
\hat{r}_3^2 = c_{J,\ell} (J(J+1)-3M^2) + \frac{1}{3},
\end{equation}
where $c_{J,\ell}$ is a constant that depends only on $J$ and
$\ell$. We have to perform  the summation over $M$, which runs
from $-J$ to $J$, which cancels the contribution of the first
term, therefore we can take {\it effectively} 1/3 as the result of
the integral.

Finally, we obtain the matrix elements for the M-modes as
\begin{equation}
Q^{(M)}_n = \frac{1}{3} \frac{\int^{X_n}_0 x^3 dx \left[ j_J^2(x)
- \frac{J}{2J+1} j_{J+1}^2(x) - \frac{J+1}{2J+1} j_{J-1}^2(x)
\right] } { X_n^3 \left[ j^2_J(X_n) - j_{J-1}(X_n) j_{J+1}(X_n)
\right] }.
\end{equation}
In the case of the E-mode, we obtain exactly the same formula
except the minus sign in front of it. (Note that the formulas for
the electric field and the magnetic field are interchanged.)

\begin{figure}
\centerline{\epsfig{file=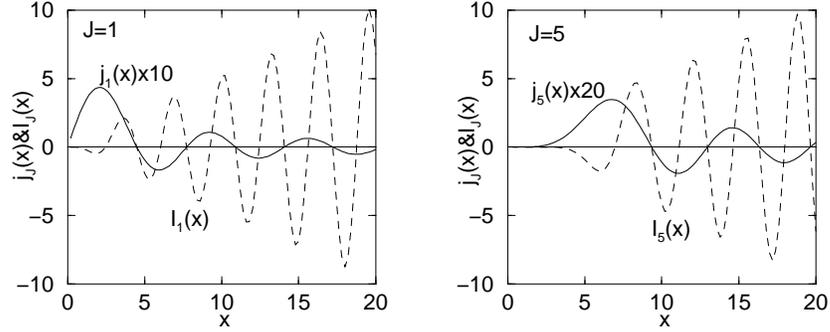, width=11cm}} \caption{$j_J(x)$
and $I(x)$ as a function of x.}\label{fig3}
\end{figure}

We have found that the matrix elements for the E-mode vanish up to
our numerical accuracy as shown in Fig.~\ref{fig3}. Here, the
solid line is the spherical Bessel function $j_J(x)$ and the
dashed line is the integral
\begin{equation}
I(x) \equiv \int^{x}_0 y^3 dy \left[ j_J^2(y) - \frac{J}{2J+1}
j_{J+1}^2(y) - \frac{J+1}{2J+1} j_{J-1}^2(y) \right]
\end{equation}
We see that the zeroes of $I(x)$ and $j(x)$ coincide, thus showing
that $Q_n^{(E)} (X_n)=0$. The analytic proof is given the
appendix. \vskip 1ex

\begin{figure}
\centerline{\epsfig{file=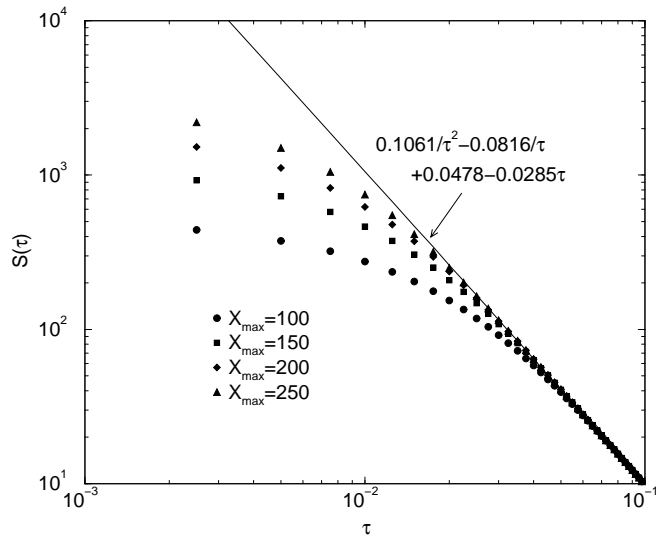, width=10cm}} \caption{Diverging
properties of $S(\tau)$ as a function of the heat kernel
regularization parameter $\tau$. All the magnetic modes up to
$\omega_n R (\equiv X_n)$=100 (solid circle), 150 (solid square),
200 (solid diamond) and 250 (solid triangle) are included in the
sum.}\label{fig4}
\end{figure}

In order to regularize the mode sum, we introduce a heat kernel
factor $\exp(-\tau X_n)$;
\begin{equation}
S(\tau) \equiv \sum_{n,J} (2J+1) Q^{(M)}_{n,J} e^{-\tau X_n},
\end{equation}
where we have carried out the trivial sum over $M$ and the
vanishing E-mode contribution is excluded.

Fig.~\ref{fig4} shows the numerical results of the sum up to
$X_{max}$=100, 150, 200, 250 for the 40 values of $\tau$ from
0.0025 to 0.1 with the step 0.0025. We can see that below $\tau <
0.06$ the convergence is poor. However, it is enough to see the
presence of an $1/\tau^2$ divergence. If we fit the data above
$\tau> 0.06$, we obtain
\begin{equation}
S(\tau) = \frac{0.1061}{\tau^2} - \frac{0.0816}{\tau} + 0.0478 -
0.0285\tau. \label{nf}\end{equation} Apart from a possible
logarithmic divergence, there are quadratic and linear divergences
as we set $\tau$ equal to zero. We shall remove these divergences
following a procedure commonly used in the Casimir
problem \cite{DivCas}. A caveat on this procedure will be highlighted
in the next chapter. Now if we neglect the logarithmic
divergence, the best way to get rid of the quadratic and linear
divergences is to evaluate
\begin{equation}
S(\tau)+2\tau S^\prime(\tau)+\frac12\tau^2 S^{\prime\prime}(\tau)
= \sum_{n,J} (2J+1) Q_{n,J} (1-2\tau X_n+0.5\tau^2 X_n^2) e^{-\tau
X_n}.
\end{equation}

\begin{figure}
\centerline{\epsfig{file=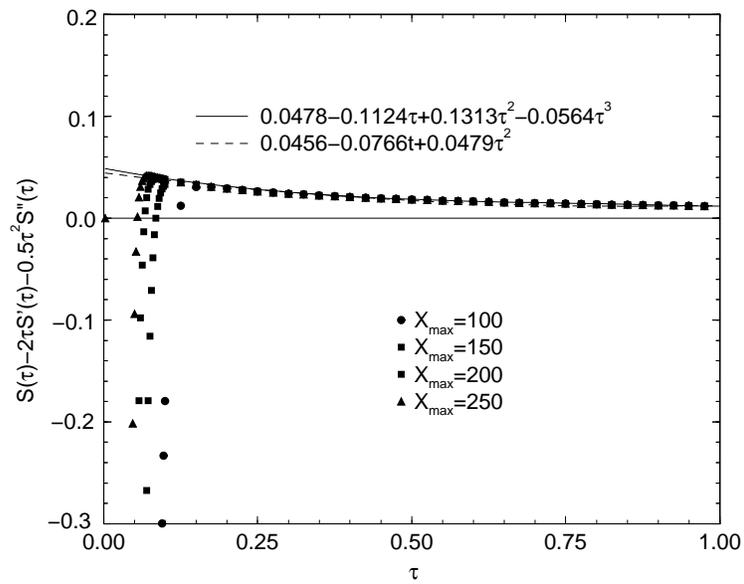, width=11cm}}
\caption{$S(\tau)-2\tau S^\prime(\tau)+\frac12\tau^2
S^{\prime\prime}(\tau)$ as a function of $\tau$. The finite term
of $S(\tau)$ is  extracted by fitting these quantities to a cubic
and quadratic curves.}\label{fig5}
\end{figure}

Fig.~\ref{fig5} show the results on this quantity for 80 values of
$\tau$ ranging from 0.0025 to 1. We see that no serious
divergences appear anymore. By fitting the convergent data with
the above expressions for $\tau$, we obtain for the finite part of
the sum 0.0478, from the cubic function fit, and 0.0456, from the
quadratic one. These results are comparable to the finite term of
the above naive fitting procedure (\ref{nf}), which yielded
0.0478.

Once we have the numerical value on the mode sum, the gluon vacuum
contribution to the FSAC can be evaluated simply as
\begin{equation}
a_{G,{\rm Cas}}^0=a_{G,vac}^0 =
-\frac{(2.10)^2}{2}\times\frac{8}{2} \times
\frac{y(R)}{122\mbox{MeV}} \times (0.0478),
\end{equation}
where $y(R)$ is related to $a^0_{B_Q}$ as
\begin{equation}
 y(R) = -\frac{(1+m_\eta R)}{[2(1+m_\eta R) + (m_\eta R)^2](m_\eta R)^2}
a^0_{B_Q}.
\end{equation}
We have used $N_F=N_c=3$, $\alpha_s=2.2$,
$f_0=\sqrt{N_F/2}f_{\eta^\prime} \sim 122$MeV and $m_\eta=958$
MeV.
\begin{figure}
\centerline{\epsfig{file=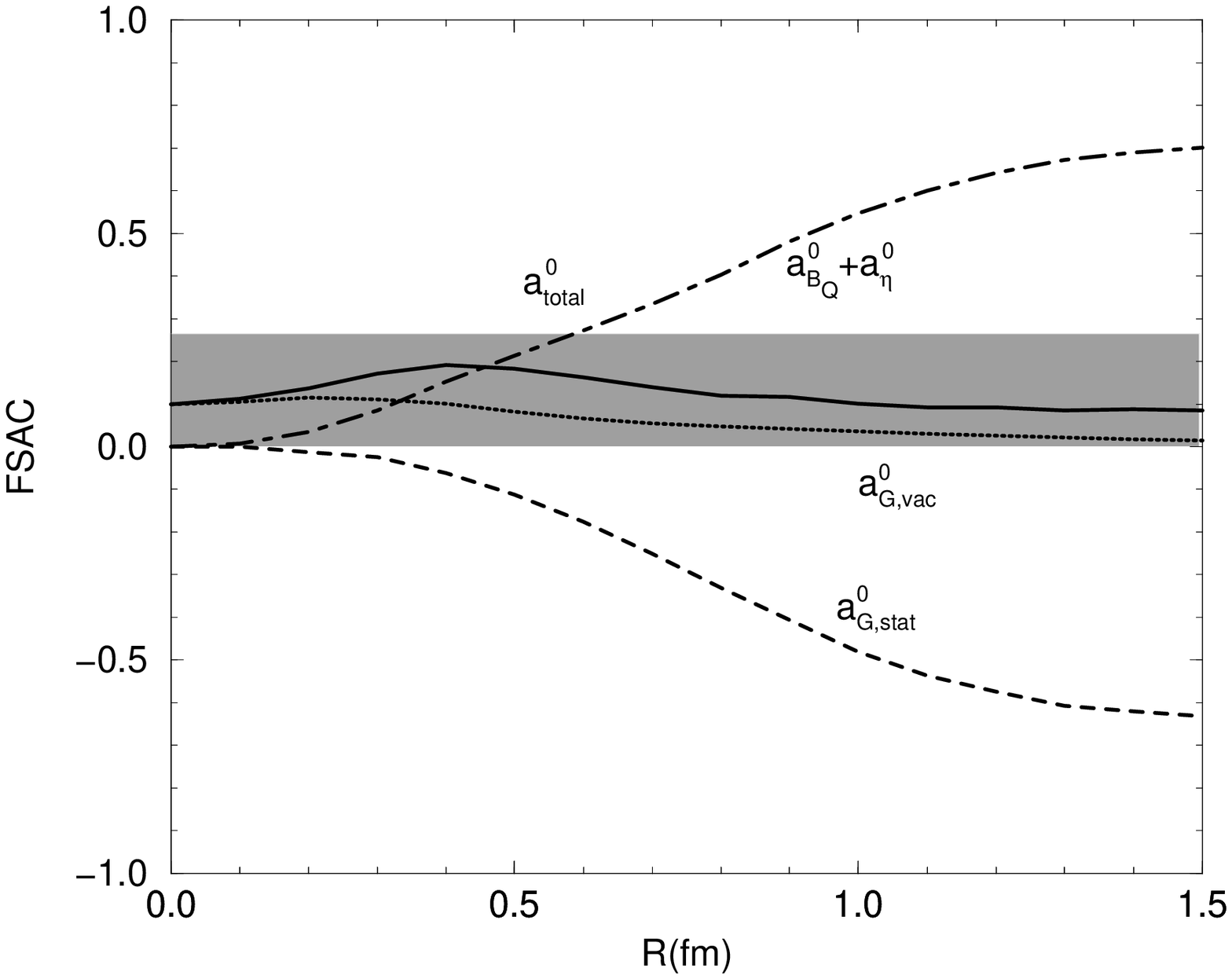, width=9cm}} \caption{Various
contributions to the flavor singlet axial current of the proton as
a function of bag radius and comparison with the experiment: (a)
quark plus $\eta$ contribution ($a^0_{B_Q} + a^0_\eta$), (b) the
contribution of the static gluons due to quark source
($a^0_{G,stat}$), (c) the gluon vacuum contribution
($a^0_{G,vac}$), and (d) their sum ($a^0_{total})$. The shaded
area corresponds to the range admitted by
experiments.}\label{result}
\end{figure}
Our numerical results are given in Fig.~\ref{result}. The quarkish
component of the FSAC is given by the sum of the quark and $\eta$
contribution, $a^0_{B_Q}+a^0_{\eta}$ and the gluonic component by
$a^0_{G,stat}+a^0_{G,vac}$. Both increase individually as the bag
radius $R$ is increased but the sum remains small,
$0<a^0_{total}<0.3$ for the whole range of radii.

\newpage

\begin{figure}
\caption{The moment of inertia associated with the collective
rotation as a function of the bag radius and the proton spin
fraction carried by each constituents. In the calculation, we have
used (a) the ``confined" color electric field with $Q_R(r)$ and
(b) the conventional one with $Q_0(r)$.}
\centerline{\epsfig{file=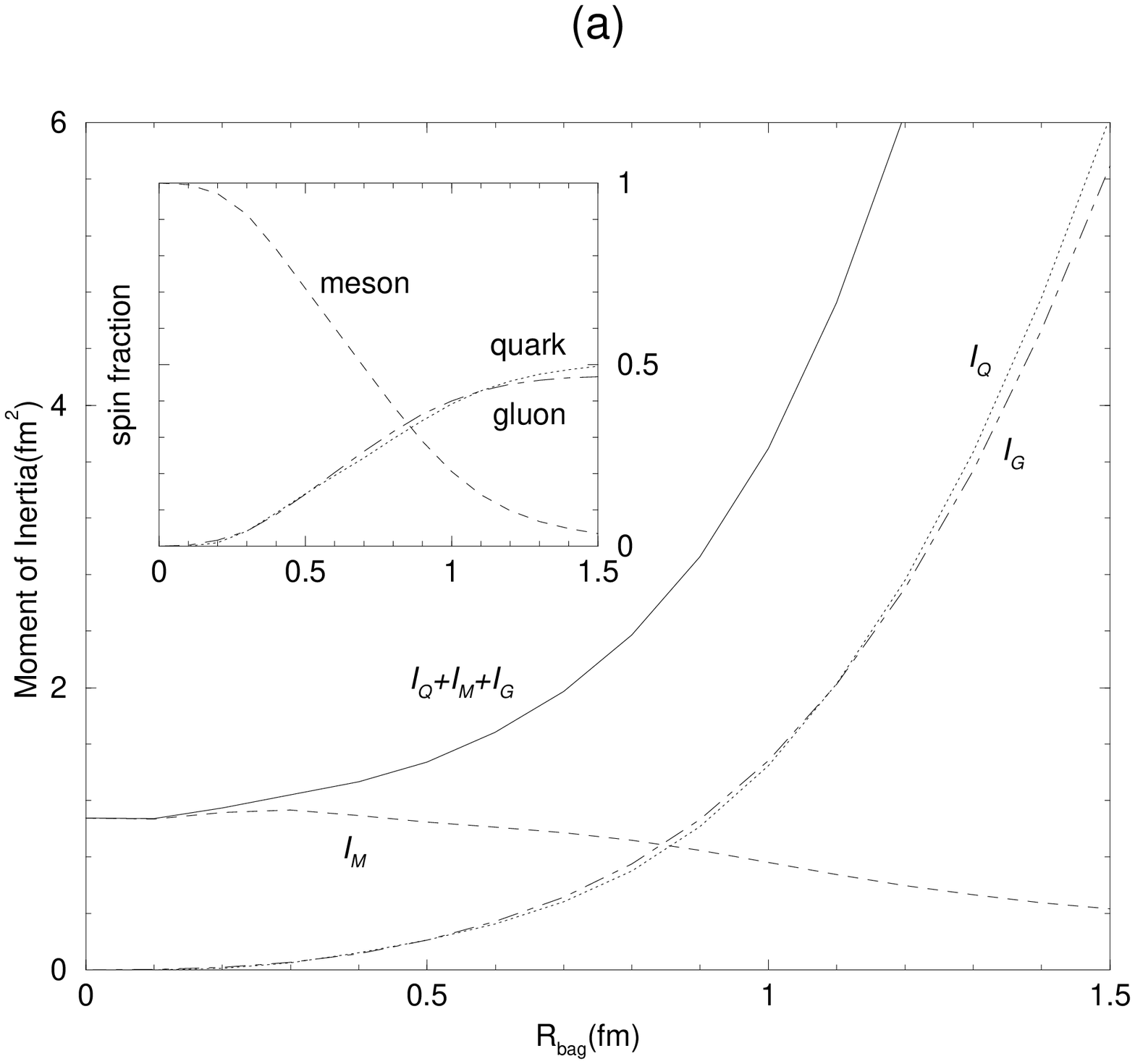, width=7.5cm, angle=270}
\epsfig{file=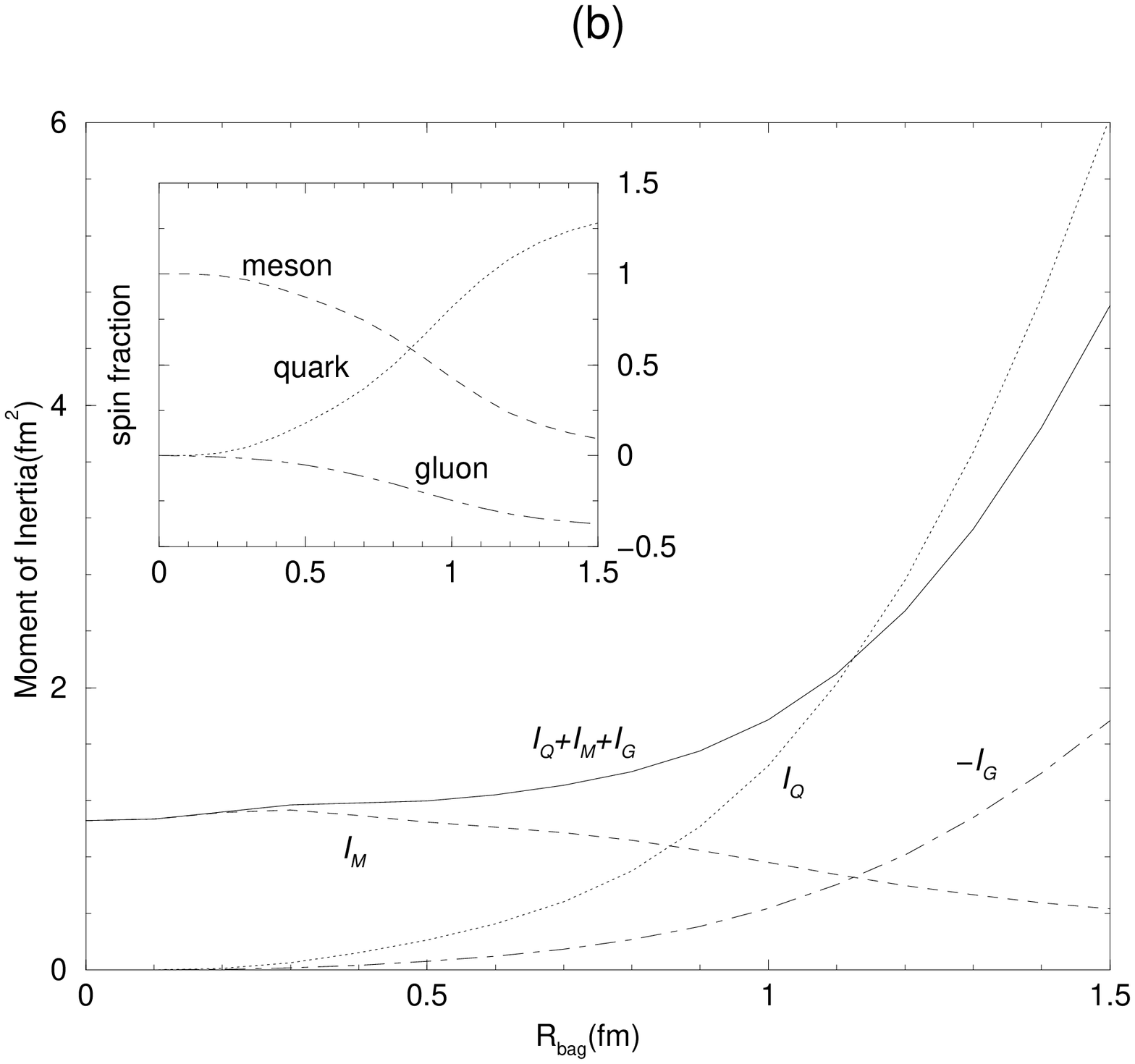, width=7.5cm, angle=270}}\label{fig7}
\end{figure}

\begin{figure}
\caption{The gluon spin $\Gamma$ as a function of the bag radius.
(a) and (b) are obtained with the color electric fields explained
in Fig.~\ref{fig7}.} \centerline{\epsfig{file=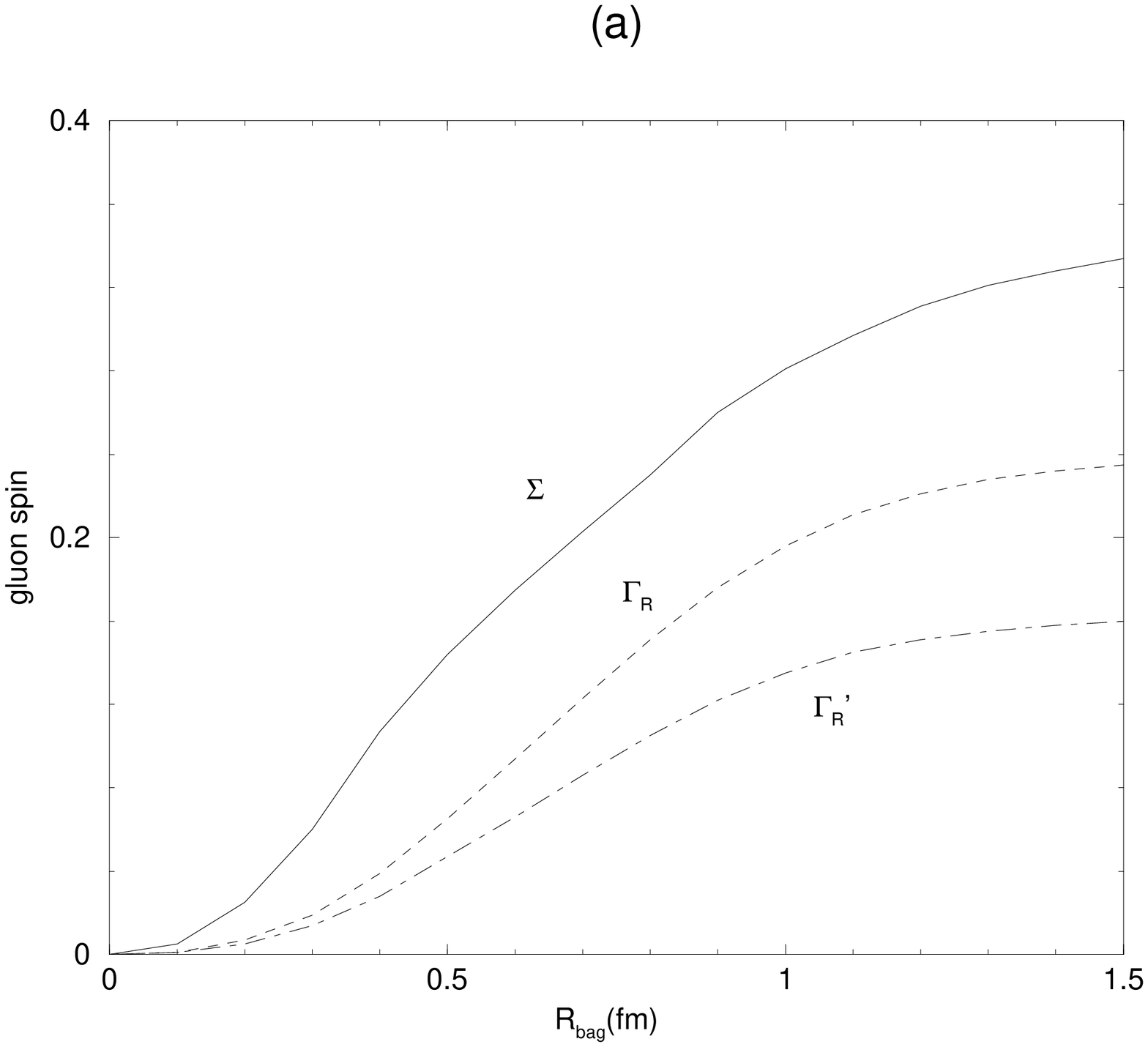,
width=7.5cm, angle=270} \epsfig{file=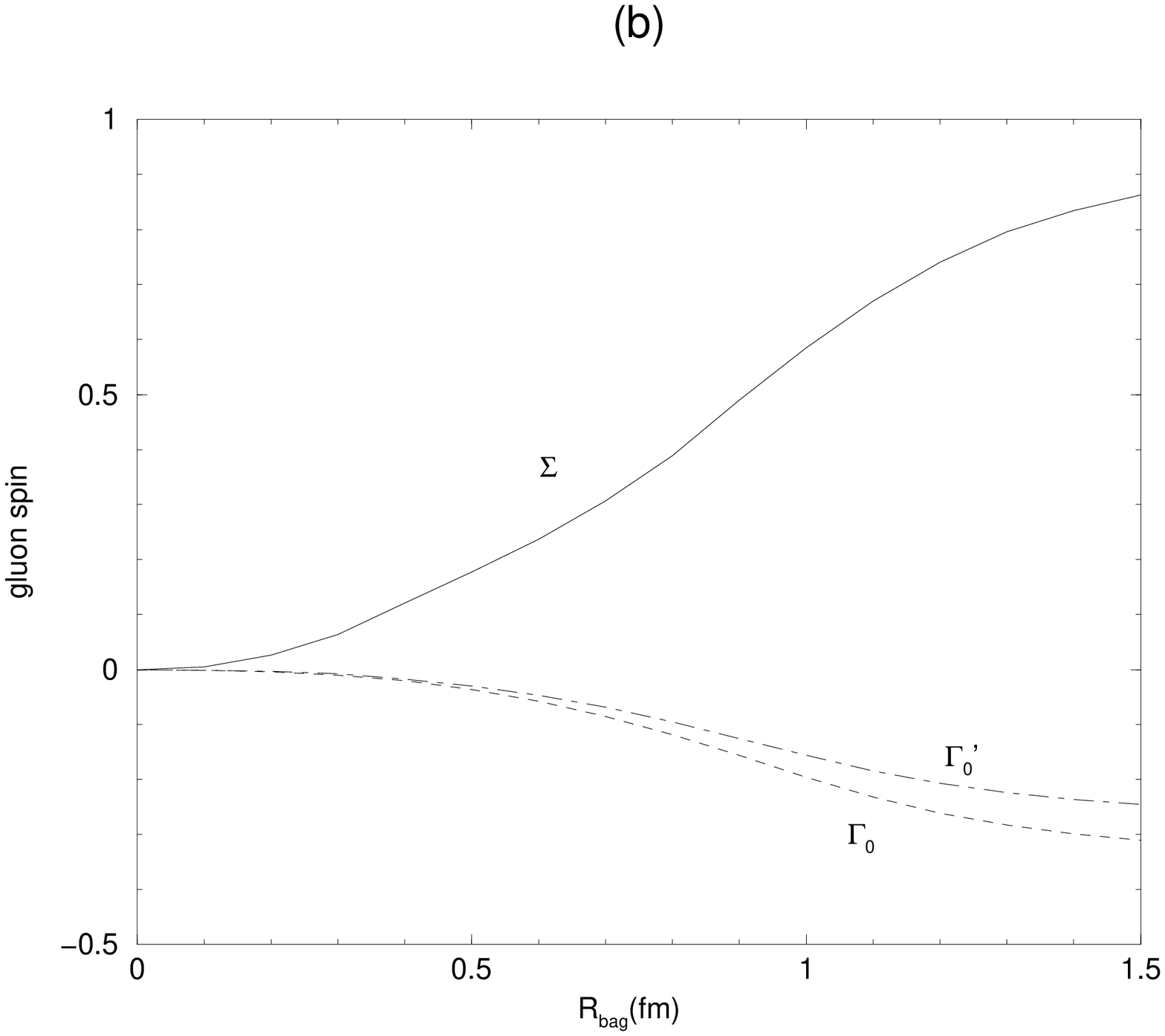, width=7.5cm,
angle=270}}\label{fig8}
\end{figure}

\begin{figure}
\caption{The flavor singlet axial current $a_0$ as a function of
the bag radius. (a) and (b) are obtained with the color electric
fields explained in Fig.~\ref{fig7}.}
\centerline{\epsfig{file=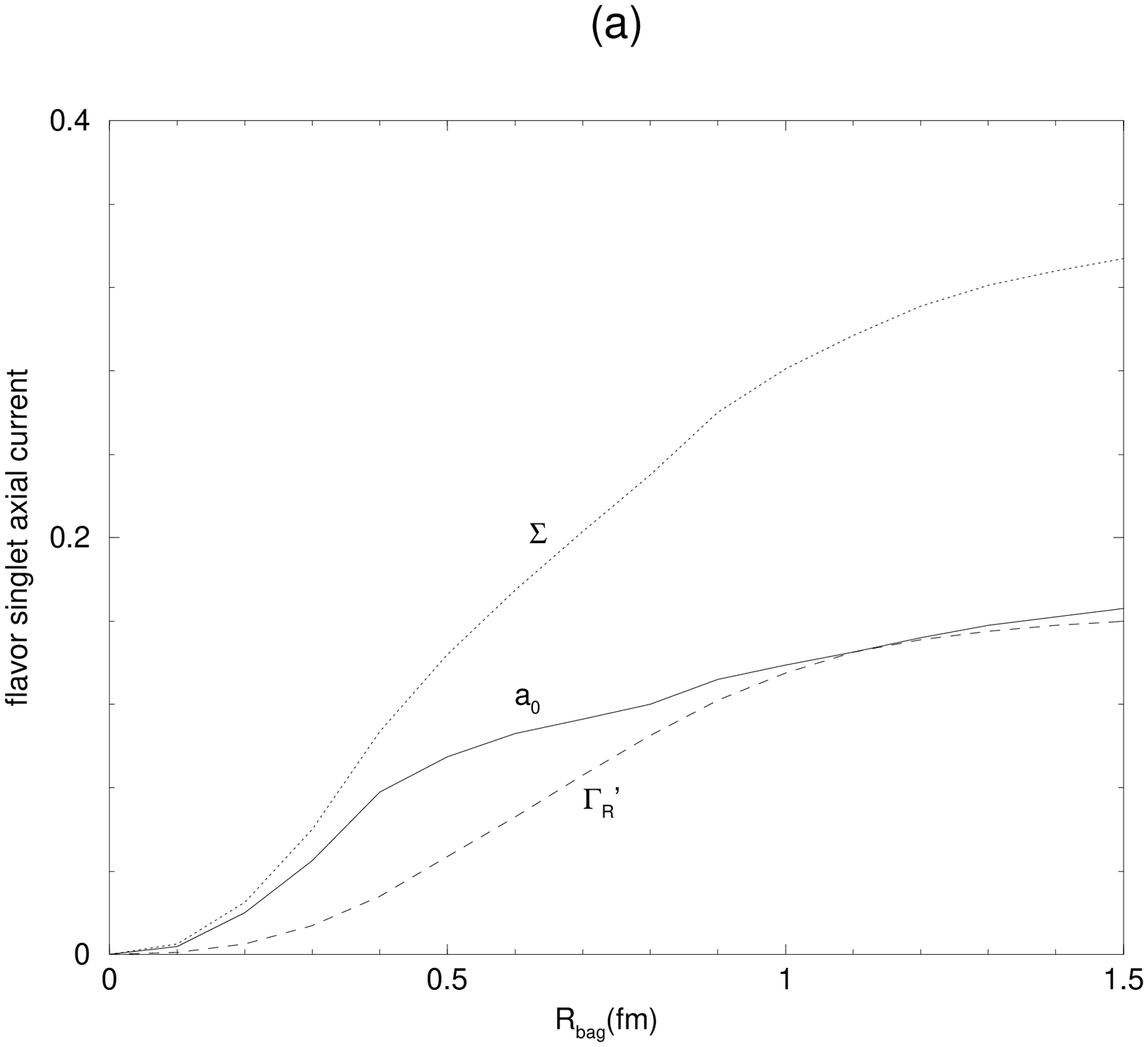, width=7.5cm, angle=270}
\epsfig{file=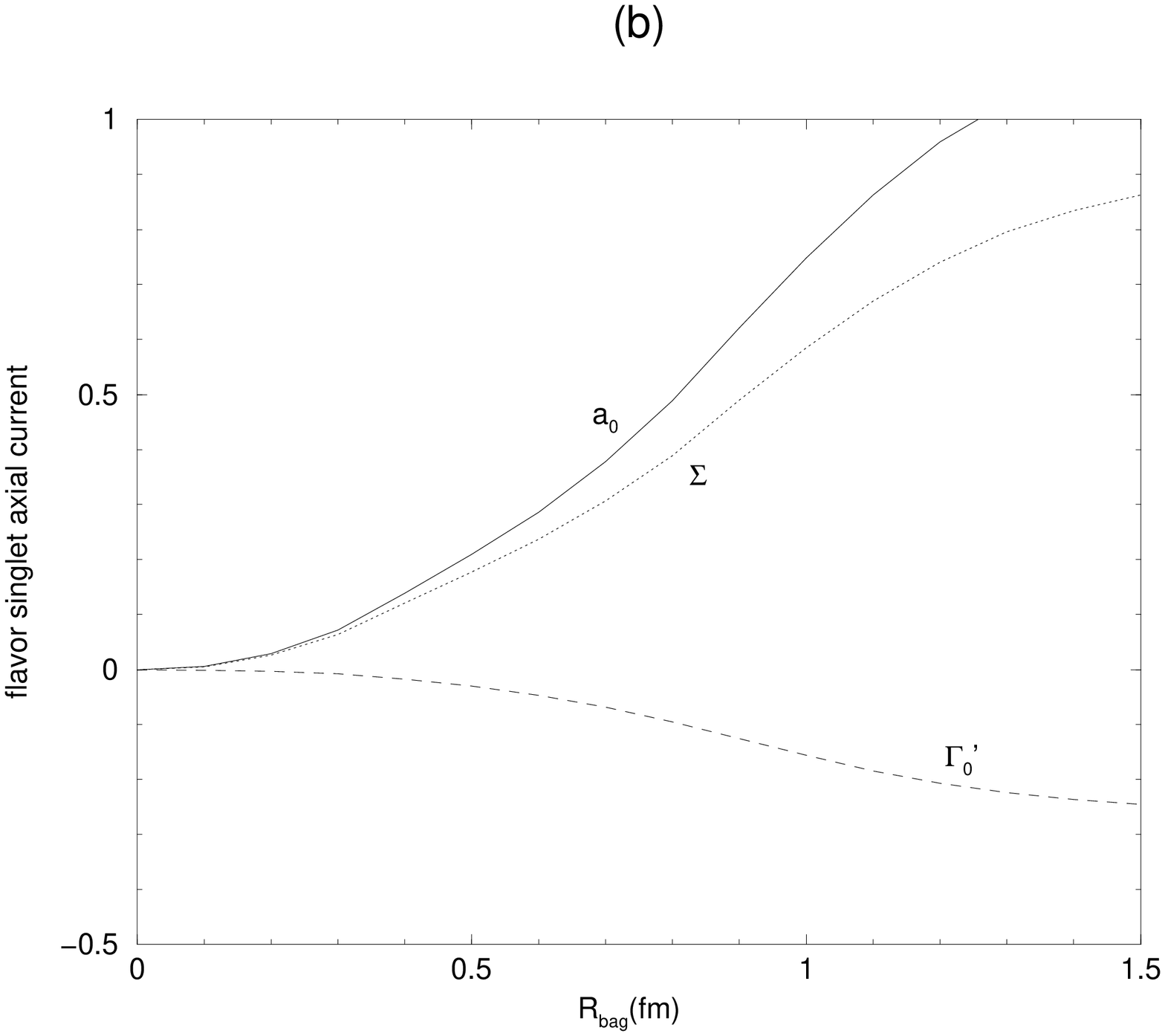, width=7.5cm, angle=270}}\label{fig9}
\end{figure}


\chapter{ Discussion}

This thesis deals with the description of hadronic phenomena from
the perspective of QCD and low energy hadronic effective theories.
The underlying goal has been the analysis of the realization of
the Cheshire Cat principle in a realistic 3+1 dimensional
scenario. We have reviewed in this thesis its formulation and have
studied an extremely subtle case, namely the FSAC, an observable
intimately related to the study of the spin of the proton, that
lends support to the notion of the Cheshire Cat Principle in QCD
governing strongly interacting systems. Before embarking onto the
discussion of the results, let us stress that the consequences of
quantum effects, through the chiral anomaly, the color anomaly and
the Casimir phenomena have played a major role in the successful
completion of this work.

In the previous chapter, we have considered  various contributions
to the FSAC of the proton. As presented in Fig. \ref{fig2}, the FSAC,
which arises from the quarks, the $\eta'$ meson and the (MIT or
monopole) static gluons, fails to fulfill the Cheshire Cat
Principle (CCP). For all cases the FSAC vanishes in  the
$R\rightarrow 0$ limit. The missing ingredient which is required
to restore the principle is the Casimir effect that accounts for
the fact that the modes are not free but strongly interacting via
the boundary conditions of the cavity. The calculated Casimir
contributions lead to a  nonzero value of the FSAC even for zero
bag radius.

Our numerical results are shown in Fig.~\ref{result}. Standard MIT
bag parameters were used for the calculation. The quarkish
component of the FSAC is given by the sum of the quark and $\eta$
contributions, $a^0_{B_Q}+a^0_\eta$, and the gluonic component by
$a^0_{G,stat}+a^0_{G,vac}$. Both contributions increase as the
confinement size $R$ is increased but their sum remains small,
$0<a^0_{total}<0.3$, for the whole range of radii, giving a value
consistent with the experiment,
$a^{exp}=a^0(\infty)=0.10^{+0.17}_{-0.10}$. It is remarkable that
$a(R=0)\simeq a(R\approx 1.5\ {\rm fm})$, while each component can
differ widely for the two extreme radii.

We have shown that the principal agent for the observed small FSAC
in the proton, in the framework of the chiral bag model, is the
CCP. It is the CCP that assures the cancellation between the
different contributions: the quarkish and the gluonic. Note that
the separation used by us is arbitrary and has no physical
meaning, only the sum is physical.

For a small bag radius, both components are small, so the small
net FSAC is inevitable. This is consistent with the observation
that in the limit as $R\rightarrow 0$ we recover the skyrmion
description, which gives vanishing FSAC at leading order, subject
to small modification by matter fields at higher order. For a
large bag radius -- a limit that corresponds to the MIT bag model
--  both the quarkish and the gluonic contributions are separately
large but they cancel each other. Our assertion is that this
cancellation is mandated by the CCP. We should recall again that
the separation between the quarkish component and the gluonic
component adopted in eqs.~(\ref{Dbag}) and (\ref{Dmeson}) is
entirely arbitrary although the sum is unique. {\it Whether the
different components by itself  are large or small has no physical
meaning. Only their  sum does.} Different separations would lead
to different scenarios leading to the same small value. These
different separations are analogs to gauge choices in gauge
theories as suggested by some authors (see, e.g., \cite{ccpph}).
It is tempting to speculate that in some limit, the FSAC is
exactly zero and the the small nonzero value corresponds to a
departure from this limit. Understanding this limit would allow a
unique separation of the components.

One of the principal results of this thesis is that {\it it is
possible to have a nonzero value for the FSAC at $R=0$ and it is
of the same size as that for large $R$.} The reason for this
nonzero value is intimately connected with the CCP, since it is
the finite part of the gluon mode sum which normalizes the value
of this contribution at the origin. Moreover, the color boundary
condition provides us with a decreasing $\eta'$ field contribution
which changes softly as a function of $R$. While the effect of the
surface color anomaly term is generally small for all radii, the
finite nonzero value of the FSAC for $R=0$ is assured by the
surface boundary term. Thus the violation of the CCP observed in
the previous calculations at $R=0$ \cite{pv} \cite{rv} is neatly
eliminated by the color anomaly boundary condition. More
importantly, the {\it monopole} structure of the color electric
field previously proposed is found to be required for the sign
that comes with the important static gluonic contribution from the
quark source. We believe that this cancellation is a manifestation
in the bag scenario of the recently discovered one for $QCD$
\cite{kochelev}. The MIT configuration would strongly violate the
CCP. We are thus led to the conclusion that the CCP {\it requires}
the monopole configuration for the color electric field. Whether
or not this configuration leaves undisturbed other -- successful
-- phenomenology was discussed in \cite{pv}.

In calculating the gluonic Casimir effect, we {\it made} the {\it
ab initio} assumption that the CCP holds, an assumption which is
expected to be valid for small bag radius. We then extended it, in
accordance with the CCP, to all bag radii. We can justify this
only \`a posteriori by showing that the CCP assumption is
consistent with what one obtains. Note, however, that the gluonic
Casimir effect is most significant for small $R$ where it is
needed for the CCP and plays little role for large $R$. Thus our
assumption is validated. It would of course be more satisfying if
one could obtain the CCP as an output of the formalism, not put in
as an input. To this end, we need to solve appropriate equations
of motion for gluons to the color boundary conditions
(\ref{colorBC}). Since the color anomaly boundary conditions are
generated by the quantum effect of quark, it is necessary to
consider an effective action for gluons which contains the quantum
effect resulting from integrating out quark fields. Although
gluons have been treated as abelianized (or Maxwell) fields in the
previous chapter, we should give special care in the
abelianization because of the nontrivial topological structure of
the QCD~\cite{cho}. Therefore, we need to construct an effective
action for the bagged QCD and see there are additional corrections
besides the Maxwell terms in the effective action.

We should mention a caveat left unspecified in Chapter 5 in
regularizing this Casimir contribution. Since $a_{G, vac}^0$
vanishes when $\eta'$ field is removed, the so-called ``vacuum
contribution" is duly subtracted in what we have computed.
However, we have also explicitly subtracted quadratic and linear
divergences appearing from the mode sum by resorting to a
procedure used in most of Casimir-type calculation
\cite{schwinger}, which, as far as we know, is physically
reasonable but has not yet been rigorously justified from the
first principles~\footnote{It is found that the specific structure
of the divergent terms depends on regularization schemes whereas
that of the finite terms does not~\cite{Ravndal00}.}. The same
caveat applies to our calculation as it does to others. The finite
term we have obtained might therefore be subject to additional
finite corrections by procedures in the renormalization.

We should also stress that our result is at best qualitative. A
better treatment (such as a more realistic gauge coupling constant
running with the bag size, a more accurate calculation of
$A^{vac}_{G}$, etc...) might modify  our results quantitatively. Even
so, we believe it to be quite robust that the overall FSAC is
small, $\lsim 0.3$ and that it is more or less independent of the
confinement size.



\appendix
\addcontentsline{toc}{chapter}{Appendices}

\chapter{Angular momentum basis of the wave functions for the strange
and the hedgehog quarks}

In this appendix, we present explicit
expressions of the angular momentum basis for the strange quark ({\it
s}-quark) wave functions, $|j,m_j\rangle_{\kappa}$, which are
eigenstates of ${\bf J}^2$ and $J_z$, and those for the hedgehog
quark ({\it h}-quark) wave functions, $|K,m_K\rangle_i$, which are
eigenstates of ${\bf K}^2$ and $K_z$. Here $\kappa=\pm1$ and $i$
runs over $1,2,3,4.$

The basis $|j,m_j\rangle_{\kappa}$ can be constructed by combining
the orbital angular momentum basis $|l,m_j\rangle$ and the spin
basis $|\frac{1}{2},m_s\rangle_s$, which are eigenstates of the
orbital angular momentum operators ${\bf L}^2$ and $L_z$ and the
spin operators ${\bf S}^2$ and $S_z$, respectively. According to
the angular momentum sum rule, there are two types.

i) $j=l+\frac{1}{2}$ and $\kappa=1$:

\begin{eqnarray}
|j,m_j\rangle_{1}&=&\bigg(\frac{j+m_j}{2j}\bigg)^{\frac{1}{2}}
|j,m_j-1/2\rangle|1/2,1/2\rangle_s\nonumber\\
&&+\bigg(\frac{j-m_j}{2j}\bigg)^{\frac{1}{2}}|j,m_j-1/2\rangle|1/2,-1/2
\rangle_s,
\end{eqnarray}

ii) $j=l-\frac{1}{2}$ and $\kappa=-1$:

\begin{eqnarray}
|j,m_j\rangle_{-1}&=&-\bigg(\frac{j-m_j+1}{2j+2}\bigg)^{\frac{1}{2}}
|j,m_j-1/2\rangle|1/2,1/2\rangle_s\nonumber\\
&&+\bigg(\frac{j+m_j+1}{2j+2}\bigg)^{\frac{1}{2}}|j,m_j-1/2\rangle|1/2,-1/2
\rangle_s.
\end{eqnarray}
The conventions of Edmonds~\cite{edmonds} for the Clebsch-Gordan
coefficients have been used. One can see that these two type of
solutions have opposite parity for a given $j$. \vskip 1ex

Using the following identities \cite{arfken},
\begin{eqnarray}
&&\cos\theta\
Y_l^m=\bigg[\frac{l-m+1}{2l+1}\cdot\frac{l+m+1}{2l+3}\bigg]^{\frac{1}{2}}Y_{l+1}^m
+\bigg[\frac{l-m}{2l-1}\cdot\frac{l+m}{2l+1}\bigg]^{\frac{1}{2}}Y_{l-1}^m,\\
&&\sin\theta \ e^{i\phi} Y_{l}^m
=-\bigg[\frac{l+m+1}{2l+1}\cdot\frac{l+m+2}{2l+3}\bigg]^{\frac{1}{2}}Y_{l+1}^{m+1}
+\bigg[\frac{l-m}{2l-1}\cdot\frac{l-m-1}{2l+1}\bigg]^{\frac{1}{2}}Y_{l-1}^{m+1}\nonumber\\
&&\sin\theta \ e^{-i\phi} Y_{l}^m
=\bigg[\frac{l-m+1}{2l+1}\cdot\frac{l-m+2}{2l+3}\bigg]^{\frac{1}{2}}Y_{l+1}^{m-1}
+\bigg[\frac{l+m}{2l-1}\cdot\frac{l+m-1}{2l+1}\bigg]^{\frac{1}{2}}Y_{l-1}^{m-1},\nonumber
\label{arf}
\end{eqnarray}
where $Y_l^m$ is the spherical harmonics, one can show that
the above states satisfy,

\begin{equation}
\mbox{\boldmath $\sigma$}\cdot\hat{\bf r}\ |j,m_j\rangle_{\pm}=-
|j,m_j\rangle_{\mp}.\label{sig}
\end{equation}
\vskip 1ex

The basis for the hedgehog quark, $|K, m_K\rangle_i$, can be
obtained by combining the total angular momentum ${\bf J}$ and the
isospin ${\bf I}$. In this case there are four types  of
states given by,\vskip 1ex

i) ${\rm i}=1\; \bigg(j=l+\frac{1}{2},\ K=j-\frac{1}{2}\bigg),$

\begin{eqnarray}
|K,m_K\rangle_1&=&-\bigg(\frac{K-m_K+1}{2K+2}\bigg)^{\frac{1}{2}}
|j,m_K-1/2\rangle_1|1/2,+1/2\rangle_t\nonumber\\
&&+\bigg(\frac{K+m_K+1}{2K+2}\bigg)^{\frac{1}{2}}
|j,m_K+1/2\rangle_1|1/2,-1/2\rangle_t,
\end{eqnarray}

ii) ${\rm i}=2\; \bigg(j=l-\frac{1}{2},\ K=j+\frac{1}{2}\bigg),$

\begin{eqnarray}
|K,m_K\rangle_2&=&\bigg(\frac{K+m_K}{2K}\bigg)^{\frac{1}{2}}
|j,m_K-1/2\rangle_{-1}|1/2,+1/2\rangle_t\nonumber\\
&&+\bigg(\frac{K-m_K}{2K}\bigg)^{\frac{1}{2}}
|j,m_K+1/2\rangle_{-1}|1/2,-1/2\rangle_t,
\end{eqnarray}

iii) ${\rm i}=3\; \bigg(j=l-\frac{1}{2},\ K=j-\frac{1}{2}\bigg),$

\begin{eqnarray}
|K,m_K\rangle_3&=&-\bigg(\frac{K-m_K+1}{2K+2}\bigg)^{\frac{1}{2}}
|j,m_K-1/2\rangle_{-1}|1/2,+1/2\rangle_t\nonumber\\
&&+\bigg(\frac{K+m_K+1}{2K+2}\bigg)^{\frac{1}{2}}
|j,m_K+1/2\rangle_{-1}|1/2,-1/2\rangle_t,
\end{eqnarray}

iv) ${\rm i}=4\; \bigg(j=l+\frac{1}{2},\ K=j-\frac{1}{2}\bigg),$

\begin{eqnarray}
|K,m_K\rangle_4&=&\bigg(\frac{K+m_K}{2K}\bigg)^{\frac{1}{2}}
|j,m_K-1/2\rangle_{1}|1/2,+1/2\rangle_t\nonumber\\
&&+\bigg(\frac{K-m_K}{2K}\bigg)^{\frac{1}{2}}
|j,m_K+1/2\rangle_{1}|1/2,-1/2\rangle_t,
\end{eqnarray}

\noindent where the subscript $t$ has been used to label states in the
isospace.\vskip 1ex

Since $K=l$ for $i=1,2$ and $K=l-1$ for $i=3,4$, one
can check that $|K,m_K\rangle_1$ and $|K,m_K\rangle_2$ have
parity $(-1)^K$, whereas $|K,m_K\rangle_3$ and $|K,m_K\rangle_4$
have parity $-(-1)^K$.

Using the identity eq.~(\ref{arf}), one can show that the states
$|K,m_K\rangle_i$ satisfy the relation
\begin{equation}
\mbox{\boldmath $\sigma$}\cdot\hat{\bf r}\ |K,m_K\rangle_i=-|K,m_K\rangle_{i+2}.
\end{equation}

Besides, applying the operator $\mbox{\boldmath $\tau$}\cdot\hat{\bf r}$ to
$|K,m_K\rangle_i$ where $\mbox{\boldmath $\tau$}$ are the Pauli matrices in
isospace, we have the following relations;
\begin{eqnarray}
&&\mbox{\boldmath $\tau$}\cdot\hat{\bf r}\
|K,m_K\rangle_1=\frac{2j-2K}{2K+1}
\ |K,m_K\rangle_3-2\frac{\sqrt{K(K+1)}}{2K+1}\ |K,m_K\rangle_4,\\
&&\mbox{\boldmath $\tau$}\cdot\hat{\bf r}\
|K,m_K\rangle_2=\frac{2j-2K}{2K+1}
\ |K,m_K\rangle_4-2\frac{\sqrt{K(K+1)}}{2K+1}\ |K,m_K\rangle_3,\\
&&\mbox{\boldmath $\tau$}\cdot\hat{\bf r}\
|K,m_K\rangle_3=\frac{2j-2K}{2K+1}
\ |K,m_K\rangle_1-2\frac{\sqrt{K(K+1)}}{2K+1}\ |K,m_K\rangle_2,\\
&&\mbox{\boldmath $\tau$}\cdot\hat{\bf r}\
|K,m_K\rangle_4=\frac{2j-2K}{2K+1} \
|K,m_K\rangle_2-2\frac{\sqrt{K(K+1)}}{2K+1}\ |K,m_K\rangle_1,
\end{eqnarray}
where we have used the fact that $K$ in $|K,m_K\rangle_{1,3}$ is
$j-{1\over 2}$ and $K$ in $|K,m_K\rangle_{2,4}$ is
$j+\frac{1}{2}$. These relations have been used to get the energy
levels for the hedgehog quarks.

\chapter{Proof that the sum over E-modes
is zero}

We have shown that the contribution of E-modes to the Casimir
effect of the FSAC vanishes  numerically  in Chapter 5. In this
appendix, we show an analytical proof of this
result~\cite{kirsten}. Let's consider $I(x)$ given in Chapter 5;
\beq I(x)=\int_0^x dy \,\, y^3 \left[j_J^2 (y) - \frac J {2J+1}
j_{J+1}^2 (y) -\frac{J+1}{2J+1} j_{J-1}^2 (y) \right], \nn \eeq
and the boundary condition for the E-modes
\begin{equation}
j_J(X_n)=0,
\end{equation}
where $X_n=\omega_n R$. To prove $I(X_n)=0$, it is convenient to
use formulae of the Bessel functions rather than the spherical
Bessel functions. From the definition \beq j_l (x) =
\sqrt{\frac{\pi}{2x}} J_{l+1/2} (x), \nn \eeq we can rewrite
$I(x)$ in terms of the Bessel functions \beq I(x) = \frac{\pi} 2
\int_0^x dy\,\, y^2 \left[J^2_{J+1/2} (y) - \frac J {2J+1}
J^2_{J+3/2} (y) -\frac{J+1}{2J+1} J^2_{J-1/2} (y) \right],
\label{1} \eeq and the boundary condition \beq J_{J+1/2} (X_n) =0.
\label{2} \eeq Using the following formula by Schafheitlin
\cite{wats44b};
\beq \lefteqn{ (\mu +2) \int_0^z dx \,\, x^{\mu
+2} J_\nu^2 (x) =
   (\mu +1) \left(\nu^2 -\frac{(\mu +1)^2} 4 \right)
       \int_0^z dx\,\, x^\mu J_\nu ^2 (x) }\label{3}\\
   & &+\frac 1 2 \left[ z^{\mu +1} \left(z J_\nu ' (z) -\frac 1 2
(\mu +1) J_\nu (z) \right)^2 +z^{\mu +1} \left( z^2 -\nu^2
 +\frac 1 4 (\mu +1) ^2\right) J_\nu ^2 (z) \right] .\nn
\eeq we can arrange $I(x)$ in terms of Bessel functions, its
derivatives and simpler integrals with a lower power in $y$.

In the next step, we use the recursion relations of the Bessel
functions in order to write $I(x)$ in terms of the Bessel
functions with an index $J+\frac{1}{2}$. In addition, to get more
useful expressions, we decompose $I(x)$ into the three terms
\beq
I_1 (x) &=& \frac \pi 2 \int_0^x dy \,\, y ^2 J_{J+1/2} ^2 (y) \label{4}\\
I_2 (x) &=& -\frac \pi 2 \frac J {2J+1}
       \int_0^x dy \,\, y^2 J_{J+3/2} ^2 (y) \label{5}\\
I_3 (x) &=& -\frac \pi 2 \frac{J+1}{2J+1} \int_0^x dy \,\, y^2
      J_{J-1/2} ^2 (y) .\label{6}
\eeq Using Schafheitlin's formula (\ref{3}) and eq.~(\ref{2}), we
have immediately \beq I_1 (X_n) = \frac \pi 4 \left\{ \frac 1 2
X_n^3 J_{J+1/2}^{\prime 2} (X_n) + J(J+1)
     \int_0^{X_n} dy \,\, J_{J+1/2}^2 (y) \right\}\label{7}
\eeq The second expression has a more complicated form in
terms of Bessel functions, i.e.

\beq I_2 (X_n) &=& -\frac
\pi 4 \frac J {2J+1} \left\{\right.
     [J^2+3J +2] \int_0^{X_n} dy \,\, J_{J+3/2} ^2 (y) \nn\\
        & &+\frac 1 2 X_n^3 J_{J+3/2}^{\prime 2}
(X_n) -\frac 1 2 X_n^2 J_{J+3/2}'
    (X_n) J_{J+3/2} (X_n) \nn\\
  & &    \left. +\frac 1  2 X_n \left[ X_n^2 -J^2 -3J -\frac 7 4 \right]
          J_{J+3/2}^2 (X_n) \right\} .\nn
\eeq From the recursion relation of the Bessel functions
\cite{grad65b}, one gets \beq J_{J+3/2} (y) = -J'_{J+1/2}(y) +
\frac {J+1/2} y J_{J+1/2} (y) ,\nn \eeq which yields by the
boundary condition eq.~(\ref{2}) \beq J_{J+3/2} (X_n) = -J'
_{J+1/2} (X_n).\nn \eeq Similarly \beq J_{J+3/2}' (y) =
-\frac{J+3/2} y J_{J+3/2} (y) +J_{J+1/2} (y) ,\nn \eeq which means
\beq J_{J+3/2} ' (X_n) = \frac{J+3/2} {X_n} J_{J+1/2} ' (X_n) \nn
\eeq So finally the following form for $I_2(X_n)$ is obtained,
\beq \lefteqn{ I_2 (X_n) = -\frac \pi 4 \frac J {2J+1} \left\{
      J_{J+1/2}^{\prime 2} (X_n) \left[\frac 1 2 X_n^3 +\frac 1 2 X_n J +X_n
         \right]\right.} \label{8}\\
 & & +[J^2+3J +2] \int_0^{X_n}dy \,\,
  \left[ J_{J+1/2}^{\prime 2} (y) -\frac{2J+1} y J_{J+1/2} (y)
      J_{J+1/2} ' (y)\right. \nn\\
   & &\left.\left.\hspace{4cm}
            +\frac{(J+1/2)^2} {y^2} J_{J+1/2} ^2 (y) \right]\right\}.   \nn
\eeq For $I_3$ an intermediate result after applying
Schafheitlin's reduction formula (\ref{2}) reads, \beq I_3(X_n)
&=& -\frac \pi 4 \frac{J+1}{2J+1} \left\{ [J^2-J]
               \int_0^{X_n} dy \,\, J_{J-1/2} ^2 (y) \right.\nn\\
      & & +\frac 1 2 X_n^3 J_{J-1/2}^{\prime 2} (X_n) -\frac 1 2 X_n^2
      J_{J-1/2} ' (X_n) J_{J-1/2} (X_n) +\frac 1 8 X_n J_{J-1/2} ^2 (X_n)
           \nn\\
      & &\left.+\frac 1 2 X_n (X_n^2 -J^2 +J ) J_{J-1/2} ^2 (X_n) \right\} .
 \nn
\eeq Using this time \beq
J_{J-1/2} (X_n) &=& J_{J+1/2} ' (X_n) \nn\\
J_{J-1/2} ' (X_n) &=& \frac{J-1/2} {X_n} J_{J+1/2} ' (X_n) \nn
\eeq it can be rewritten in the form \beq I_3 (X_n) &=& -\frac
\pi 4 \frac{J+1}{2J+1} \left\{
    {J_{J+1/2} '}^2 (X_n) \left[ \frac 1 2 X_n^3 -\frac 1 2 X_n J +\frac 1 2
          X_n \right] \right.\label{9}\\
  & & + [J^2-J] \int_0^{X_n} dy \,\,
     \left[ J_{J+1/2}^{\prime 2}
    (y) + \frac{2J+1} y J_{J+1/2} ' (y)\jj (y) \right.\nn\\
   & &  \left. \left.
          \hspace{4cm} + \frac{(J+1/2)^2}{y^2} \jj^2 (y)
             \right] \right\}.\nn
\eeq Before actually adding up all three pieces, it is better to
have a further consideration. The integrals involve three
different types of terms, namely $\jj ^{\prime 2}$, $\jj \jj '$
and $\jj ^2$. However, these are not independent from each other
due to the differential equation they fulfill. The differential
equation for the Bessel functions reads \beq \frac{d^2 \jj (y)
}{dy^2} +\frac 1 y \frac{d\jj (y)}{dy } +
 \left( 1-\frac{(J+1/2)^2}{y^2}\right) \jj (y)=0, \nn
\eeq and therefore we have \beq \int_0^{X_n} dy \,\, \jj ^{\prime
2} (y) &=& -\int_0^{X_n}dy\,\,
            J_{J+1/2} (y)
          \frac {d^2}{dy^2} \jj (y) \nn\\
 &=& \int_0^{X_n} \left\{ \frac 1 y \jj (y) \jj ' (y)
         +\left[ 1-\frac{(J+1/2)^2}{y^2}\right] \jj ^2 (y) \right\} .\nn
\eeq Using this identity, $I_2(X_n), I_3(X_n)$
simplify to, \beq I_2( X_n)&=& -\frac \pi 4 \frac J {2J+1} \left\{
            \jj ^{\prime 2} (X_n) \left[\frac 1 2 X_n^3 +\frac 1 2 X_n J
       +X_n \right]\right. \label{10} \\
      & &\left.+[J^2 +3J +2] \int_0^{X_n} dy \left[ \jj ^2 (y) -
       \frac {2J} y \jj (y) \jj ' (y) \right] \right\}  \nn\\
I_3 (X_n)&=& -\frac \pi 4 \frac{J+1}{2J+1} \left\{
            \jj ^{\prime 2} (X_n) \left[\frac 1 2 X_n^3 -\frac 1 2 X_n J
       +\frac 1 2 X_n\right]\right. \label{11} \\
      & &\left.+[J^2 -J] \int_0^{X_n} dy \left[ \jj ^2 (y)
      +\frac {2J+2}{y }\jj (y) \jj ' (y) \right] \right\}.  \nn
\eeq Collecting the results, we have \beq I(X_n) &=& \frac \pi 4
\left\{ 2J(J+1)
    \int_0^{X_n} dy \,\, \frac 1 y \jj (y) \jj ' (y)
            -\frac {X_n} 2 \jj ^{\prime 2} (X_n) \right\}.\nn
\eeq

In the final step, we use the recursion relation again to rewrite
the derivative of the Bessel function in the first term of the
equation above in terms of the Bessel function as, \beq \jj ' (y)
= \frac 1 2 \left( J_{J-1/2} (y) -J_{J+3/2} (y) \right), \nn \eeq
and the integral formula \cite{grad65b}
\begin{equation}
\int dy \frac{1}{y}J_p(\alpha y)J_q(\alpha y)=\alpha y
\frac{J_{p-1}(\alpha y)J_q(\alpha y)-J_{p}(\alpha y)J_{q-1}(\alpha
y)}{p^2-q^2}-\frac{J_{p}(\alpha y)J_q(\alpha y)}{p+q}.
\end{equation} Then, the first term becomes
 \beq \int_0^{X_n} dy \,\,
\frac 1 y \jj (y) \jj ' (y) = \frac 1 {4J(J+1)} X_n
            \jj ^{\prime 2} (X_n) ,\nn
\eeq and can be cancelled by the second term. Therefore, we have
the final result \beq I(X_n) = 0.\label{12} \eeq


\addcontentsline{toc}{chapter}{Acknowledgments}
\chapter*{Acknowledgments} First of all, I wish to tender my
gratitude with my whole heart to Prof. Dong-Pil Min and Prof.
Mannque Rho for their guidance, suggestions, encouragement, and
helps throughout my graduate years. I would like to thank Prof.
Vicente Vento for his concerns. How can I forget many discussions
with him, his kindness during the stay in Spain, and suggestions
on this manuscript! And I also wish to thank Prof. Byung-Yoon Park
for his comments and helps during the collaborations.

I wish to give my thank to Prof. Yongmin Cho and Prof. Hee Sung Song for
their critiques and suggestions on this manuscript.

I feel grateful to members of Nuclear Theory Group in Seoul
National University, Gwan Soo Shin, Joon-Il Kim, Byoung-Uk Sohn,
and Young-Ho Song. We had many useful discussions on the
contemporary Nuclear Physics. My colleagues in room 27-318A and
316, Zaeyoung Ghim, Bumseok Kyae, Jung-Tay Yee, Ishwaree Neupane,
Seong Chan Park, Dong-Won Jung, Won Suk Bae, O-Kab Kwon, Chaehyun
Yu, Ju-Ho Kim, Jong Dae Park, Sangheon Yi, and Hyun Min Lee, gave
me much information and pleasure in my life. My thanks also goes
to them.

I indebted to my family for their love and prayer in silence during graduate
years. I dedicate my thesis to my parents, specially to my father in Heaven.

\end{document}